\documentclass[twocolumn,aps,prd,longbibliography]{revtex4-2}
\usepackage{amsmath}
\usepackage{amssymb}
\usepackage{graphicx}
\usepackage{dcolumn}
\usepackage{bm}
\usepackage{bbold}
\usepackage{simplewick}
\usepackage{braket}
\usepackage[autostyle]{csquotes}
\usepackage{upgreek}
\usepackage{hyperref}

\begin{document}


\title{Local pairing of Feynman histories in many-body Floquet models}

\author{S. J. Garratt}
\author{J. T. Chalker}
\affiliation{Theoretical Physics, University of Oxford, Parks Road, Oxford OX1 3PU, United Kingdom}

\date{\today}

\begin{abstract}
We study many-body quantum dynamics using Floquet quantum circuits in one space dimension as simple examples of systems with local interactions that support ergodic phases. Physical properties can be expressed in terms of multiple sums over Feynman histories, which for these models are paths or many-body orbits in Fock space. A natural simplification of such sums is the diagonal approximation, where the only terms that are retained are ones in which each path is paired with a partner that carries the complex conjugate weight. We identify the regime in which the diagonal approximation holds, and the nature of the leading corrections to it. We focus on the behaviour of the spectral form factor (SFF) and of matrix elements of local operators, averaged over an ensemble of random circuits, making comparisons with the predictions of random matrix theory (RMT) and the eigenstate thermalisation hypothesis (ETH). We show that properties are dominated at long times by contributions to orbit sums in which each orbit is paired locally with a conjugate, as in the diagonal approximation, but that in large systems these contributions consist of many spatial domains, with distinct local pairings in neighbouring domains. The existence of these domains is reflected in deviations of the SFF from RMT predictions, and of matrix element correlations from ETH predictions; deviations of both kinds diverge with system size. We demonstrate that our physical picture of orbit-pairing domains has a precise correspondence in the spectral properties of a transfer matrix that acts in the space direction to generate the ensemble-averaged SFF. In addition, we find that domains of a second type control non-Gaussian fluctuations of the SFF. These domains are separated by walls which are related to the entanglement membrane, known to characterise the scrambling of quantum information.
\end{abstract}

\maketitle

\section{Introduction}

This paper is concerned with understanding generic features of spectral and eigenstate correlations for a class of ergodic many-body systems. Random matrix theory (RMT) and the eigenstate thermalisation hypothesis (ETH) provide a baseline description of these features \cite{mehta2004random,haake1991quantum,gutzwiller1990chaos,bohigas1984characterization,polkovnikov2011colloquium,dallesio2016from,deutsch2018eigenstate,deutsch1991quantum,srednicki1994chaos,rigol2008thermalization}.
Our objective is to identify when these approximations are accurate, and to characterise the behaviour when they break down, focussing on the regime of large times and distances for which one can expect a degree of universal behaviour. As a route to doing so, we start from expressions for physical properties in terms of sums over Feynman histories, or paths in Fock space, and aim to identify the dominant contributions to these sums.

Analogous approaches involving sums over histories have been applied very successfully to single-particle quantum systems in a variety of settings. In particular, the Gutzwiller trace formula provides a connection between the periodic orbits of a classical system and the spectrum of its quantum counterpart \cite{gutzwiller1990chaos,gutzwiller1980classical}. Contributions from pairs of orbits that carry opposite phases are dominant in the semiclassical limit, and the restriction to these pairs is known as the diagonal approximation \cite{berry1985semiclassical}. RMT spectral correlations are then a consequence of the nature of periodic orbits in classical systems that are chaotic, and the diagonal approximation is the starting point for a systematic semiclassical expansion \cite{argaman1993correlations,bogomolny1996gutzwiller,sieber2001correlations}.  As a second example, in the theory of mesoscopic conductors the spectral and transport properties are expressed in terms of paths of electrons undergoing multiple scattering by impurities \cite{altshuler1985electron,lee1985disordered,altshuler1991mesoscopic}. Pairs of paths with opposite phases, known as diffusons and Cooperons, survive a disorder average and determine long distance and low energy properties. They are also the basis for an expansion in inverse powers of the mean free path that captures weak localisation effects in disordered conductors. Here we will refer quite generally to the approximation of retaining only those pairs of paths with opposite phases as the diagonal approximation. 


The many-body systems we discuss are quantum circuits. These offer a minimal description of time evolution in lattice models: each lattice site carries a `spin' with Hilbert space dimension $q$, and pairs of sites are coupled by unitary gates during discrete time steps of the evolution. The examples we treat are one-dimensional brickwork models, in which every site is coupled alternately to its nearest neighbours on either side in successive time steps. In an obvious way, matrix elements of the evolution operator for a single time step define amplitudes for the steps of a path in Fock space. The transition amplitude between particular initial and final states under evolution over multiple time steps is a sum of contributions from multiple paths, labelled by states at intermediate times and weighted by products of amplitudes for the individual steps. These sums over Feynman histories will be at the centre of our discussion.

Following the spirit of RMT, many useful insights have been obtained recently by considering quantum circuits in which gates are drawn from a random distribution and properties of a circuit are averaged over an ensemble. In this way, two classes of model arise: random unitary circuits (RUCs) \cite{nahum2017quantum,nahum2018operator,keyserlingk2018operator,khemani2018operator,rakovsky2018diffusive,claeys2020maximum}, in which gates are chosen independently at different time steps, and random Floquet circuits (RFCs) \cite{kos2018analytic,chan2018solution,chan2018spectral,bertini2018exact,bertini2019entanglement,bertini2019exact}. RFCs may equally be viewed as examples of kicked spin chains, which have been studied extensively in their own right \cite{prosen1998time,prosen1999ergodic,pineda2007universal,akila2016particle,akila2017semiclassicalprl}. In RFCs the evolution operator is the same over each whole period, and is known as a Floquet operator. RUCs with gates chosen from the Haar distribution are particularly simple to analyse because the diagonal approximation is exact following an ensemble average. By contrast, RFCs, even with Haar-distributed gates, are not amenable to an exact analysis except in the large-$q$ limit \cite{chan2018solution}. There is nevertheless a strong interest in understanding their properties, because time evolution in a RFC provides information on the spectral and eigenstate correlations of the evolution operator for a simple form of local many-body dynamics. This motivates the study of RFCs that we present here.

The spectral statistics of a Floquet operator are characterised by the spectral form factor (SFF). Denoting the Floquet operator by $W$ and its $t$-th power for integer time $t$ by $W(t)$, the SFF is $K(t) = |\text{Tr}W(t)|^2$. Similarly the matrix elements of a local operator $\tau$ in the basis of eigenstates of $W$ are characterised, in the off-diagonal case, by the autocorrelation function ${\rm Tr}[ \tau W(t) \tau W^\dagger(t)]$ and in the diagonal case by $|{\rm Tr} [\tau W(t)]|^2$. A key technical fact is that the ensemble average of the SFF can be generated using a transfer matrix that acts in the spatial direction. One of the central ideas that we present in this paper is that questions about the pairings of Feynman histories that contribute to the ensemble-averaged SFF and matrix element correlators can be rephrased as questions about the eigenvectors associated with the leading eigenvalues of this average transfer matrix. Similarly, deviations of the SFF from RMT predictions are controlled by the behaviour of these eigenvalues as a function of $t$. Although the dimension of this matrix grows very rapidly with $t$ (for a brickwork RFC it is of dimension $q^{4t} \times q^{4t}$), we are able to probe its properties numerically for sufficiently large $t$ that we believe we have established its asymptotic behaviour. On this basis we argue that the character of the relevant pairs of histories is determined by the relative size of $t$ and of the system length $L$. In large systems the diagonal approximation becomes exact for times which are large but much smaller than the inverse level spacing. On the other hand, at sufficiently large $L$ for fixed large $t$, the contributing Feynman histories are locally paired, with multiple domains in the orbit pairing, and distinct pairings in neighbouring domains.

This work builds on and complements other recent research in a number of ways. Most directly, the possibility that domains arise in the pairing of Feynman histories was shown for a model solvable in the large-$q$ limit by one of the present authors and others \cite{chan2018spectral}. The current paper shows how to formulate and test this idea in a generic setting, demonstrates that it has consequences beyond the behaviour of the SFF, and links it to the notion of the entanglement membrane, which characterises the scrambling of quantum information in chaotic quantum systems \cite{jonay2018coarse,zhou2019emergent,zhou2020entanglement}. The idea of using a transfer matrix to generate the SFF and to calculate other quantities has been applied previously in several settings \cite{bertini2018exact,bertini2019entanglement,piroli2020exact,akila2016particle,akila2017semiclassicalprl,akila2018semiclassicalannals,gopalakrishnan2019unitary,braun2020transition,lerose2020influence,sonner2020characterizing}.
This approach arose from discussions of periodic orbits in many-body systems \cite{gutkin2016classical}, was developed as a method for treating kicked spin chains \cite{akila2016particle}, and has been elaborated further as a way of accessing the semiclassical limit and making connections with periodic orbit theory \cite{akila2017semiclassicalprl,akila2018semiclassicalannals}. Such a transfer matrix also forms the basis for the analysis of self-dual kicked spin chains \cite{bertini2018exact,flack2020statistics}, which display exact RMT behaviour of the SFF. Away from the self-dual point, recent work has investigated the evolution of this transfer matrix from ergodic to many-body localised behaviour \cite{braun2020transition}. The important distinction between those works and ours is that we are concerned with generic behaviour unrelated to self-duality, and at late times. 

Both the RFCs we consider and the kicked spin chains investigated by other groups lead naturally to a description in terms of Feynman histories in Fock space. This perspective has also been adopted in studies of thermalisation, through the introduction of an influence matrix \cite{lerose2020influence,sonner2020characterizing}. We note that there have been complementary efforts to study many-body quantum dynamics in the semiclassical limit by relating the orbits of classical models to properties of their quantum counterparts. These include studies of the spectral properties, many-body versions of coherent backscattering, and the out-of-time-order correlator \cite{engl2014coherent,dubertrand2016spectral,waltner2017trace,rammensee2018manybody}.

The models we study here differ in two ways from standard random-matrix systems, such as the Gaussian and circular ensembles \cite{mehta2004random}. In our case the non-zero matrix elements are local in space. There is a long history of work on systems with two-body random interactions \cite{french1970validity,bohigas1971two}, but without the restriction of locality. The behaviour of the SFF in ensembles of that type has been examined from the current perspective in Ref.~\cite{gharibyan2018onset} and we comment on this further in Sec.~\ref{sec:moderate}.

The remainder of this paper is organised as follows. In Sec.~\ref{sec:overview} we first gather definitions of the models and physical quantities we consider and describe how sums over histories appear in our calculations. We then give a brief overview of our results. In Sec.~\ref{sec:local} we show how to construct the transfer matrices generating the SFF of a RFC, Haar-average these transfer matrices, and then study the spectral properties of the average. Our analysis reveals a picture of local pairings of paths, in this case closed orbits in Fock space, and correspondingly deviations of the SFF from RMT which diverge in the thermodynamic limit. Through the transfer matrices we will also highlight a connection between the spectral statistics of chaotic many-body systems and their behaviour under local coupling to a bath. Then, in Sec.~\ref{sec:observables}, we show that local orbit-pairing implies strong correlations between the diagonal matrix elements of local observables. Relative to the predictions of ETH these correlations grow without bound in the thermodynamic limit, and we verify this growth numerically. Following this in Sec.~\ref{sec:fluctuations} we consider the statistical fluctuations of the SFF. By writing the higher moments of the SFF in terms of multiple sums over histories, we show that a distinct freedom in their local pairing gives rise to non-Gaussian statistics. We also highlight a connection with the entanglement membrane. In Sec.~\ref{sec:moderate} we extend aspects of our discussion to a class of models with gates whose distribution can be tuned continuously from Haar to the identity, and discuss how their behaviour differs away from the Haar case. In Sec.~\ref{sec:discussion} we provide a summary, discussion and outlook. Various technical details are described in a series of appendices.

\begin{figure}
	\includegraphics[width=0.3\textwidth]{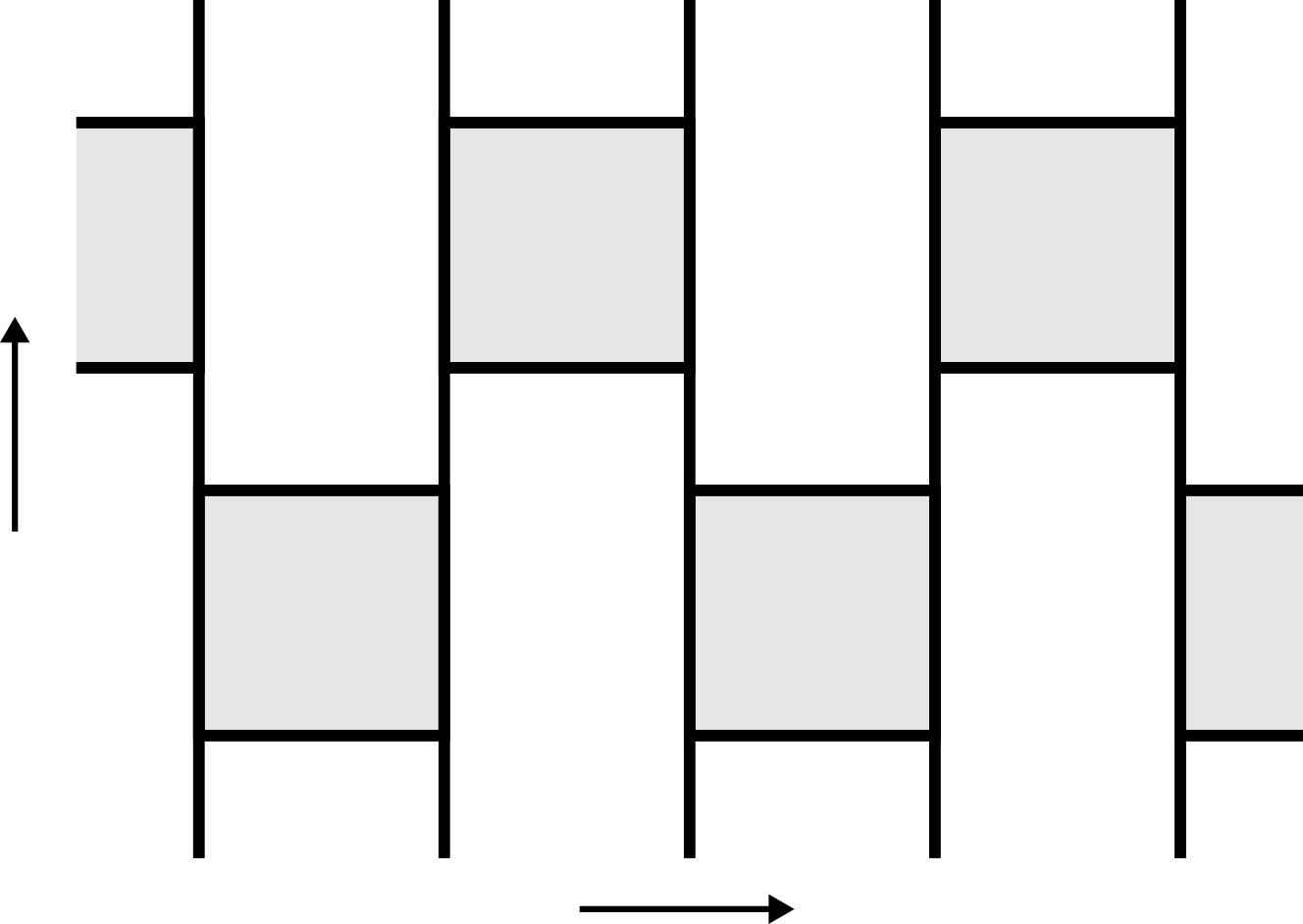}
	\put(5,75){$W_2$}
	\put(5,32){$W_1$}
	\put(-120,32){$U_{0,1}$}
	\put(-62,32){$U_{2,3}$}
	\put(-91,76){$U_{1,2}$}
	\put(-33,76){$U_{3,4}$}
	\put(-76,-7){$x$}
	\put(-160,54){$t$}
	\caption{Diagram of the Floquet operator of a brickwork model with periodic boundary conditions, with time running vertically. The vertical lines mark positions of sites, and the Floquet operator is $W = W_2 W_1$. $W_1$ and $W_2$ are tensor products of two-site unitary gates $U_{x,x+1}$, where $U_{x,x+1}$ couples the sites $(x,x+1)$.}
\label{fig:brickwork}
\end{figure}
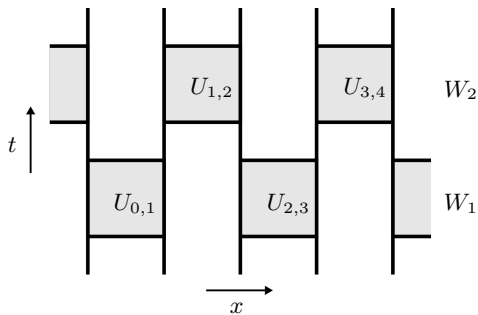

\section{Overview}\label{sec:overview}

Before discussing our work in detail, we first introduce the Floquet models and the physical quantities that we consider throughout the paper. Following this we discuss sums over histories and the diagonal approximation in the context of quantum circuits. We then give an overview of our main results.

\subsection{Models and correlators}\label{sec:definitions}
We focus on one-dimensional Floquet circuits with brickwork structure \cite{chan2018solution}, illustrated in Fig.~\ref{fig:brickwork}. The unitary evolution operator over integer time $t$ is $W(t) \equiv W^t$, where $W = W_2 W_1$ is the Floquet operator. With local Hilbert space dimension $q$, each of $W_1$ and $W_2$ are tensor products of two-site ($q^2 \times q^2$) unitary matrices $U_{x,x+1}$ coupling alternate pairs of neighbouring sites $(x,x+1)$. We label sites $x=0,1 \ldots (L-1)$ for a system of length $L$, and so with Fock-space dimension $q^L$. For periodic boundary conditions, which necessitates $L$ even, the two half-steps $W_{1,2}$ are given by
\begin{align}
\begin{split}
	W_1 &= U_{0,1} \otimes U_{2,3} \otimes \ldots \otimes U_{L-2,L-1} \\
	W_2 &= U_{1,2} \otimes U_{3,4} \otimes \ldots \otimes U_{L-1,0}.
\end{split}
\end{align}
With open boundary conditions and $L$ even we replace $U_{L-1,0}$ with the $q^2 \times q^2$ identity matrix. With open boundary conditions and odd $L$ we instead have
\begin{align}
\begin{split}
	W_1 &= U_{0,1} \otimes U_{2,3} \otimes \ldots U_{L-3,L-2} \otimes \mathbb{1}\\
	W_2 &= \mathbb{1} \otimes U_{1,2} \otimes U_{3,4} \otimes \ldots U_{L-2,L-1},
\end{split}
\end{align}
where now $q \times q$ identity matrices act on site $x=(L-1)$ in $W_1$ and $x=0$ in $W_2$. With the exception of Sec.~\ref{sec:moderate} we are concerned with circuits constructed from Haar-random gates, or Haar-RFCs. For all numerical investigations we focus on a local Hilbert space dimension $q=2$ (a form of kicked spin-$\frac{1}{2}$ chain). To demonstrate that our choice of Haar-random gates is not crucial to the results we obtain, in Appendix~\ref{sec:heisenberg} we compare with behaviour for a kicked Heisenberg model.

The central quantity considered in this work is the spectral form factor (SFF), which probes correlations in the level density. The SFF is defined, for integer $t$, by
\begin{align}
\begin{split}
	K(t) &= |\text{Tr}W(t)|^2 \\
	&= \sum_{nm} e^{i(\theta_n-\theta_m)t}.
\end{split}
\end{align}
Here $\theta_n$ is the quasienergy of the Floquet operator $W$ associated to eigenstate $\ket{n}$, $W\ket{n}=e^{i\theta_n}\ket{n}$.

RMT, in the standard Wigner-Dyson sense \cite{mehta2004random}, will be a key reference point. For $W$ a Haar-random $N \times N$ unitary matrix, $\text{Tr}W(t)$ is normally and isotropically distributed in the complex plane in the limit of large $N$ \cite{kunz1999probability}. In RMT for all $N$ we have the average SFF for integer $t$
\begin{align}
\begin{split}
	\overline{K}_{\text{RMT}}(t) = \begin{cases} N^2 \quad &t = 0 \\
	t \quad &1 \leq t \leq N \\
	N \quad &N \leq t. \end{cases}
\label{eq:sffRMT}
\end{split}
\end{align}
An important timescale is the Heisenberg time ${t_{\rm{H}}=N}$, set by the mean level spacing $2\pi /N$. For circuit models $N=q^L$, and except where stated explicitly we consider times below $t_{\rm{H}}$.

For many examples of chaotic quantum systems it is known that RMT behaviour of the SFF sets in only beyond a characteristic timescale. In some circumstances this scale reflects specific microscopic features of the system, but for spatially extended models it may diverge with system size and arise from physical processes that are common to a class of systems. Diffusive mesoscopic conductors provide a key example, for which this timescale is known as the Thouless time, and is given in terms of the diffusion constant $D$ and the linear system size $L$ by $t_{\rm Th} = L^2/D$. For the chaotic many-body systems we consider, we will also refer to the timescale at which RMT behaviour of the SFF sets in as the Thouless time \cite{bertrand2016anomalous,chan2018spectral,friedman2019spectral,sierant2020thouless,moudgalya2020spectral} (see also Ref.~\cite{gharibyan2018onset}). We note however that there are a number of alternative definitions, based on the behaviour of local observables \cite{serbyn2017thouless}, the sensitivity to boundary conditions \cite{monthus2017many}, and on many-body return probabilities \cite{schiulaz2019thouless}.

The characterisation of the behaviour of a system involves the matrix elements of local observables, and here the predictions of the ETH provide a useful reference. In Floquet systems the ETH is as follows: for a set of eigenstates drawn from a sufficiently narrow window of quasienergies, the statistical properties of the matrix elements of local observables are as for random vectors. 

The diagonal matrix elements of the operator $\tau$ at site $x$ can be written in terms of the reduced density matrices at this site, $\rho_x(n) = \text{Tr}_x' \ket{n}\bra{n}$ for the eigenstate $\ket{n}$, in the form $\text{Tr}[\tau\rho_x(n)]$. Here and throughout this paper, we use $\text{Tr}_x$ to denote a trace over site $x$ and $\text{Tr}_x'$ to denote a trace over its complement, the other $(L-1)$ sites. A correlator between the eigenstates $\ket{n}$ and $\ket{m}$ can then be defined as $\text{Tr}_x[\tau \rho_x(n)]\text{Tr}_x[\tau \rho_x(m)]$. To avoid referencing a specific observable it is useful to average over a complete orthonormal set of operators (see Sec.~\ref{sec:observables}). This leads us to consider the correlator of reduced density matrices $\text{Tr}_x[\rho_x(n)\rho_x(m)]$. It is convenient to work in the time domain, and so we define the reduced form factor (RFF) at site $x$,
\begin{align}
	R_x(t) &= \sum_{nm} \text{Tr}_x[\rho_x(n)\rho_x(m)] e^{i(\theta_n-\theta_m)t} \notag\\
	&= \text{Tr}_x [\text{Tr}'_x W(t)[\text{Tr}'_x W(t)]^{\dag}]. \label{eq:RFF}
\end{align}
We will see also that a straightforward generalisation of this,
\begin{align}
	R_x'(t) = \text{Tr}_x'[ \text{Tr}_x W(t) [\text{Tr}_x W(t)]^{\dag}],
\end{align}
describes spectral structure in the off-diagonal matrix elements of operators local to site $x$.

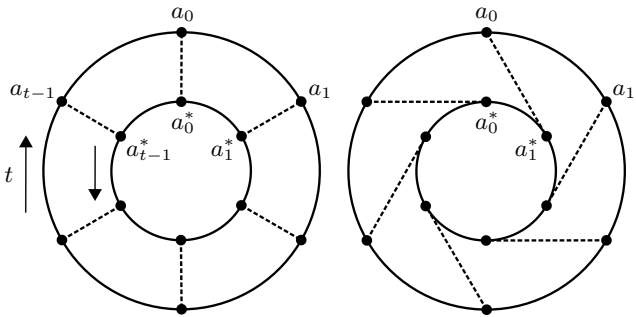
\begin{figure}
	\includegraphics[width=0.45\textwidth]{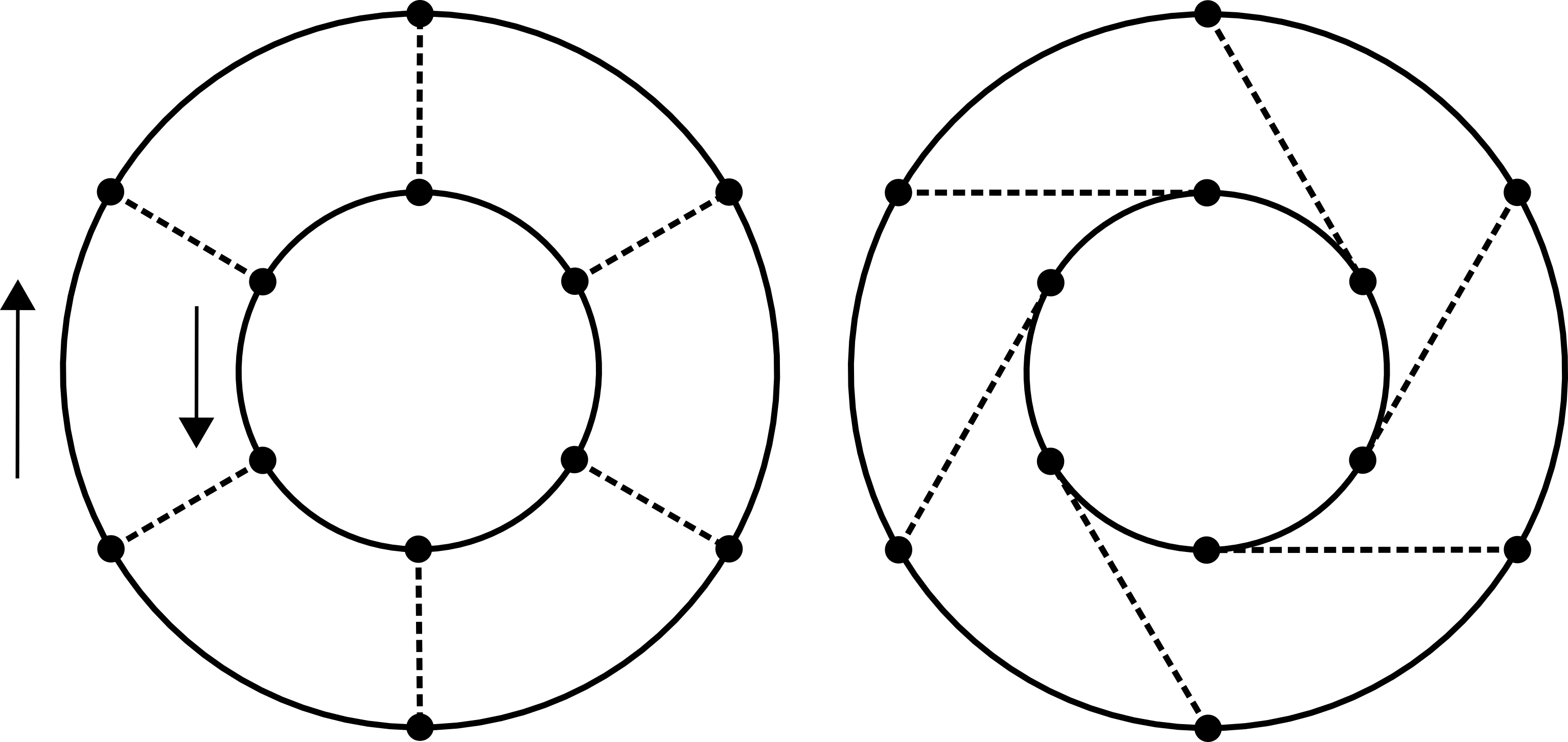}
	\put(-172,112){$a_0$}
	\put(-172,70){$a^*_0$}
	\put(-120,83){$a_1$}	
	\put(-157,60){$a^*_1$}
	\put(-233,83){$a_{t-1}$}	
	\put(-189,60){$a^*_{t-1}$}
	\put(-57,112){$a_0$}
	\put(-57,70){$a^*_0$}
	\put(-5,83){$a_1$}	
	\put(-42,60){$a^*_1$}
	\put(-235,50){$t$}
	\caption{Two of the diagonal orbit pairings that contribute to the spectral form factor $K(t) = |\text{Tr}W(t)|^2$. The outer circles represent the forward orbit, and the inner the backward, corresponding to terms in the sum-over-histories representations of $\text{Tr}W(t)$ and $\text{Tr}W^*(t)$, respectively. The dashed lines represent the pairing of forward and backward orbits. Diagonal pairings have $a_r=a_{r+s}^*$, with addition defined modulo $t$. On the left and right we show the $s=0$ and $s=1$ pairings, respectively.}
\label{fig:diagonal}
\end{figure}

\subsection{Histories in circuits}

We now discuss sums over histories in quantum circuits, and how they appear in the quantities of interest. Consider the probability for a Floquet system to evolve from an initial state $\ket{a_0}$ to final state $\ket{a_t}$ over time $t$,
\begin{align}
\begin{split}
	|\braket{a_t|W(t)|a_0}|^2 =& \sum_{a_1 \ldots a_{t-1}} W_{a_t a_{t-1}} \ldots W_{a_1 a_0} \\ &\times \sum_{a^*_1 \ldots a^*_{t-1}}   W^*_{a_t a^*_{t-1}} \ldots W^*_{a^*_1 a_0}.
\label{eq:transition_prob}
\end{split}
\end{align}
This is a discrete sum over all pairs of paths between state $a_0$ and $a_t$. The forward $(a_0 a_1 \dots a_t)$ and backward $(a_0 a_1^* \dots a_t)$ paths are labelled by integers $a_r,a_r^* = 0 \ldots (N-1)$. The transition amplitudes are simply the matrix elements of the Floquet operator.

The weights of the $N^{t-1}$ diagonal pairs of paths, with $a_r = a_r^*$ for all $r$, give real non-negative contributions to Eq.~\eqref{eq:transition_prob}. If the weights of the $N^{t-1}(N^{t-1}-1)$ off-diagonal pairs behave like a set of independent random variables, their sum gives a real contribution of the same order as the sum of the diagonal pairs. The off-diagonal contributions may, however, vanish after a suitable average. We then arrive at the diagonal approximation. The dynamics within the diagonal approximation is described by the diagonal propagator, which could also be thought of as the analogue of a diffuson in Fock space,
\begin{equation}
	\mathcal{P}_{a_{r+1} a_r} = W_{a_{r+1}a_r}W_{a_{r+1}a_r}^*.
\label{eq:average_propagator}
\end{equation}
For example, provided the ensemble average can be performed independently for each step, the transition probability Eq.~\eqref{eq:transition_prob} becomes $\overline{\mathcal{P}}^t_{a_t a_0}$ (see Sec.~\ref{sec:moderate} for a discussion of the distinction between $\overline{\mathcal{P}}^t$ and $\overline{\mathcal{P}^t}$). For definiteness in the following we use the term diagonal approximation to refer to calculations based on the average diagonal propagator $\overline{\mathcal{P}}$. Unitarity constrains $\mathcal{P}$ to be a doubly stochastic matrix (the sum along any row or column is unity) and so it has a leading eigenvalue of unity.

To construct the SFF in the diagonal approximation we write $\text{Tr}W(t)$ as a sum over all closed paths of $t$ steps,
\begin{align}
	\text{Tr}W(t) &= \sum_{a_0 \ldots a_{t-1}} W_{a_0 a_{t-1}} \ldots W_{a_1 a_0}.
\label{eq:TrWt}
\end{align}
We will refer to these closed paths as orbits. The SFF is a sum over all pairs of forward and backward orbits,
\begin{align}
\begin{split}
	K(t) = \sum_{a_0 \ldots a_{t-1}}& W_{a_0 a_{t-1}} \ldots W_{a_1 a_0} \\ &\times \sum_{a^*_0 \ldots a^*_{t-1}} W^*_{a^*_0 a^*_{t-1}} \ldots W^*_{a^*_1 a^*_0}.
\end{split}
\label{eq:sffpaths}
\end{align}
The amplitude of an orbit $(a_0,a_1 \ldots a_t)$ is invariant under a cyclic permutation. Consequently, in a system without time-reversal symmetry, for a typical forward orbit in Eq.~\eqref{eq:sffpaths} there are $t$ backward orbits with the complex conjugate amplitude. In Fig.~\ref{fig:diagonal} we illustrate two of these diagonal orbit pairings. The diagonal approximation to the average SFF is
\begin{align}
	\overline{K}(t) = t \text{Tr} \overline{\mathcal{P}}^t.
	\label{eq:sffdiagonal}
\end{align}
For a propagator with just one unit-modulus eigenvalue, $\text{Tr} \overline{\mathcal{P}}^t \to 1$ at late times. In the regime $t<t_H$, the diagonal approximation to the SFF therefore approaches the RMT result of Eq.~\eqref{eq:sffRMT}.

As a concrete example consider again the SFF within RMT. For $W$ Haar-random, and in the limit of large $N$, the ensemble average of Eq.~\eqref{eq:sffpaths} is \cite{samuel1980integrals,brouwer1996diagrammatic}
\begin{align}
	\overline{K}(t) = \sum_{a_0 \ldots a_{t-1} \atop a^*_0 \ldots a^*_{t-1}}
	\frac{1}{N^t}\sum_{s=0}^{t-1} \prod_{r=0}^{t-1} \delta_{a_r a^*
_{r+s}} = t,
\label{eq:large_q_average}
\end{align}
for nonzero integer $t \ll N$. The sum over $s=0\ldots (t-1)$ is the sum over $t$ diagonal orbit pairings. Here the diagonal approximation coincides with the exact result.

In circuit models we are concerned with sums over paths in Fock space. Choosing our Fock-space basis states $\ket{a_r}$ to be product states, the sum in for example Eq.~\eqref{eq:TrWt} can be recast as a multiple sum over the local orbits of individual sites. $\text{Tr}W(t)$ and the SFF can then be written in terms of transfer matrices acting on these orbits (see Sec.~\ref{sec:local} for details). For the Haar-RFCs we study, all choices of site basis are statistically equivalent. 

These transfer matrices are associated with individual gates $U_{x,x+1}$. Writing the transfer matrices generating $K(t)$ as $\mathcal{T}_{x,x+1}(t)$, with open boundary conditions we find for example
\begin{equation}
	K(t) = \bra{\mathcal{B}_L} \mathcal{T}_{0,1} \mathcal{T}_{2,3} \ldots \mathcal{T}_{L-2,L-1} \ket{\mathcal{B}_R}.
\label{eq:Kobc_overview1}
\end{equation}
Here $\bra{\mathcal{B}_L}$ and $\ket{\mathcal{B}_R}$ are vectors encoding the boundary conditions at the left and right ends of the system, respectively. For independently and identically distributed gates the average SFF is determined by powers of the average transfer matrix $\overline{\mathcal{T}}(t)$,
\begin{equation}
	\overline{K}(t) = \bra{\mathcal{B}_L}\overline{\mathcal{T}}^{L-1}(t)\ket{\mathcal{B}_R}.
\label{eq:Kobc_overview2}
\end{equation}
Fixing $t$ and taking the limit of large $L$, the average SFF is dominated by contributions from the leading eigenvalues of $\overline{\mathcal{T}}(t)$. We show, for a system without time-reversal symmetry, that there are exactly $t$ leading eigenvalues, and we denote these by $\lambda(\omega,t)$, where $\omega$ is a $t$-valued symmetry label associated with time-periodicity.

\subsection{Results}\label{sec:results}

In this work we study behaviour beyond RMT in chaotic many-body Floquet systems with local interactions. For Haar-RFCs the deviations of the SFF from RMT arise from a particular class of off-diagonal pairings of paths.

\begin{figure}
	\hspace{-0.1in}
	\includegraphics[width=0.47\textwidth]{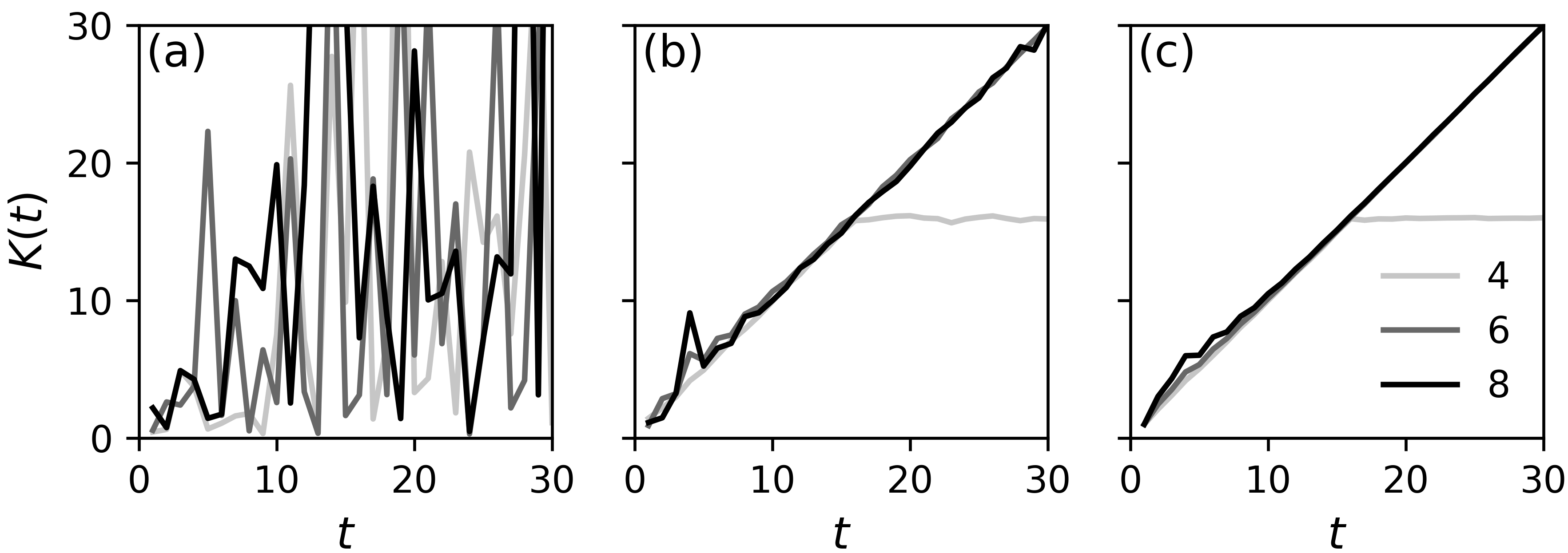}
	\caption{SFF of the Haar-RFC with different degrees of averaging. All data is for periodic boundary conditions with the number of sites $L$ shown on the legend. (a)~No averaging: for each $L$ this is the SFF for an individual realisation of the gates. (b)~Average over only a single gate, with all others fixed. (c)~Average over all of the gates.}
\label{fig:sff_averaging}
\end{figure}

Simplicity, in the form of the diagonal approximation or something beyond, is only to be expected after a degree of averaging. Moreover, the SFF is not self-averaging \cite{prange1997spectral}, and typically exhibits large system-dependent fluctuations around the RMT result \cite{kunz1999probability}. Averaging can in principle be approached in various ways. For a specific model or a specific realisation drawn from an ensemble, one can average the SFF over a time window. This window should be narrow on the scale set by its mid-point, to avoid distortion, but wide enough to contain many Floquet periods, to ensure efficient averaging \cite{prange1997spectral}. Alternatively, one can average over a small region in the space of possible models. In this case the average may be representative of all systems in this region, but system-dependent fluctuations will only be washed out at late times. A final alternative, which is the one we follow, is to average over a wide variety of systems: in our case the Haar distribution for gates. In practice however quite limited averaging is sufficient. In Fig.~\ref{fig:sff_averaging} we show that an average over just one gate is enough to dramatically suppress fluctuations of the SFF.

In Sec.~\ref{sec:local} we show that the spectral structure in Haar-RFCs, with local Hilbert space dimension $q=2$, is not captured by the diagonal approximation. To understand the form of the SFF we develop the picture of local orbit pairing which first appeared in the large-$q$ limit in Ref.~\cite{chan2018spectral}, demonstrating here its appearance in a generic setting. To do this we show how to express the SFF of a Haar-RFC in terms of transfer matrices $\mathcal{T}(t)$ acting in the space direction. For independently and identically distributed gates the average SFF $\overline{K}(t)$ is then expressible in terms of powers of the average transfer matrix $\overline{\mathcal{T}}(t)$. Through this average transfer matrix the accuracy of the diagonal approximation, as well as the corrections arising in large systems, acquire a sharply defined meaning in the language of local orbit pairing.

The transfer matrices are too large to study directly in the regimes of interest. To probe the spectral properties of $\overline{\mathcal{T}}(t)$ we instead impose a variety of boundary conditions on the model, thereby coupling to the eigenvectors in controlled ways. By analysing the length-scaling of the SFF and related objects, we separate the leading eigenvalues of $\overline{\mathcal{T}}(t)$ according to their symmetry sector and determine their magnitudes. In practice our approach requires only very small systems of $L \leq 8$ sites for the times of interest, and our results allow us to reconstruct the SFF for arbitrarily large $L$. By studying the corresponding eigenvectors we then directly probe the local orbit pairing.

\begin{figure}
	\centering
	\includegraphics[width=0.4\textwidth]{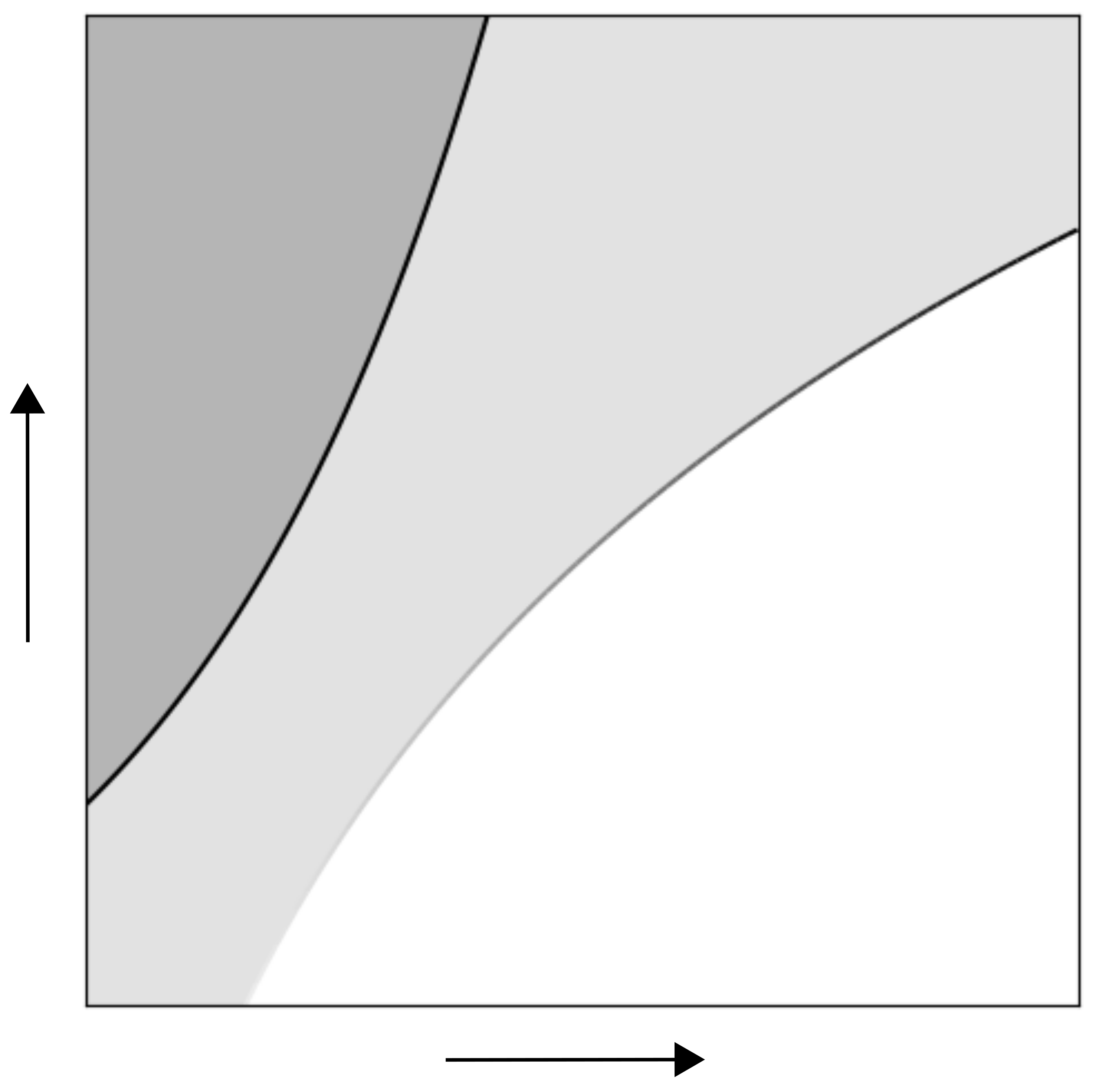}
	\put(-100,-5){$L$}
	\put(-205,100){$t$}
	\put(-110,75){$t_{\text{Th}}$}
	\put(-155,120){$t_{\text{H}}$}
	\put(-115,140){Diagonal}
	\put(-115,125){$\overline{K}(t) \simeq t$}
	\put(-55,100){Domains}
	\put(-55,85){$\overline{K}(t) \gg t$}
	\put(-175,165){$\overline{K}(t) = q^L$}
	\caption{The different regimes of behaviour for the average SFF. See main text for discussion.}
\label{fig:tLplane}
\end{figure}
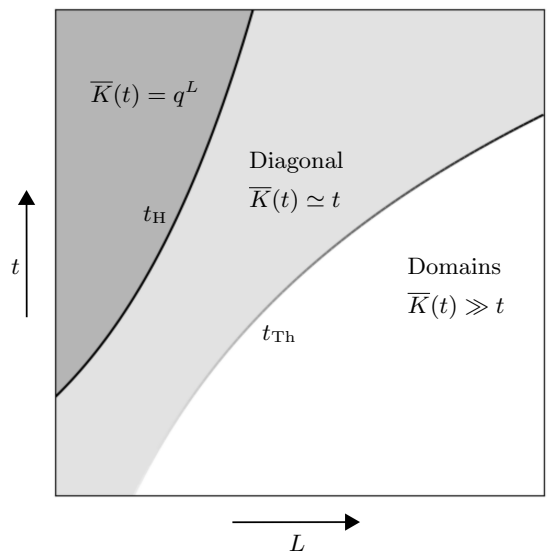

In Fig.~\ref{fig:tLplane} we illustrate the different regimes of the many-body spectrum for the Haar-RFC and relate these regimes to the transfer matrix spectrum. At fixed $L$, increasing $t$ brings us into the diagonal regime (for $t<t_{\text{H}}=q^L$). Here $\overline{K}(t)$ is dominated by the sum over the $t$ diagonal orbit pairings, each contributing unity, so we find the RMT result $\overline{K}(t)=t$. We will refer to these contributions as the global diagonal orbit pairings. By contrast, taking the limit of large $L$ at fixed $t$ the SFF is dominated by orbits which are diagonally paired only locally, with distinct diagonal pairings in neighbouring domains. These two regimes can be understood by considering only the leading eigenvalues $\lambda(\omega,t)$ of the average transfer matrix, of which there are exactly $t$. The deviations of these eigenvalues from unity control the contributions of domain walls to the SFF. For example, increasing $t$ these eigenvalues approach unity and we move from a picture of local orbit pairing to one of global orbit pairing. Conversely, taking the limit of large $L$ at fixed $t$, the largest of these close-lying eigenvalues dominates. Small deviations of the eigenvalues from unity are then responsible for large deviations of the SFF from RMT. We show that under certain assumptions the timescale for the crossover between these two regimes, which we refer to as the Thouless time, scales logarithmically with system size $L$. A third regime is entered on increasing $t$ beyond the Heisenberg time. There the subleading eigenvalues of $\overline{\mathcal{T}}(t)$ control the SFF.

A natural question is whether this picture of local orbit pairing has implications for local observables and eigenstate correlations, and we investigate this in Sec.~\ref{sec:observables}. The trace structure of the RFF $R_x(t)$ in Eq.~\eqref{eq:RFF} is, away from the site $x$, identical to that of the SFF, and we show that this implies exponential growth of $\overline{R}_x(t)$ with increasing $L$. Remarkably, on transforming this behaviour from the time domain to that of quasienergies we find that the deviations from ETH are most prominent on small quasienergy-scales, where the ETH is conventionally expected to be most accurate. We find enhancements of the correlations relative to RMT by two orders of magnitude even in systems of $L \leq 14$ sites, so our results represent a substantial correction to the ETH for one-dimensional chaotic Floquet systems.

We investigate the statistical fluctuations of the SFF in Sec.~\ref{sec:fluctuations}. Focussing on its second moment we highlight the presence of another kind of domain wall, in the pairing of the multiple copies of orbits which appear. While these are distinct from the domain walls in the orbit pairing considered in earlier sections, their role also becomes more prominent with increasing system size. These domain walls in the pairing of copies of orbits are closely related to entanglement membranes. Our results here also provide an understanding of the striking effect of the single-gate average in Fig.~\ref{fig:sff_averaging}(b). 

In Sec.~\ref{sec:moderate} we discuss deviations of the SFF from RMT which arise within the diagonal approximation. These deviations result from the subleading eigenvalues of the diagonal propagator \cite{kos2018analytic,friedman2019spectral,roy2020random,moudgalya2020spectral}. We consider this effect both for Haar-RFCs and more generally.

\section{Local orbit pairing}\label{sec:local}
The diagonal approximation to sums over pairs of orbits is blind to locality. In this section, which contains the central arguments of the paper, we demonstrate how the picture of paired orbits must be extended in one-dimensional many-body systems with local interactions. First in Sec.~\ref{sec:breakdown} we show that the diagonal approximation fails to describe the spectral correlations in Haar-RFCs. We then discuss the model of Ref.~\cite{chan2018spectral}, where the breakdown of the diagonal approximation is evident even in the large-$q$ limit. There an exact treatment is possible, and this reveals the picture of local orbit pairing. Motivated by this, in Sec.~\ref{sec:transfer} we express the SFF of a brickwork model (with arbitrary $q$) in terms of transfer matrices for the local orbits which act in the spatial direction. On averaging these transfer matrices the picture of local orbit pairing emerges.

Through Secs.~\ref{sec:open} and \ref{sec:twisted} we extract information on the average transfer matrix by imposing various boundary conditions on the orbits. This allows us to determine the leading eigenvalues, as well as the relationship between the corresponding eigenvectors and the local orbit pairing. As a test of our results, we show that this information is sufficient to reconstruct the average SFF accurately in large systems (see Fig.~\ref{fig:sff_reconstruction}). In Sec.~\ref{sec:pairingdomains} we further explore the connection between the eigenvectors of the average transfer matrix and the local orbit pairing, and then in Sec.~\ref{sec:bath} discuss the behaviour of individual circuit realisations.

\subsection{Breakdown of the diagonal approximation}\label{sec:breakdown}

\begin{figure}
	\includegraphics[width=0.45\textwidth]{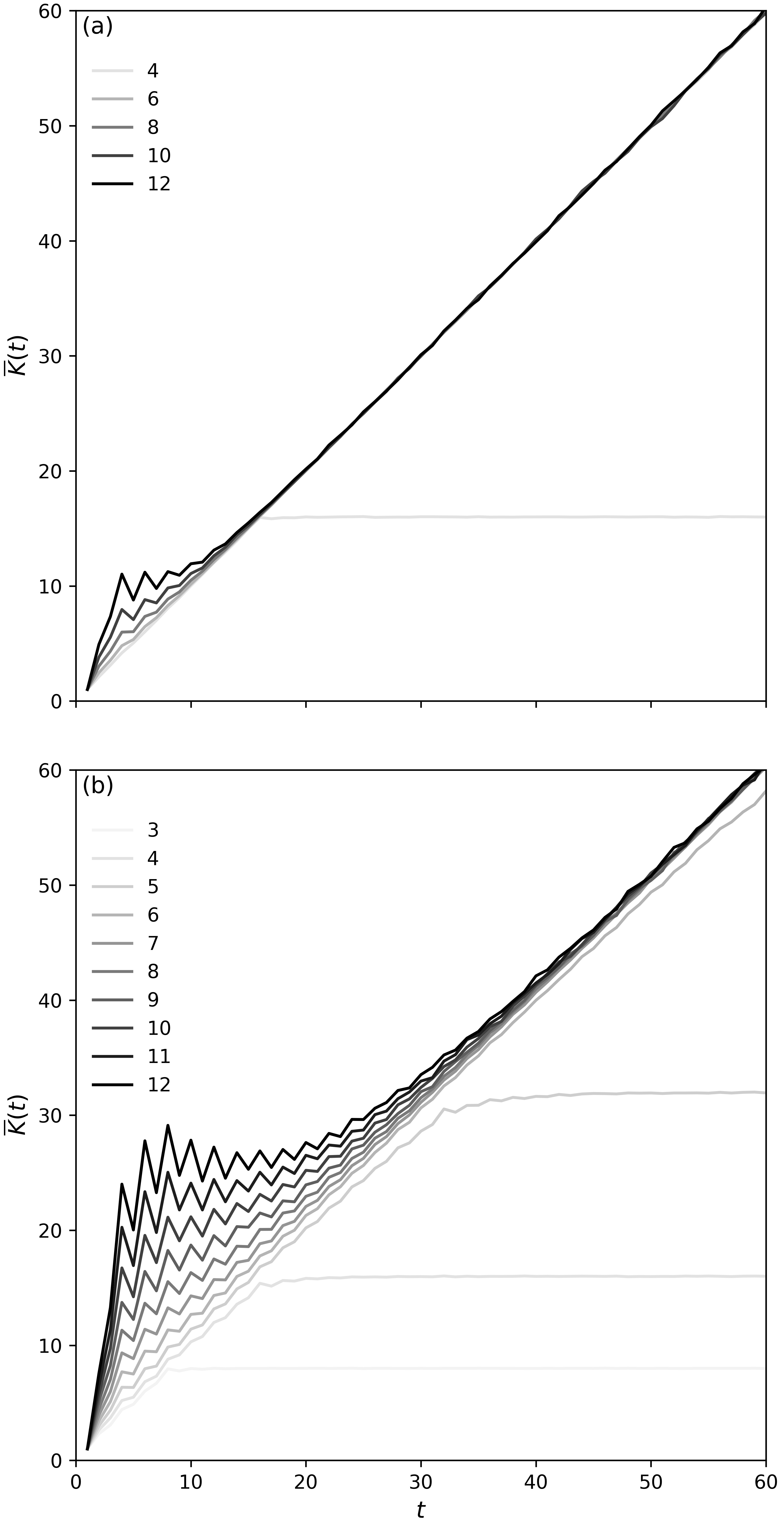}
	\caption{Average SFF $\overline{K}(t)$ in the $q=2$ brickwork model with (a) periodic and (b) open boundary conditions. The system size $3 \leq L \leq 12$ is shown on the legend. $L$ is necessarily even with periodic boundary conditions. In the diagonal regime $\overline{K}(t) \simeq t$, and beyond the Heisenberg time $t_{\text{H}}=q^L$ we have $\overline{K}(t) \simeq q^L$.} 
\label{fig:sff}
\end{figure}

First we demonstrate the necessity of moving beyond the diagonal approximation. In particular, we will show that for Haar-RFCs $\overline{K}(t)=t$ within the diagonal approximation. This is also the large-$q$ result \cite{chan2018solution}, and should be compared with numerical results for $q=2$ in Fig.~\ref{fig:sff}. The diagonal propagator for brickwork models takes the form
\begin{align}
	\mathcal{P}_{ab} = \sum_{cc^*} (W_2)_{ac} (W_1)_{cb} (W^*_2)_{ac^*} (W^*_1)_{c^*b},
\label{eq:brickwork_propagator}
\end{align}
where all subscripts are many-body indices (taking $q^L$ values). Each independent gate appears once in $\mathcal{P}$, as does its conjugate. $\overline{\mathcal{P}}$ is given by averaging over the gates. For example, the average over the gate $U_{0,1}$ acting on sites $0$ and $1$ in the first half-step $W_1$ is
\begin{align}
	\overline{[U_{0,1}]_{c_0 c_1,b_0 b_1}[U^*_{0,1}]_{c^*_0 c^*_1,b_0 b_1}} = \frac{1}{q^2} \delta_{c_0 c_0^*}\delta_{c_1 c_1^*},
\label{eq:HaarUUaverage}
\end{align}
where we have made the single-site indices explicit. The result is the same for gates acting in the second half-step. Multiplying the expressions Eq.~\eqref{eq:HaarUUaverage} for each independent gate, and summing over the internal $c,c^*$ indices, we find the average propagator
\begin{align}
	\overline{\mathcal{P}}_{a_0 \ldots a_{L-1},b_0 \ldots b_{L-1}} = \frac{1}{q^L},
\label{eq:HaarPmatrix}
\end{align}
a matrix with constant entries. The leading eigenvalue of $\overline{\mathcal{P}}$ is unity and all others are zero. As a result, the diagonal approximation to the SFF $\overline{K}(t)=t$ as in RMT. By contrast in Fig.~\ref{fig:sff} there are obvious deviations of the SFF from RMT. These deviations grow with increasing system size, and furthermore are significantly larger with open boundary conditions than with periodic. Through this section we set out to understand this behaviour.

The diagonal approximation to the SFF involves an overall factor of $t$, the number of diagonal orbit pairings. The breakdown of the diagonal approximation in a chaotic Floquet system with local interactions was demonstrated in Ref.~\cite{chan2018spectral}; there a calculation of the SFF in the large-$q$ limit revealed deviations from RMT which grow exponentially with increasing $L$. Here we briefly review certain aspects of that work.

The Floquet operator $W=W_2 W_1$ of Ref.~\cite{chan2018spectral} is made up of two half-steps. The first, $W_1$, consists of independent $q \times q$ Haar-random gates $u_x$ acting on individual sites of a chain, $W_1 = u_0 \otimes \ldots \otimes u_{L-1}$, for $L$ sites. The second, $W_2$, consists of weak interactions between adjacent sites with coupling strength $\varepsilon$.

In $K(t)$ each of the $q \times q$ gates $u_x$ appears $t$ times in $\text{Tr}W(t)$ and its conjugate appears $t$ times in $[\text{Tr}W(t)]^*$. To calculate $\overline{K}(t)$ we must average independently over each $u_x$, and this gives a sum over orbit pairings $s_x=0\ldots (t-1)$ as in Eq.~\eqref{eq:large_q_average}. This immediately promotes the orbit pairing to a local degree of freedom. $\overline{K}(t)$ is then given by a sum over the orbit pairings $s_x$ at each site $x=0 \ldots (L-1)$, and takes the form of a partition function
\begin{equation}
	\overline{K}(t) = \sum_{s_0 \ldots s_{L-1}=0}^{t-1} \prod_{x=0}^{L-1} \Big[ \delta_{s_x s_{x+1}} + (1-\delta_{s_x s_{x+1}})e^{-\varepsilon t} \Big].\label{eq:phasecouplingK}
\end{equation}
Here and throughout this paper, we will refer to a configuration $s_x \neq s_{x+1}$ as a domain wall. The statistical weights of domain walls are suppressed by factors $e^{-\varepsilon t}$; the coupling strength $\varepsilon$ appears here as the domain wall line tension. In this way $\overline{K}(t)$ has been expressed in terms of the transfer matrix of a ferromagnetic $t$-state Potts model. At late times the domain walls are suppressed, and we recover the RMT result $\overline{K}(t) = t$, corresponding to a sum over $t$ global diagonal orbit pairings.

The first corrections to RMT with open boundary conditions come from configurations with one domain wall,
\begin{align}
	\overline{K}(t) &= t + t(t-1)(L-1)e^{-\varepsilon t} + \ldots \label{eq:Kobcphase}
\end{align}
and with periodic boundary conditions they come from configurations with two domain walls,
\begin{align}
	\overline{K}(t) &= t + \frac{1}{2}t(t-1)L(L-1)e^{-2\varepsilon t} + \ldots
\label{eq:Kpbcphase}
\end{align}
The factors $t(t-1)$ correspond to the choices of orbit pairing in the two domains. With open boundary conditions $(L-1)$ is the translational entropy of one domain wall, and with periodic $\frac{1}{2}L(L-1)$ is the translational entropy of two domain walls. It is clear that with open boundary conditions, where it is possible to have just one domain wall with weight $e^{-\varepsilon t}$ rather than $e^{-2 \varepsilon t}$ for two domain walls, the deviations of $\overline{K}(t)$ from RMT are larger.

Although these results were derived in the limit of large $q$, in this work we show that the picture of local orbit pairing is more general. For example, it describes the spectral statistics of Haar-RFCs at small $q$. Indeed, the deviations from RMT displayed in Fig.~\ref{fig:sff}, including their dependence on both the system size and on the boundary conditions, illustrate the phenomenology of domain walls in the orbit pairing. Specifically, for fixed $t$ the deviations from RMT behaviour grow with $L$ and are larger with open than with periodic conditions.

\subsection{Transfer matrix for orbits}\label{sec:transfer}

Here we show how to construct the transfer matrices generating the SFF of a brickwork model, discuss their spectral properties, and connect the behaviour of their average to the behaviour observed in Fig.~\ref{fig:sff}. Similar approaches have appeared in the study of kicked Ising models \cite{akila2016particle}. By writing the average spectral form factor of the brickwork model in terms of a transfer matrix, we make the local orbit pairing degrees of freedom explicit. 

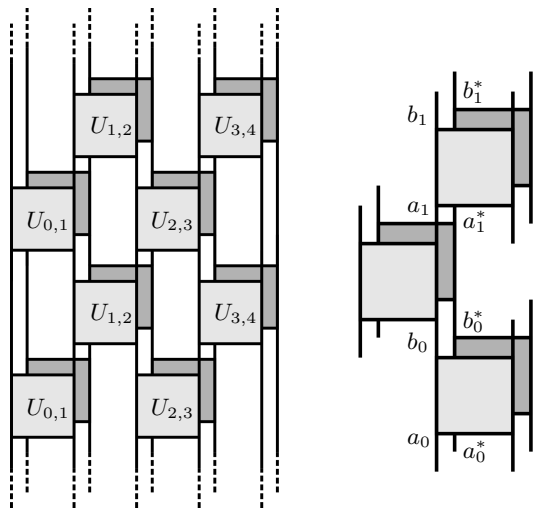
\begin{figure}
	\includegraphics[width=0.4\textwidth]{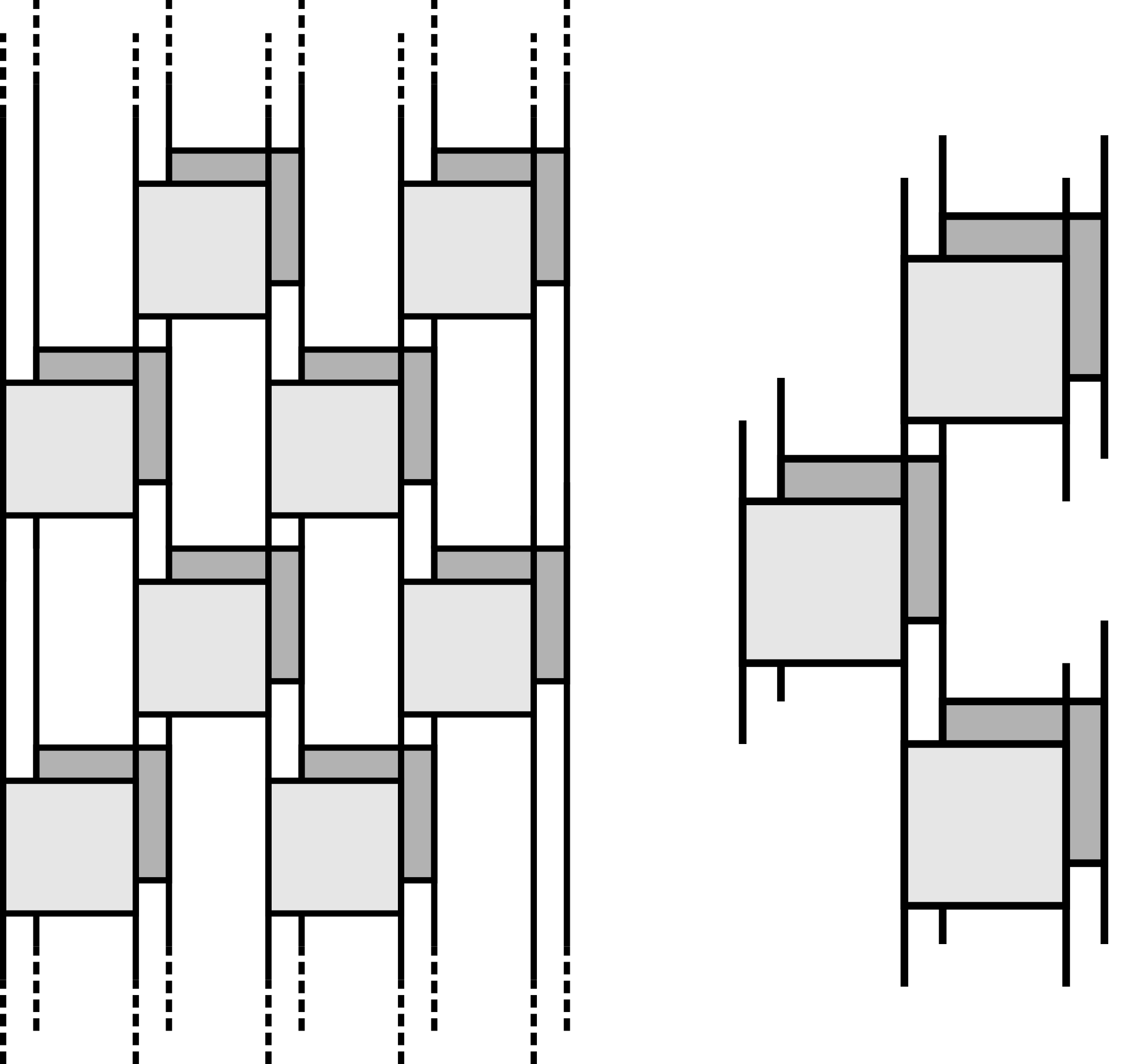}
	\put(-198,35){$U_{0,1}$}
	\put(-198,106){$U_{0,1}$}
	\put(-174,71){$U_{1,2}$}
	\put(-174,142){$U_{1,2}$}
	\put(-151,35){$U_{2,3}$}
	\put(-151,106){$U_{2,3}$}
	\put(-127,71){$U_{3,4}$}
	\put(-127,142){$U_{3,4}$}
	\put(-54,24){$a_0$}
	\put(-54,60){$b_0$}
	\put(-33,19){$a^*_0$}
	\put(-33,68){$b^*_0$}
	\put(-54,112){$a_1$}
	\put(-54,146){$b_1$}
	\put(-33,106){$a^*_1$}
	\put(-33,155){$b^*_1$}
	\caption{Circuit diagram for the SFF of a brickwork model, with time $t$ running vertically. Left: $K(t)$ with $t=2$ and open boundary conditions. The foreground (light) shows $\text{Tr}W(t)$ and the background (dark) its complex conjugate. Dashed lines exiting at the top of the figure are connected to those entering from below, giving independent traces $\text{Tr}W(t)$ and $\text{Tr}W^*(t)$. Right: segment of the doubled time evolution, highlighting notation for forward $(a_0 b_0 \ldots)$ and backward $(a^*_0 b^*_0 \ldots)$ orbits}
\label{fig:sff_diagram}
\end{figure}

Throughout this section we will work with single-site orbits, and so here we introduce some notation. Two unitary gates act on each site during a time step, so at time $t$ each single-site orbit is a string of $2t$ integers, each taking a value $0 \ldots (q-1)$. We denote the forward orbits appearing in $\text{Tr}W(t)$ by $(a_0 b_0 \ldots a_{t-1} b_{t-1})$. For integer $r$, $a_r$ represents the state of the site at time $r$, and $b_r$ represents the state of the site at time $(r+1/2)$, in the middle of the step. We denote the backward orbits appearing in $[\text{Tr}W(t)]^*$ by strings $(a^*_0 b^*_0 \ldots a^*_{t-1} b^*_{t-1})$. In Fig.~\ref{fig:sff_diagram} we illustrate the SFF of a brickwork model, as well as a pair of forward and backward single-site orbits.

\subsubsection{Construction of the transfer matrix}

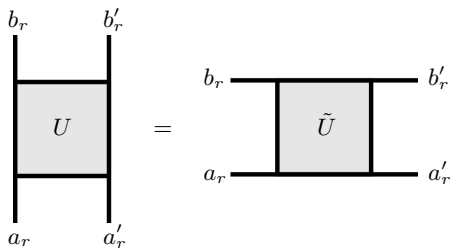
\begin{figure}
\includegraphics[width=0.3\textwidth]{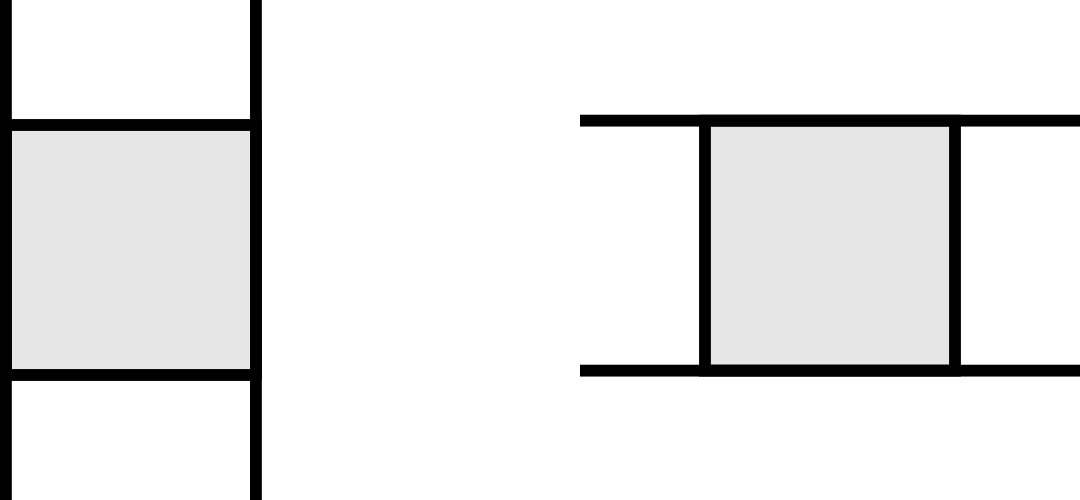}
\put(-138,33){$U$}
\put(-38,33){$\tilde U$}
\put(-155,-8){$a_r$}
\put(-155,74){$b_r$}
\put(-119,-8){$a_r'$}
\put(-119,74){$b_r'$}
\put(-100,33){$=$}
\put(-81,16){$a_r$}
\put(-81,52){$b_r$}
\put(4,16){$a_r'$}
\put(4,52){$b_r'$}
\caption{Components $U_{b_r b_r',a_r a_r'}$ of the $q^2 \times q^2$ unitary matrix $U$, and components $\tilde U_{a_r b_r,a_r'b_r'}$ of the $q^2 \times q^2$ non-unitary matrix $\tilde U$. Time runs vertically and space horizontally. Here, $U$ acts in the first half-step of the Floquet operator, so it describes evolution from time step $r$ to $(r+\frac{1}{2})$. Where $U$ acts on the state of a pair of sites $(x,x+1)$ at time $r$, $\tilde U$ acts on the state of site $(x+1)$ at times $(r,r+\frac{1}{2})$.}
\label{fig:spacegate}
\end{figure}

Consider a unitary gate $U_{x,x+1}$ acting on sites $(x,x+1)$ in the first half of the Floquet step and so with $x$ even (see Fig.~\ref{fig:brickwork}). This matrix evolves the state of the two sites from a time $r$ to $(r+\frac{1}{2})$, with $r$ integer. We can also think of $U_{x,x+1}$ as a non-unitary matrix acting on the orbit of site $(x+1)$ at the times $(r,r+\frac{1}{2})$, and refer to this matrix as $\tilde U_{x,x+1}$. Writing the components of $U_{x,x+1}$ as $[U_{x,x+1}]_{b_r b_r',a_r a_r'}$, where unprimed indices $a_r,b_r$ correspond to site $x$ and primed $a_r',b_r'$ to site $(x+1)$, the components of $\tilde U_{x,x+1}$ are ${[\tilde U_{x,x+1}]_{a_r b_r,a_r' b_r'} = [U_{x,x+1}]_{b_r b_r',a_r a_r'}}$, as illustrated in Fig.~\ref{fig:spacegate}.

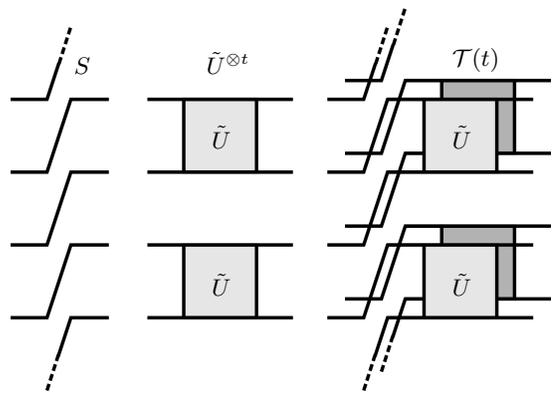
\begin{figure}
	\includegraphics[width=0.4\textwidth]{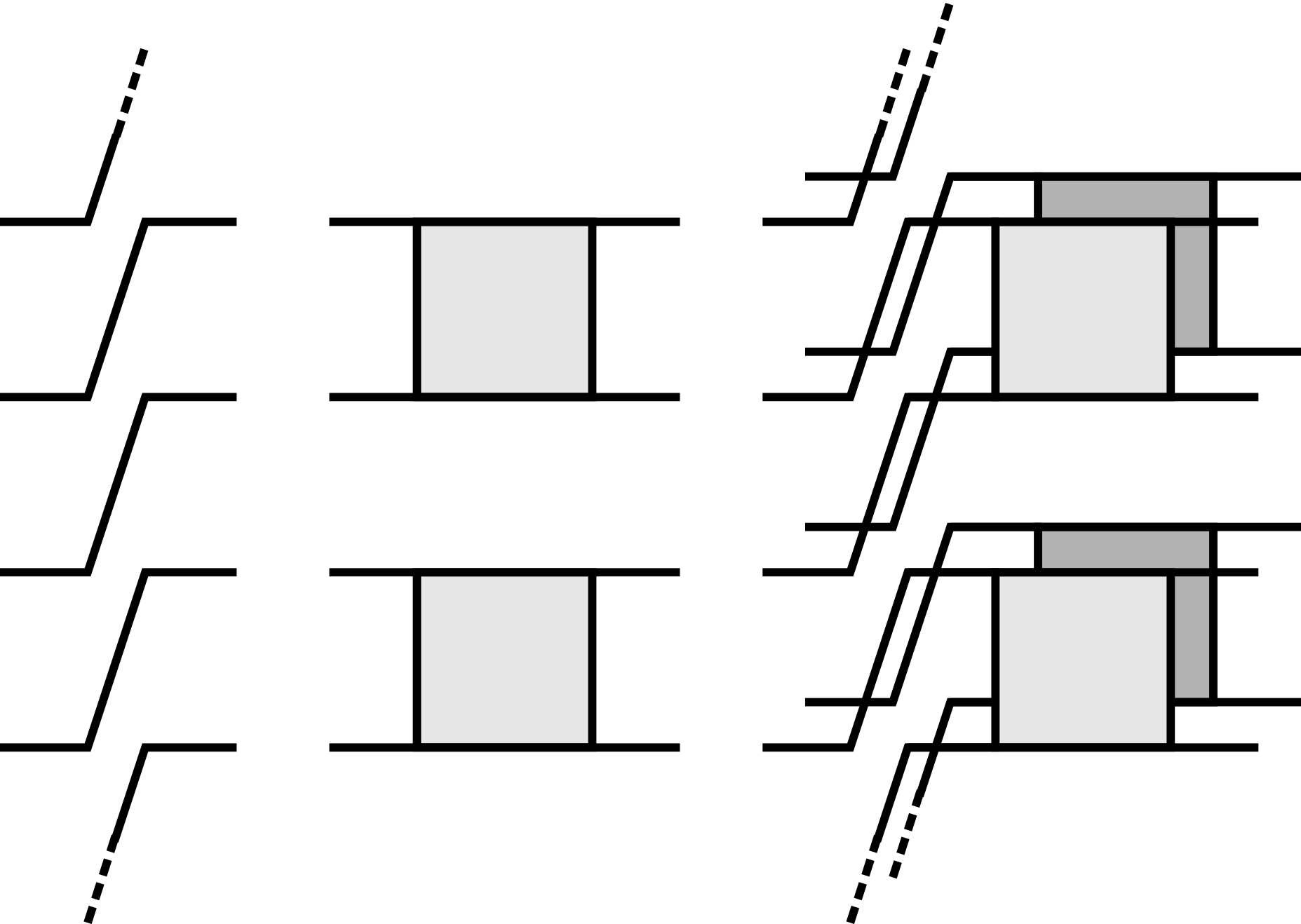}
	\put(-180,120){$S$}
	\put(-130,120){$\tilde U^{\otimes t}$}
	\put(-37,123){$\mathcal{T}(t)$}
	\put(-128,92){$\tilde U$}
	\put(-128,36){$\tilde U$}
	\put(-37,92){$\tilde U$}
	\put(-37,36){$\tilde U$}
	\caption{Construction of the transfer matrix generating $K(t) = |\text{Tr}W(t)|^2$, here for $t=2$. Dashed lines exiting above and entering below are connected. Left: the orthogonal matrix $S$. Centre: the tensor product of identical matrices $\tilde U$ acting in the space direction, $\tilde U^{\otimes t}$. Right: the transfer matrix $\mathcal{T}(t) = [S \tilde U^{\otimes t}] \otimes [S \tilde U^{\otimes t}]^*$. In the foreground we show the transfer matrix for the forward orbits, $S \tilde U^{\otimes t}$, and in the background the transfer matrix for backward orbits, $[S\tilde U^{\otimes t}]^*$.}
\label{fig:transfer}
\end{figure}

In $\text{Tr}W(t)$ each gate $U_{x,x+1}$ appears $t$ times. By taking a tensor product of the $t$ copies of $\tilde U_{x,x+1}$, we create a $q^{2t} \times q^{2t}$ matrix $\tilde U_{x,x+1}^{\otimes t}$ acting on the entire forward orbit of site $(x+1)$. To construct $\text{Tr}W(t)$ from the standard matrix multiplication of these operators, we introduce an orthogonal matrix $S$ which acts on orbits as a translation of one half-step in time,
\begin{align}
	S\ket{a_0 b_0 a_1 b_1 \ldots b_{t-1}} = \ket{b_0 a_1 b_1 \ldots a_0}.
\end{align}
The matrices $S$ and $\tilde U^{\otimes t}$ are illustrated in Fig.~\ref{fig:transfer}. With periodic boundary conditions
\begin{align}
	\text{Tr}W(t) = \text{tr}\Big[\tilde U^{\otimes t}_{0,1} S \tilde U^{\otimes t}_{1,2} S^{\rm{T}} \ldots S U^{\otimes t}_{L-1,0} S^{\rm{T}} \Big],
\end{align}
where $L$ is the number of sites, which is necessarily even with these boundary conditions. Here $\text{tr}$ is a trace over single-site orbits, to be distinguished from $\text{Tr}$, a trace over many-body states. Since the evolution operator is periodic we have $[S^2,\tilde U^{\otimes t}]=0$, where $S^2$ translates an orbit one full-step in time. Clearly $S^{2t} = \mathbb{1}$, where $\mathbb{1}$ is the $q^{2t} \times q^{2t}$ identity acting in the space of single-site orbits. Using these properties we can write $\text{Tr}W(t)$ in terms of one type of transfer matrix, $S \tilde U^{\otimes t}$,
\begin{align}
	\text{Tr}W(t) = \text{tr}[S \tilde U^{\otimes t}_{0,1} \ldots S \tilde U^{\otimes t}_{L-1,0} (S^{\rm{T}})^L].
\label{eq:TrWtperiodic}
\end{align}
A similar expression can be derived for open boundary conditions,
\begin{align}
	\text{Tr}W(t) = \bra{B} S^{\rm{T}} S \tilde U^{\otimes t}_{0,1} \ldots S \tilde U^{\otimes t}_{L-2,L-1} \ket{B},
\label{eq:TrWtopen}
\end{align}
where the boundary state $\ket{B}$ has components
\begin{align}
	\braket{a_0 b_0 a_1 b_1 \ldots a_{t-1}b_{t-1}|B} = \prod_{r=0}^{t-1} \delta_{b_r a_{r+1}},
\label{eq:boundarystate}
\end{align}
and is invariant under time translation by a full step, $S^2\ket{B} = \ket{B}$.

The transfer matrices which generate the SFF ${K(t)=|\text{Tr}W(t)|^2}$ are straightforwardly expressed as tensor products of the transfer matrices which generate $\text{Tr}W(t)$ and $[\text{Tr}W(t)]^*$,
\begin{align}
	\mathcal{T}_{x,x+1}(t) = [S \tilde U_{x,x+1}^{\otimes t} ]\otimes [S \tilde U_{x,x+1}^{\otimes t}]^*,
\end{align}
and are also illustrated in Fig.~\ref{fig:transfer}. The transfer matrix $\mathcal{T}_{x,x+1}(t)$ acts on the product space of forward and backward orbits at site $(x+1)$, which has dimension $q^{4t}$. These matrices have some obvious symmetries associated with translation in the time direction, which we now discuss (see also \cite{braun2020transition} for a related treatment in the context of the kicked Ising model).

The transfer matrices $\mathcal{T}_{x,x+1}(t)$ commute with the translation operators for a full time step of either of the forward and backward orbits,
\begin{align}
	[S^2 \otimes \mathbb{1},\mathcal{T}_{x,x+1}] = [\mathbb{1} \otimes S^2,\mathcal{T}_{x,x+1}] = 0.
\end{align}
Since $(S^2)^t = \mathbb{1}$, $\mathcal{T}_{x,x+1}$ can be block-diagonalised into $t^2$ sectors, having eigenvalues $e^{i \omega_+}$ under $S^2 \otimes \mathbb{1}$ and $e^{i \omega_-}$ under $\mathbb{1} \otimes S^2$. The frequencies are $\omega_{\pm} = 2\pi n/t$ for $n = 0 \ldots (t-1)$ defined modulo $t$. The transfer matrices for each bond $(x,x+1)$ are different. Our focus, however, is on the full ensemble average. Following such an average all the transfer matrices are the same.

\subsubsection{Average transfer matrix}
The notion of orbit pairing takes on a concrete meaning in the ensemble average of the SFF. For identically and independently distributed gates, $\overline{K}(t)$ is determined by powers of a single averaged transfer matrix, $\overline{\mathcal{T}}(t)$. From here on, $\overline{\mathcal{T}}(t)$ will be at the centre of our discussion.

To calculate $\overline{\mathcal{T}}(t)$ we must Haar-average the tensor product $\tilde U^{\otimes t} \otimes [\tilde U^{\otimes t}]^*$. Writing the first indices of $\tilde U^{\otimes t}$ and $[\tilde U^{\otimes t}]^*$ as orbits $(a_0 b_0 \ldots a_{t-1} b_{t-1})$ and $(a^*_0 b^*_0 \ldots a^*_{t-1} b^*_{t-1})$, respectively, the only non-vanishing matrix elements of their average tensor product have $a_r = a^*_{\sigma(r)}$ and $b_r = b^*_{\tau(r)}$, for $r=0\ldots (t-1)$, where $\sigma$ and $\tau$ denote permutations of $t$ objects \cite{samuel1980integrals,brouwer1996diagrammatic}. The same is true of the second indices. To arrive at a succinct expression for the average transfer matrix, it is convenient to introduce vectors $\ket{\sigma,\tau}$ in the product space of forward and backward orbits which have nonzero entries only where indices are paired in this way:
\begin{align}
	\braket{a_0 b_0 \ldots a^*_0 b^*_0 \ldots|\sigma,\tau} = \prod_{r=1}^{t-1} \delta_{a_r a^*_{\sigma(r)}}\delta_{b_r b^*_{\tau(r)}}.
\label{eq:permutation_states}
\end{align}
Note that the vectors $\ket{\sigma,\tau}$ are neither normalised nor orthogonal. The average of $\tilde U^{\otimes t} \otimes [\tilde U^{\otimes t}]^*$ can be expressed in terms of the vectors $\ket{\sigma,\tau}$ and the ($q^2$-dependent) Weingarten functions $\text{Wg}(\sigma \tau^{-1})$, here taking the composed permutation $\sigma \tau^{-1}$ as an argument.

From the Haar average over the gate we find the average transfer matrix
\begin{align}
	\overline{\mathcal{T}}(t) = \mathcal{S} \sum_{\sigma \tau} \text{Wg}(\sigma \tau^{-1})\ket{\sigma,\tau}\bra{\sigma, \tau},
\label{eq:Tmatrixaverage}
\end{align}
where the sum is over all pairs of permutations of $t$ objects. We have also introduced the doubled half-step time-translation operator $\mathcal{S} = S \otimes S$. Only $t$ terms in the sum in Eq.~\eqref{eq:Tmatrixaverage} contribute in the large-$q$ limit, and these are essentially the local orbit pairings discussed in Sec.~\ref{sec:breakdown}. We elaborate on this in Appendix~\ref{sec:largeq}. 

As discussed above, the transfer matrices have a block structure associated with symmetry under time translation. Focussing on the average $\overline{\mathcal{T}}(t)$, we write the left and right eigenvectors in the block $\omega_+,\omega_-$ as $\bra{\omega_+ \omega_- \alpha_L;t}$ and $\ket{\omega_+ \omega_- \alpha_R;t}$, respectively, and the corresponding eigenvalues as $\lambda(\omega_+\omega_-\alpha;t)$. Here $\alpha=0,1,\ldots$ label eigenstates in descending order of magnitude, so $\lambda(\omega_+\omega_-,0;t)$ is the leading eigenvalue in the block $\omega_+,\omega_-$. The spectral decomposition of the average transfer matrix $\overline{\mathcal{T}}(t)$ is then
\begin{align}
\begin{split}
	\overline{\mathcal{T}}(t) = \sum_{\omega_+ \omega_- \alpha}&\lambda(\omega_+ \omega_- \alpha;t) \\ &\times \ket{\omega_+\omega_- \alpha_R;t} \bra{\omega_+\omega_- \alpha_L;t}.
\label{eq:T_spectral_decomposition}
\end{split}
\end{align}

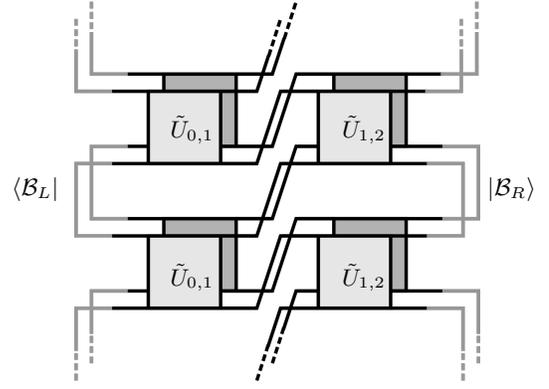
\begin{figure}\includegraphics[width=0.3\textwidth]{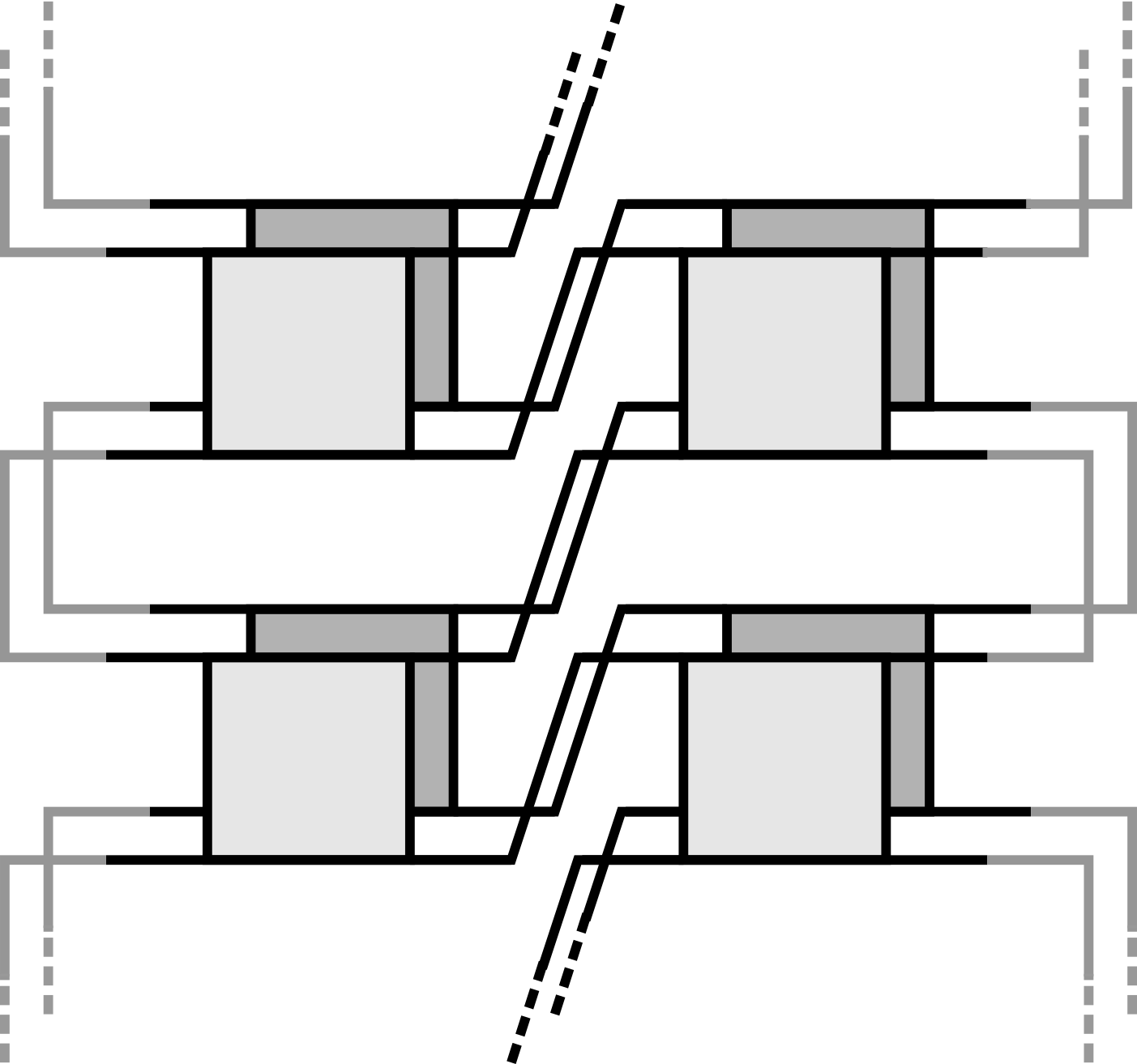}
\put(-177,70){$\bra{\mathcal{B}_L}$}
\put(3,70){$\ket{\mathcal{B}_R}$}
\put(-117,36){$\tilde U_{0,1}$}
\put(-117,92){$\tilde U_{0,1}$}
\put(-52,36){$\tilde U_{1,2}$}
\put(-52,92){$\tilde U_{1,2}$}
\caption{The SFF with open boundary conditions in terms of the transfer matrix (for $t=2$ and $L=3$ sites), $K(t) = \bra{\mathcal{B}_L} \mathcal{T}_{0,1} \mathcal{T}_{1,2} \ket{\mathcal{B}_R}$. The vectors $\ket{\mathcal{B}_{L,R}}$ encode the boundary conditions.}
\label{fig:opencircuit}
\end{figure}

We are now in a position to write down expressions for the average SFF in terms of the average transfer matrix. With periodic boundary conditions, where $L$ is necessarily even,
\begin{align}
\begin{split}
	\overline{K}(t) &= \text{tr}[\overline{\mathcal{T}}^L(t) (\mathcal{S}^{\rm{T}})^L] \\
	&= \sum_{\omega_+ \omega_- \alpha} [\lambda(\omega_+ \omega_- \alpha;t)]^L e^{-i(L/2)(\omega_+ + \omega_-)}.
\label{eq:Kpbc}
\end{split}
\end{align}
For the case of open boundary conditions we define the doubled boundary vectors $\ket{\mathcal{B}_R} \equiv \ket{B} \otimes \ket{B}$ and $\bra{\mathcal{B}_L} \equiv \bra{\mathcal{B}_R}\mathcal{S}^T$, for the right and left ends of the chain, respectively. These vectors are in the $\omega_+ = \omega_- =0$ sector. From Eq.~\eqref{eq:TrWtopen} we then have
\begin{align}
\begin{split}
	\overline{K}(t) &= \bra{\mathcal{B}_L}\overline{\mathcal{T}}^{L-1}\ket{\mathcal{B}_R},\\
	 &= \sum_{\alpha} [\lambda(0,0,\alpha;t)]^{L-1}  \\ &\times \braket{ \mathcal{B}_L |0,0,\alpha_R;t} \braket{0,0,\alpha_L;t|\mathcal{B}_R},
\label{eq:Kobc}
\end{split}
\end{align}
and this is illustrated in Fig.~\ref{fig:opencircuit}.

An immediate question is how, for general $L$, RMT level statistics emerge from the average transfer matrix $\overline{\mathcal{T}}(t)$ at late times. We anticipate that in the diagonal regime the result $\overline{K}(t) = t$ is expressible as a sum over the $t$ global diagonal orbit pairings, each contributing unity. In the language of the average transfer matrix $\overline{\mathcal{T}}(t)$ this corresponds to having $t$ leading eigenvalues which at late times approach unity.

Different behaviour sets in beyond $t_{\text{H}}=q^L$, where the average SFF plateaus at $\overline{K}(t)=q^L$. To see how this behaviour could arise, note that the number of nonzero subleading eigenvalues of $\overline{\mathcal{T}}(t)$ grows very rapidly with $t$, but that the contributions of these eigenvalues to the SFF are suppressed for large $L$. In Appendix \ref{sec:blockdiagonalisation} we show, based on exact diagonalisation of $\overline{\mathcal{T}}(t)$ for $t \leq 5$, that the proliferation of subleading eigenvalues with increasing $t$ is responsible for the plateau.

Our analysis is focused on the regime $1 \ll t \ll t_{\rm{H}}$ for large $L$, and therefore on the late-time behaviour of the $t$ leading eigenvalues of $\overline{\mathcal{T}}(t)$. General features then emerge because the only microscopic timescale in $\overline{\mathcal{T}}(t)$ is $q^2$, the Heisenberg time for a single gate. As we show in Sec.~\ref{sec:twisted}, each of the sectors $\omega = \omega_+ = -\omega_-$ contains one of the $t$ leading eigenvalues, and since our attention is effectively limited to these and the corresponding eigenvectors, it is convenient to introduce the shorthand notation
\begin{align}
\begin{split}
	\lambda(\omega,t) &\equiv \lambda(\omega,-\omega,0;t) \\
	\ket{\omega,t;R} &\equiv \ket{\omega,-\omega,0_R;t} \\
	\bra{\omega,t;L} &\equiv \bra{\omega,-\omega,0_L;t}.
\end{split}
\end{align}
In the regime of interest the transfer matrix is too large to compute directly, so in the next two sections we study the length-scaling of the average SFF and related objects with a variety of boundary conditions. Through this we will determine $\lambda(\omega,t)$ as well as certain properties of the corresponding eigenvectors.

\subsection{Open boundary conditions}\label{sec:open}
In this section we study the length-scaling of the average SFF $\overline{K}(t)$ with open boundary conditions. From Eq.~\eqref{eq:Kobc} this probes the $\omega_+=\omega_-=0$ block of the average transfer matrix. Based on the discussion in Sec.~\ref{sec:breakdown} we anticipate that the leading correction to the SFF relative to RMT can be understood in terms of a domain wall, and so here we aim also to infer an effective domain wall tension.

If the leading eigenvalue in the $\omega_+=\omega_-=0$ sector dominates the SFF in Eq.~\eqref{eq:Kobc}, we have
\begin{equation}
	\overline{K}(t) = \lambda^{L-1}(0,t) \braket{ \mathcal{B}_L |0,t;R} \braket{0,t;L|\mathcal{B}_R} + \ldots,
\label{eq:Kobc_leadingterm}
\end{equation}
where the ellipses indicate contributions from subleading eigenvalues. By following the $L$-dependence of $\overline{K}(t)$ for each time $t$ we extract the leading eigenvalue $\lambda(0,t)$ and the overlap of the corresponding eigenvector with the boundary states, $\braket{ \mathcal{B}_L |0,t;R} \braket{0,t;L|\mathcal{B}_R}$. The results are shown in Fig.~\ref{fig:Kopen_eigenvalue_overlap}, and details of the analysis are presented in Appendix \ref{sec:Lscaling}. We find that $\lambda(0,t)$ approaches unity from above at large $t$. We find also that the overlap, shown in the inset, approaches $t$ at late times. This is exactly the value expected if the leading eigenvector of $\overline{\mathcal{T}}(t)$ represents a locally diagonal orbit pairing, as we discuss in Sec.~\ref{sec:pairingdomains}.

\begin{figure}
	\hspace{-0.1in}
	\includegraphics[width=0.47\textwidth]{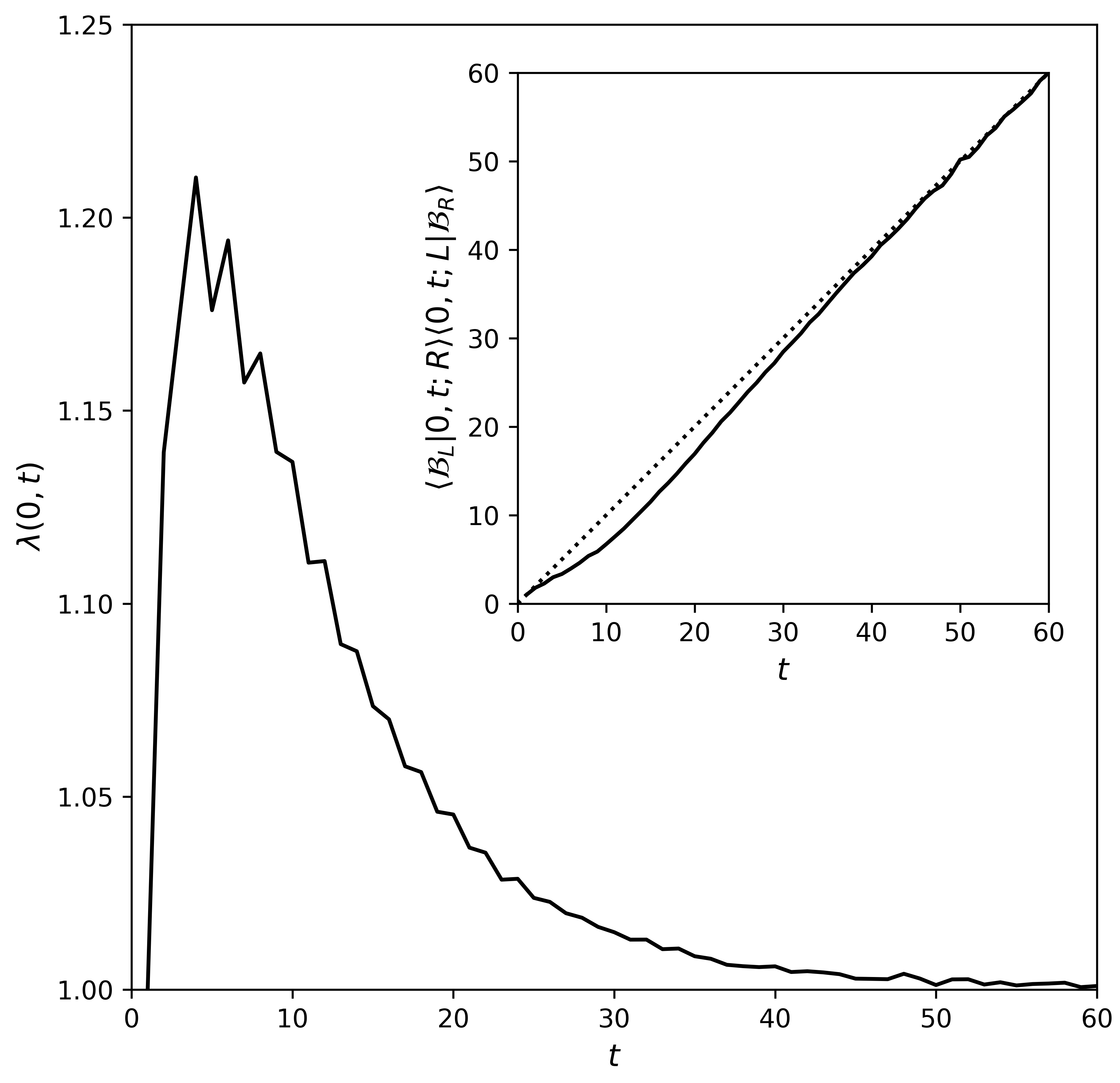}
	\caption{The leading eigenvalue $\lambda(0,t)$ of the average transfer matrix $\overline{\mathcal{T}}(t)$. In the inset the solid line shows the overlap of the corresponding eigenvectors with the boundary states $\ket{\mathcal{B}_{L,R}}$, which approaches $t$ (dotted) at late times.}
	\label{fig:Kopen_eigenvalue_overlap}
\end{figure}

\begin{figure}
	\hspace{-0.1in}
	\includegraphics[width=0.47\textwidth]{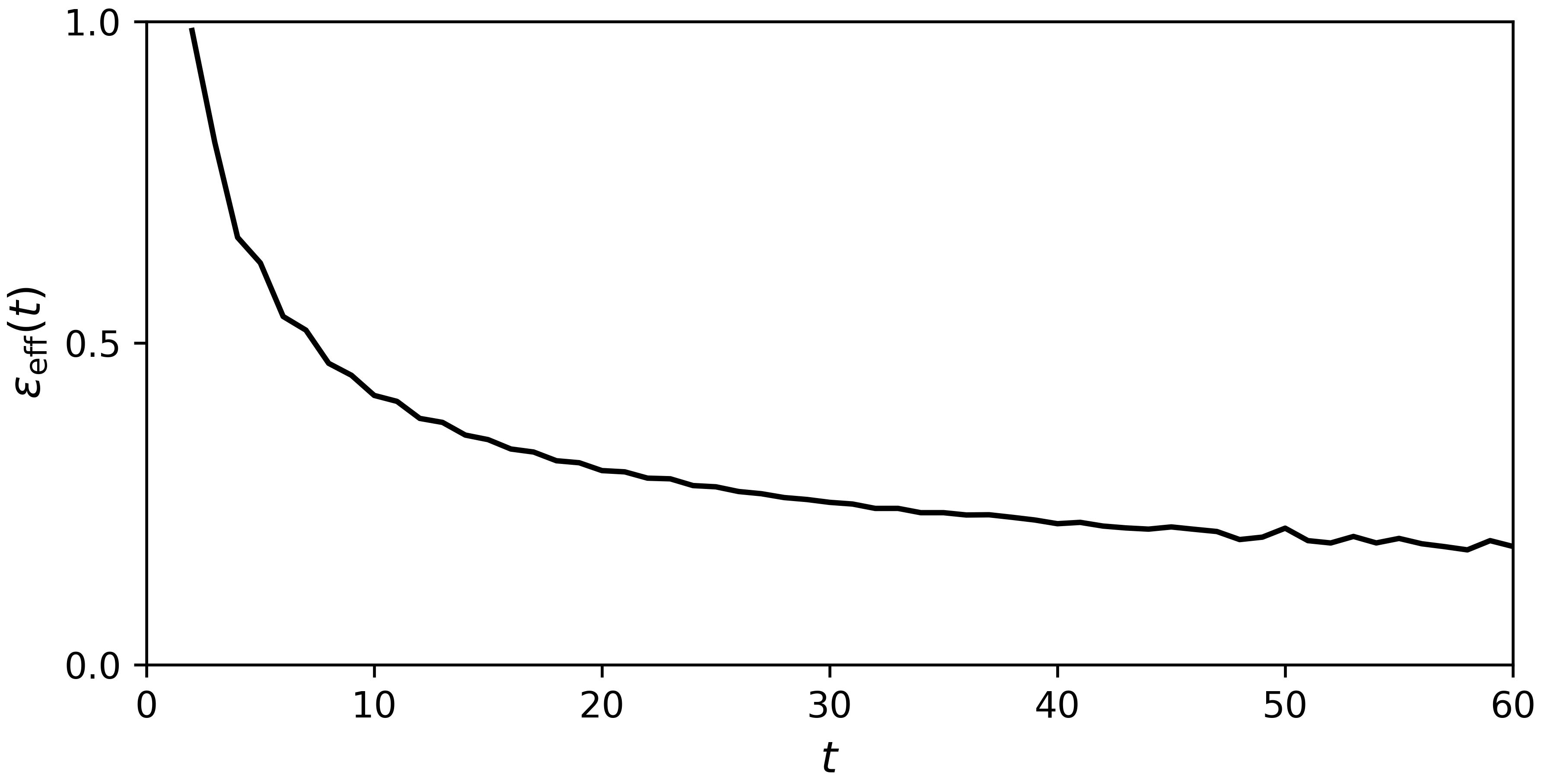}
	\caption{Evolution of the effective tension $\varepsilon_{\rm{eff}}(t)$ defined in Eq.~\eqref{eq:effectivetension}, calculated using numerical results for the leading eigenvalue $\lambda(0,t)$ in Fig.~\ref{fig:Kopen_eigenvalue_overlap}.}
\label{fig:effectivetension}
\end{figure}

Expanding the average SFF around the RMT result, at large $t$ we then have
\begin{equation}
	\overline{K}(t) = t + t(L-1)\delta \lambda(0,t) + \ldots,
\label{eq:Kobc_expansion}
\end{equation}
where we have defined $\delta \lambda(0,t)=\lambda(0,t)-1$. We can interpret the deviation from RMT as a domain wall contribution in analogy with Eq.~\eqref{eq:Kobcphase}. Here $(L-1)$ is the translational entropy, and $\delta \lambda(0,t)$ in Eq.~\eqref{eq:Kobc_expansion} plays the role of $(t-1)e^{-\varepsilon t}$, the sum of statistical weights of the $(t-1)$ different domain walls. This interpretation motivates the definition of an effective domain wall tension $\varepsilon_{\text{eff}} = \varepsilon_{\text{eff}}(t)$ via
\begin{equation}
	e^{-\varepsilon_{\text{eff}}(t) t} = \frac{\delta \lambda(0,t)}{t-1}.
\label{eq:effectivetension}
\end{equation}
In the model of Ref.~\cite{chan2018spectral} all domain walls have the same tension $\varepsilon$, so there $\varepsilon_{\text{eff}} = \varepsilon$, which is also time-independent. This is not the case in general \cite{chan2018spectral,mace2019quantum}. We show in Fig.~\ref{fig:effectivetension} that the effective domain wall tension in the $q=2$ Haar-RFC decreases monotonically with time. To understand this decrease, suppose that a domain wall can be located on a particular bond with corresponding gate $U$, and that we can associate a tension $\varepsilon(U)$ with this gate. The average of the statistical weight $e^{-\varepsilon(U)t}$ over $U$ is at late times increasingly dominated by those gates with small $\varepsilon(U)$. Consequently, it is natural for the effective tension derived from this average to decrease with time.

We can relate this behaviour to the Thouless time $t_{\text{Th}}$ as follows. From Eqs.~\eqref{eq:Kobc_expansion} and \eqref{eq:effectivetension} we see that the crossover to RMT occurs when the decreasing statistical weight $e^{-\varepsilon_{\text{eff}}(t)t}$ associated with a domain wall overwhelms the translational entropy $(L-1)$. If the effective domain wall tension approaches a nonzero constant $\varepsilon_{\text{eff}}(\infty)$ at late times, the timescale for the crossover of the SFF to RMT behaviour is $t_{\text{Th}} = \ln L/ \varepsilon_{\text{eff}}(\infty)$.

In this section we have determined one of the $t$ leading eigenvalues of the average transfer matrix. Next we will determine the other $(t-1)$ and provide further evidence for domain walls in the orbit pairing.

\subsection{Twisted boundary conditions}\label{sec:twisted}

In order to access $t$ different symmetry blocks of $\overline{\mathcal{T}}(t)$ here we impose local diagonal orbit pairings at the two ends of the system. This approach allows us to force domain walls into the many-body orbit pairing, and gives us information on $\lambda(\omega,t)$, $\bra{\omega,t;L}$ and $\ket{\omega,t;R}$ for all $\omega$. By determining all of these leading eigenvalues, we will show that they alone are sufficient to calculate the SFF with periodic boundary conditions to a remarkable degree of accuracy.

\subsubsection{Domain walls}

\begin{figure}
\includegraphics[width=0.35\textwidth]{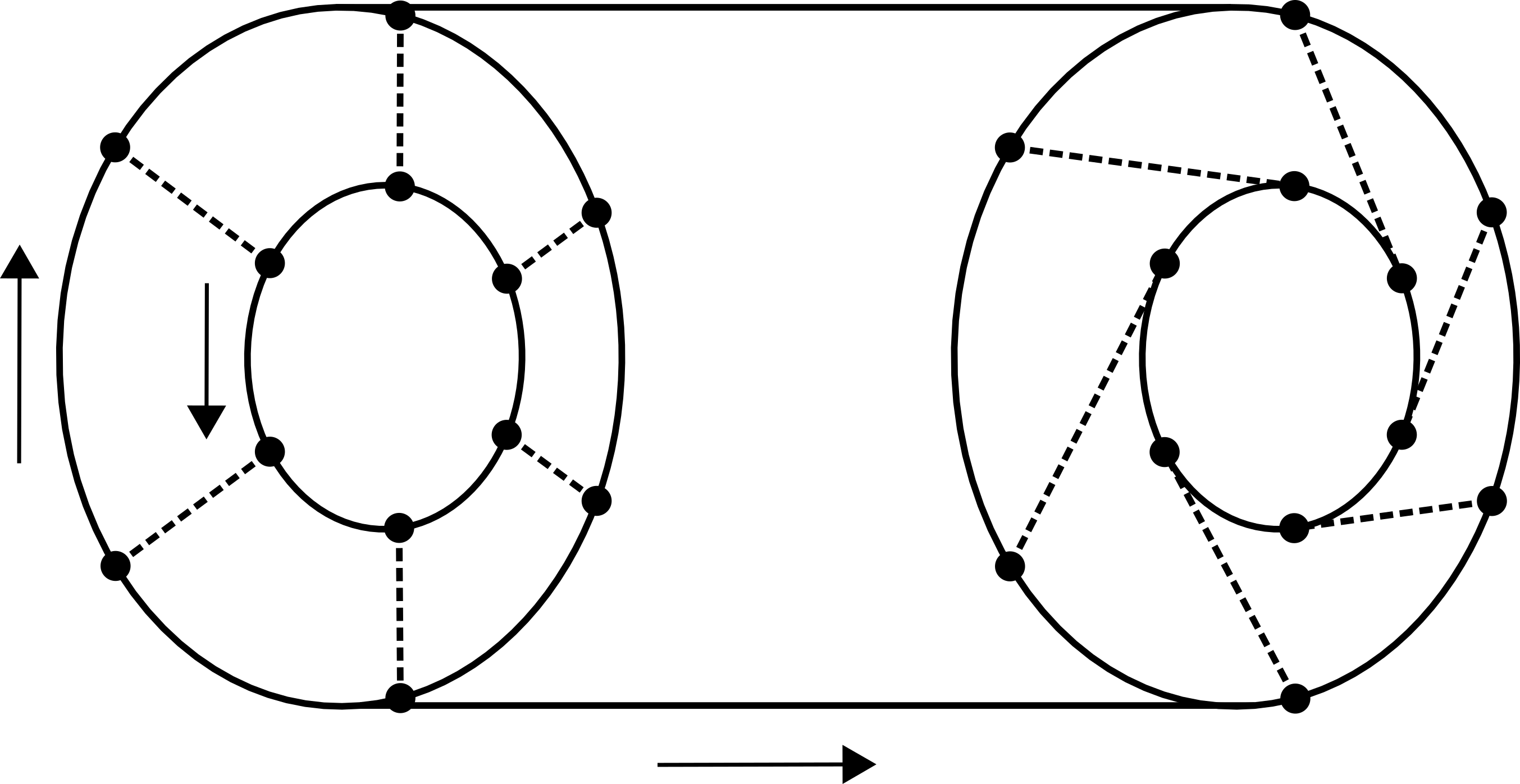}
\put(-138,95){$a_{r+1}$}
\put(-175,78){$a_{r}$}
\put(-144,56){$a^*_{r}$}
\put(-185,48){$t$}
\put(-93,-5){$x$}
\\
\vspace{0.2in}
\includegraphics[width=0.3\textwidth]{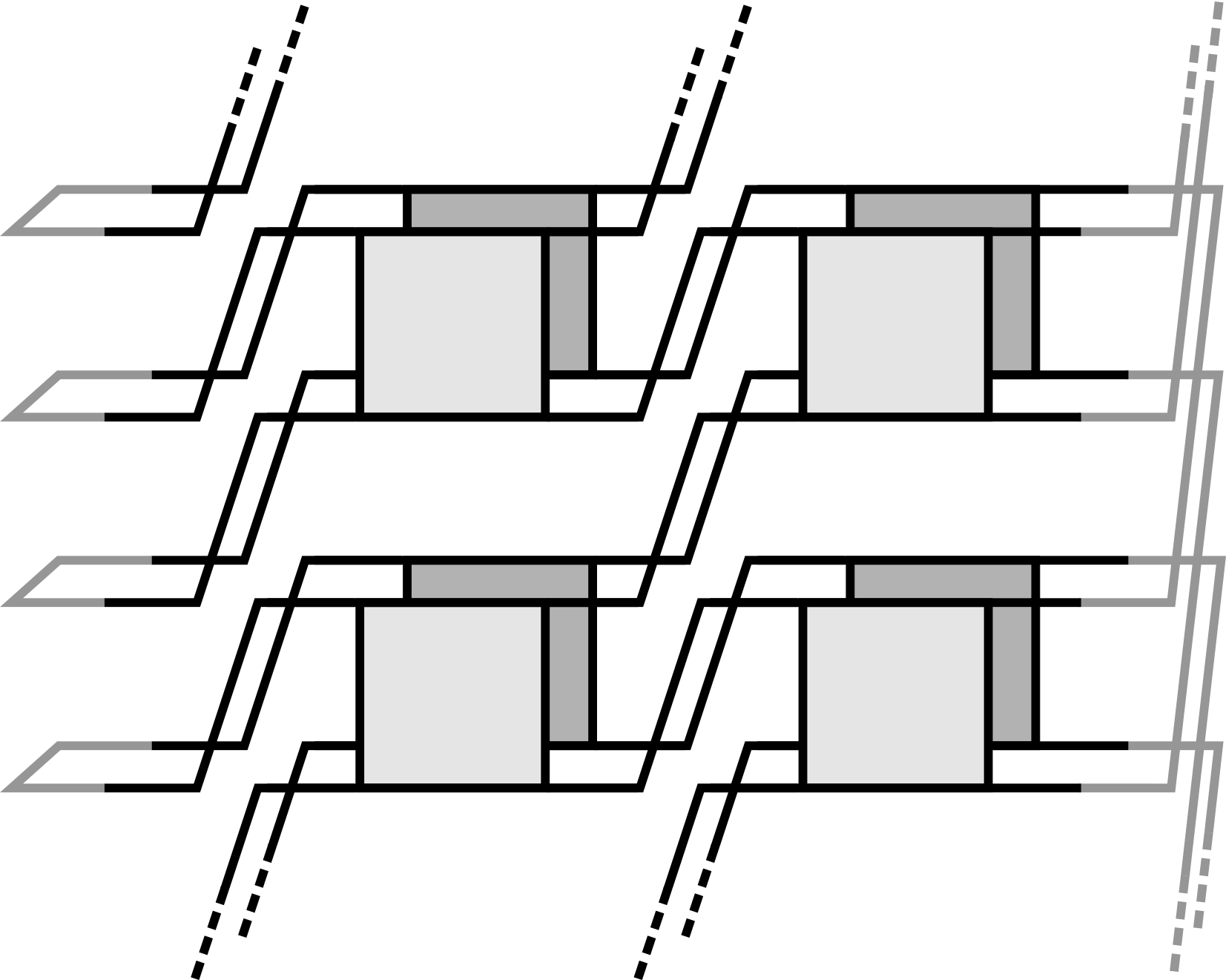}
\put(-175,50){$\bra{s_L}$}
\put(5,50){$\ket{s_R}$}
\put(-102,31){$\tilde U_{0,1}$}
\put(-102,78){$\tilde U_{0,1}$}
\put(-47,31){$\tilde U_{1,2}$}
\put(-47,78){$\tilde U_{1,2}$}
\caption{Twisted boundary conditions on a pair of forward and backward orbits, $Z(s,t)$. At the top we show the two orbit pairings of Fig.~\ref{fig:diagonal} imposed on the left and right sites, $s_L=0$ and $s_R=1$ so $s=s_R-s_L=1$. Below this we show $Z(1,2)$ for $L=3$ sites in terms of the transfer matrices $\mathcal{T}$ for bonds $(0,1)$ and $(1,2)$, $Z(1,2) = \bra{s_L=0} \mathcal{T}_{0,1}(2) \mathcal{T}_{1,2}(2) \ket{s_R=1}$. The states $\bra{s_L}$ and $\ket{s_R}$ are represented by grey lines.}
\label{fig:twistdiagram}
\end{figure}

The local diagonal orbit pairings are represented by vectors $\ket{s}$ in the product space of forward and backward orbits. The pairing $\ket{s}$ is a normalised member of the set of vectors $\ket{\sigma,\tau}$ introduced in Eq.~\eqref{eq:permutation_states}, with $\sigma=\tau$ the permutation mapping $r \to (r+s)$ modulo $t$, for $s=0,1 \ldots (t-1)$. The components are
\begin{equation}
	\braket{a_0 b_0 \ldots; a^*_0 b^*_0 \ldots |s} = \frac{1}{q^{t}}\prod_{r=0}^{t-1} \delta_{a_r a_{r+s}^*}\delta_{b_r b_{r+s}^*}.
\label{eq:spairedstate}
\end{equation}
Imposing the orbit pairing $s_L$ on the left and $s_R$ on the right of an $L$-site system we find
\begin{align}
	Z(s_R-s_L,t) &= \braket{s_L|\mathcal{T}_{0,1}(t)\ldots \mathcal{T}_{L-2,L-1}(t)|s_R}.
\label{eq:Zsnoaverage}
\end{align}
For $s_L=0$ and $s_R=1$ these boundary conditions correspond to the two pairings shown in Fig.~\ref{fig:diagonal}. We illustrate $Z(s,t)$ in Fig.~\ref{fig:twistdiagram}. For $s_L = s_R$ we force equal-time pairings, and for $s_L \neq s_R$ we force a domain wall into the orbit pairing. In practice we impose these boundary conditions using a Monte-Carlo method, and give details on the implementation in Appendix \ref{sec:montecarlopairing}. For the particular case of $Z(0,t)$ these boundary conditions can also be thought of as local couplings to Markovian baths. We discuss this further in Sec.~\ref{sec:bath}.

\begin{figure}
\hspace{-0.1in}
\includegraphics[width=0.47\textwidth]{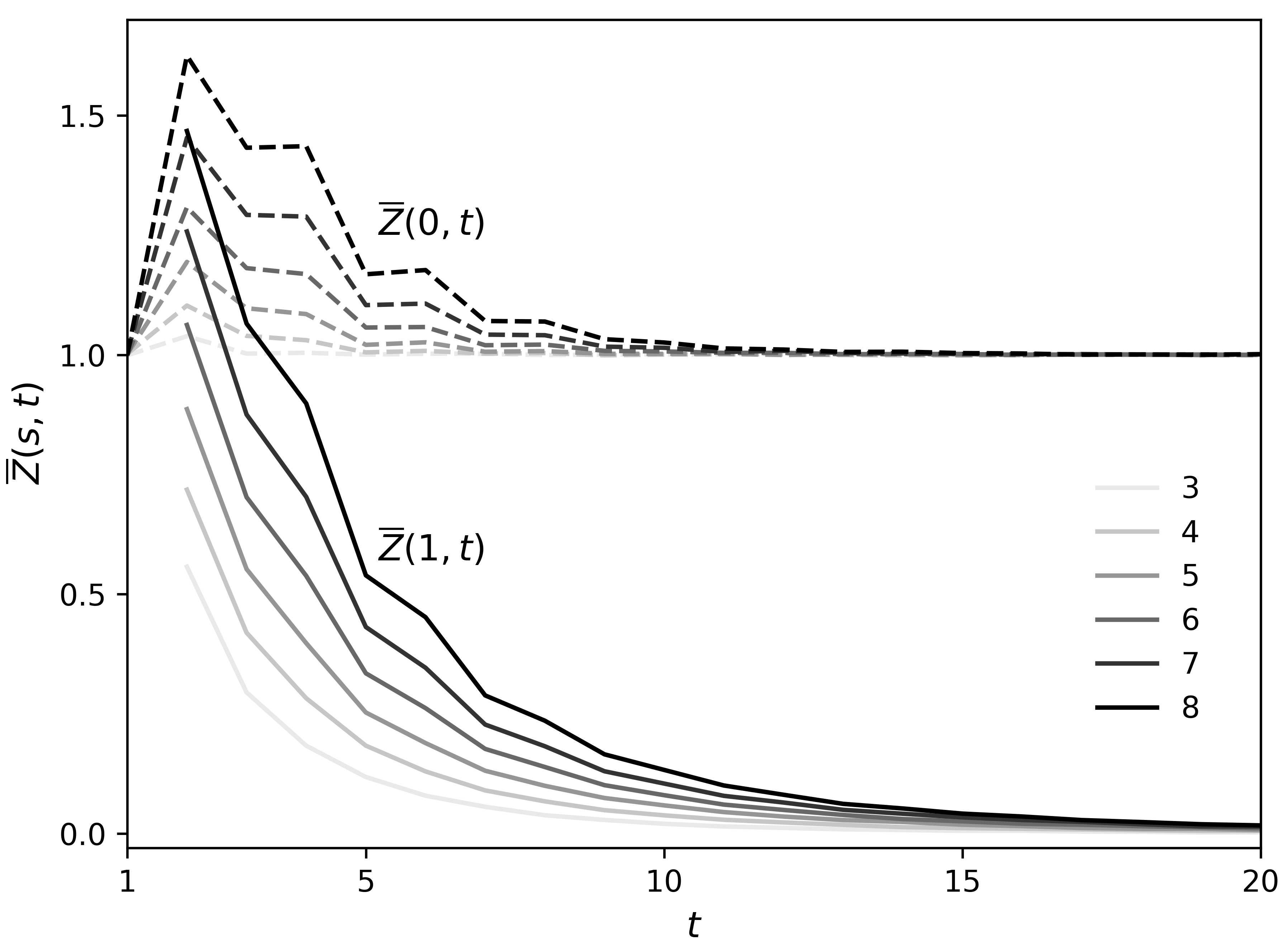}
\caption{$\overline{Z}(s,t)$, defined in Eq.~\eqref{eq:Zs}, with system size $L$ shown on the legend. For $s=1$ (solid) the boundary conditions force a domain wall into the orbit pairing, whereas for $s=0$ (dashed) they do not. See main text for discussion.}
\label{fig:domain_wall}
\end{figure}

Via the ensemble average,
\begin{equation}
	\overline{Z}(s,t) = \braket{0|\overline{\mathcal{T}}^{L-1}(t)|s},
\label{eq:Zs}
\end{equation}
we probe the average transfer matrix $\overline{\mathcal{T}}(t)$. Suppose that the RMT behaviour $\overline{K}(t)=t$ at late times arises from the sum over the $t$ possible global diagonal orbit pairings. The boundary conditions imposed for $Z(0,t)$ are compatible with one of these pairings, whereas those for $Z(s \neq 0,t)$ are not compatible with any of them. We therefore expect that $\overline{Z}(0,t) \to 1$ and $\overline{Z}(s \neq 0,t) \to 0$ with increasing $t$, on the grounds that contributions from pairs of many-body orbits with domain walls should vanish at late times. We show that this is indeed the case in Fig.~\ref{fig:domain_wall} for $s=0$ and $s=1$ for various $L$, and in Fig.~\ref{fig:Ksw}(a) for $L=8$ and various $s$. Making a connection with the model of Ref.~\cite{chan2018spectral} at large $q$ (see also Sec.~\ref{sec:breakdown}), there we expect $\overline{Z}(s \neq 0,t) \simeq (L-1)e^{-\varepsilon t}$ at late times, where $(L-1)$ is the translational entropy of the domain wall. The observed increase of $\overline{Z}(s,t)$ with $L$ in Fig.~\ref{fig:domain_wall} is then understood as a consequence of this entropy, and furthermore demonstrates the failure of the diagonal approximation. In principle, and in contrast to Ref.~\cite{chan2018spectral}, we expect the decay rate of $\overline{Z}(s \neq 0,t)$ to depend on $s$ (see Appendix~\ref{sec:largeq}), and behaviour of this kind is evident in Fig.~\ref{fig:Ksw}(a).

\subsubsection{Leading eigenvalues}

Using our numerics on $\overline{Z}(s,t)$ we can extract the leading eigenvalues $\lambda(\omega,t)$. The local diagonal pairings $\ket{s}$ defined in Eq.~\eqref{eq:spairedstate} are eigenvectors of $\mathcal{S}^2$ with unit eigenvalue, so they are linear combinations of vectors in sectors with $\omega_+ = -\omega_-$. We define their Fourier transform in sector $\omega = \omega_+ = -\omega_-$ as
\begin{equation}
	\ket{\omega} = \frac{1}{\sqrt{t}}\sum_{s=0}^{t-1} e^{-i \omega s} \ket{s}.
\label{eq:fourier_local_diagonal}
\end{equation}
Using Eq.~\eqref{eq:fourier_local_diagonal} and the spectral decomposition of $\overline{\mathcal{T}}(t)$ in Eq.~\eqref{eq:T_spectral_decomposition} we have the Fourier transform of $\overline{Z}(s,t)$,
\begin{align}
\begin{split}
 \label{eq:Kw}
	\overline{\mathcal{Z}}(\omega,t) &= \sum_{s=0}^{t-1} e^{-i \omega s} \overline{Z}(s,t) \\&= \sum_{\alpha} [\lambda(\omega,-\omega,\alpha;t)]^{L-1} \\ &\times \braket{\omega|\omega,-\omega,\alpha_R;t}\braket{\omega,-\omega,\alpha_L;t|\omega}.
\end{split}
\end{align}
For fixed $t$ and large $L$,
\begin{align}
	\overline{\mathcal{Z}}(\omega,t) = \lambda^{L-1}(\omega,t) \braket{\omega|\omega,t;R}\braket{\omega,t;L|\omega} + \ldots,
\label{eq:Kw_expansion}
\end{align}
where the ellipses represent the contributions of subleading eigenvalues.

\begin{figure}
\hspace{-0.1in}
\includegraphics[width=0.47\textwidth]{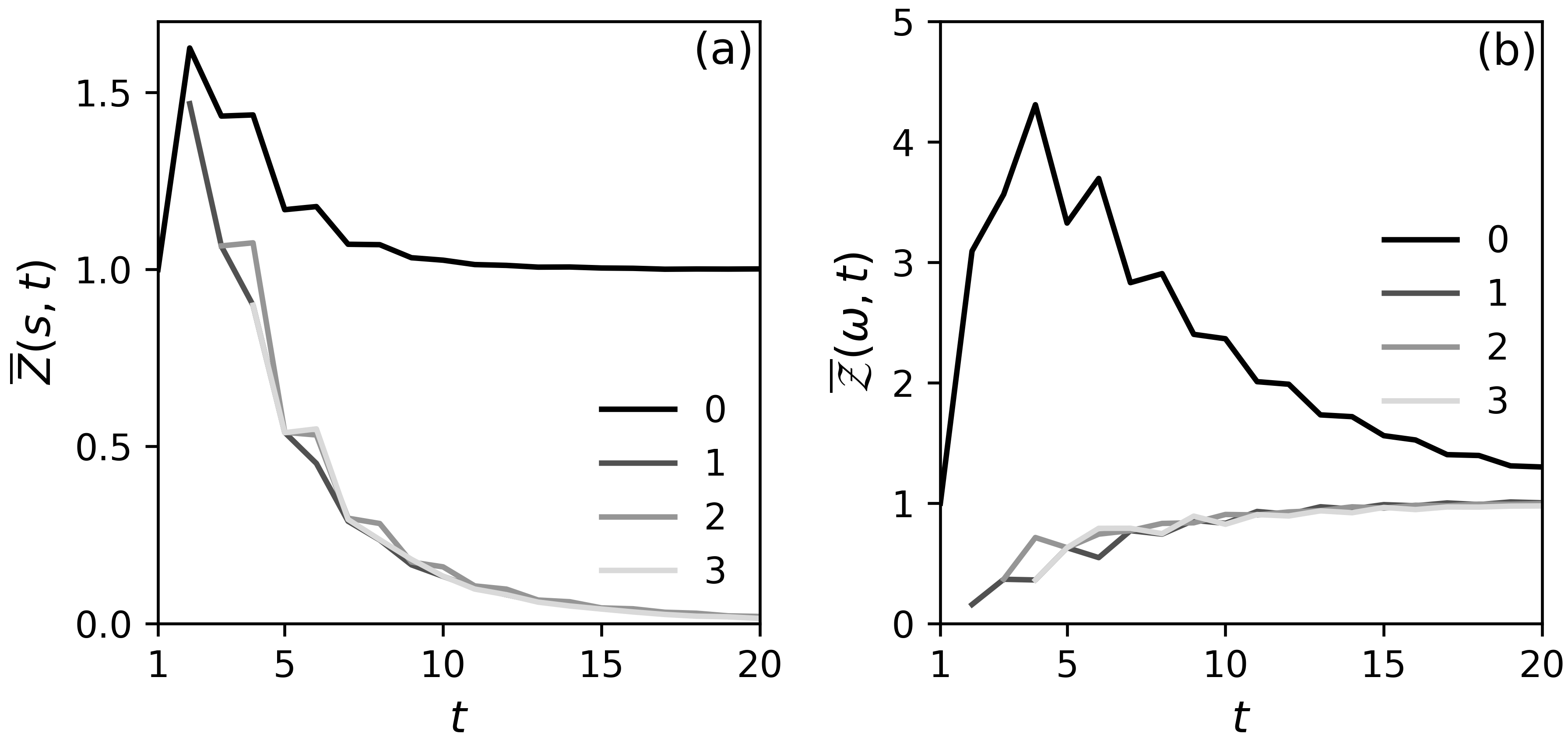}
\caption{$\overline{Z}(s,t)$ and $\overline{\mathcal{Z}}(\omega,t)$ for $q=2$ and $L=8$. (a) $\overline{Z}(s,t)$, with $s$ on the legend. $\overline{Z}(0,t)$ tends to unity and $\overline{Z}(s \neq 0,t)$ decays approximately exponentially as in Fig.~\ref{fig:domain_wall}. (b) $\overline{\mathcal{Z}}(\omega,t)$, defined in Eq.~\eqref{eq:Kw} as the Fourier components of $\overline{Z}(s,t)$ with respect to $s$; the legend shows $n=\omega t/2 \pi$.}
\label{fig:Ksw}
\includegraphics[width=0.47\textwidth]{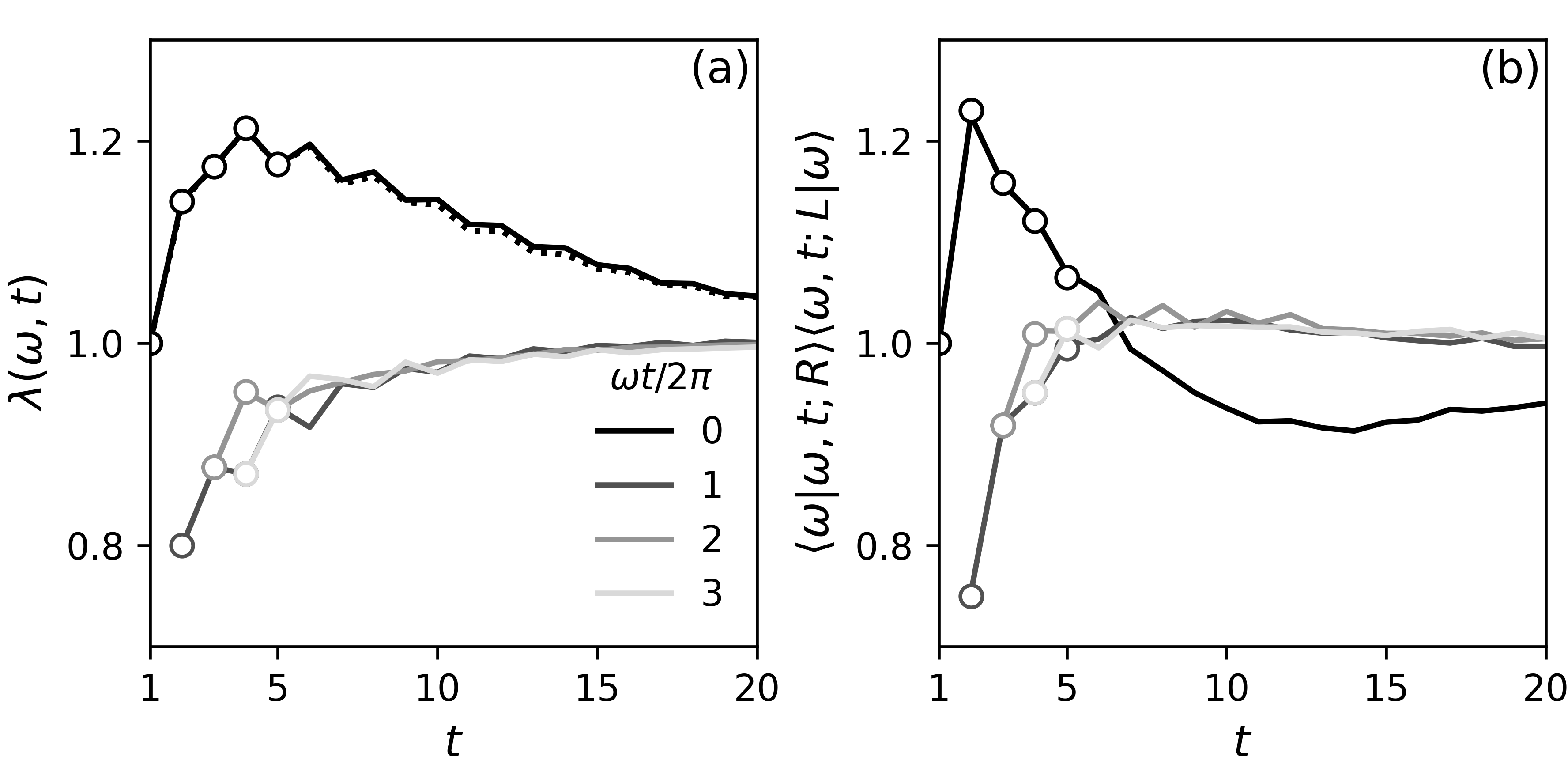}
\caption{Spectral properties of $\overline{\mathcal{T}}(t)$. (a) Leading eigenvalues in the sectors $\omega=\omega_+=-\omega_-$. The solid lines are extracted from $L$-scaling of $\overline{\mathcal{Z}}(\omega,t)$ (see Appendix~\ref{sec:Lscaling}) with $n = \omega t / (2 \pi)$ on the legend. The dotted line shows the $\omega=0$ data in Fig.~\ref{fig:Kopen_eigenvalue_overlap} for comparison, and the white data points are from exact diagonalisation of $\overline{\mathcal{T}}(t)$ for $t \leq 5$, detailed in Appendix \ref{sec:blockdiagonalisation}. (b) Overlaps of the leading eigenvectors with the vectors $\ket{\omega}$.}
\label{fig:Kw_eigenvalue_overlap}
\end{figure}

We show $\overline{\mathcal{Z}}(\omega,t)$ for $L=8$ and various $\omega$ as a function of $t$ in Fig.~\ref{fig:Ksw}(b). The qualitative behaviour of the leading eigenvalues and corresponding eigenvectors at large $t$ can be deduced as follows. First note that $\overline{Z}(0,t) \to 1$ and $\overline{Z}(s \neq 0,t) \to 0$ at large $t$ for all $L$. Substituting this behaviour into the first line of Eq.~\eqref{eq:Kw}, we find $\overline{\mathcal{Z}}(\omega,t) \simeq 1$ for large $t$, for all $L$ and $\omega$. From Eq.~\eqref{eq:Kw_expansion}, this implies that the leading eigenvalues $\lambda(\omega,t) \simeq 1$, and the overlaps $\braket{\omega|\omega,t;R}\braket{\omega,t;L| \omega} \simeq 1$, at large $t$.

Moving beyond this qualitative analysis, the leading eigenvalues $\lambda(\omega,t)$ and the overlaps $\braket{\omega|\omega,t;R}\braket{\omega,t;L| \omega}$ can be extracted from the length-scaling of $\overline{\mathcal{Z}}(\omega,t)$, and the results are shown in Figs.~\ref{fig:Kw_eigenvalue_overlap}(a) and (b), respectively. We give details of the analysis in Appendix~\ref{sec:Lscaling}, and repeat it for another model in the same symmetry class in Appendix~\ref{sec:heisenberg}. For comparison with length-scaling of $\overline{\mathcal{Z}}(\omega,t)$ we show (i) the exact results for $t \leq 5$, determined from the exact diagonalisation of $\overline{\mathcal{T}}(t)$ (see Appendix~\ref{sec:blockdiagonalisation}), and (ii) the calculation of $\lambda(0,t)$ from Fig.~\ref{fig:Kopen_eigenvalue_overlap}, and find excellent agreement between the different approaches. We see that $\lambda(0,t) \geq 1$ and $\lambda(\omega \neq 0,t) \leq 1$, with all $t$ of the leading eigenvalues approaching unity at late times. In Fig.~\ref{fig:Kw_eigenvalue_overlap}(b), $\braket{\omega|\omega,t;R}\braket{\omega,t;L|\omega}$ also approaches unity, and we discuss this in Sec.~\ref{sec:pairingdomains}.

\begin{figure}
\hspace{-0.1in}
	\includegraphics[width=0.47\textwidth]{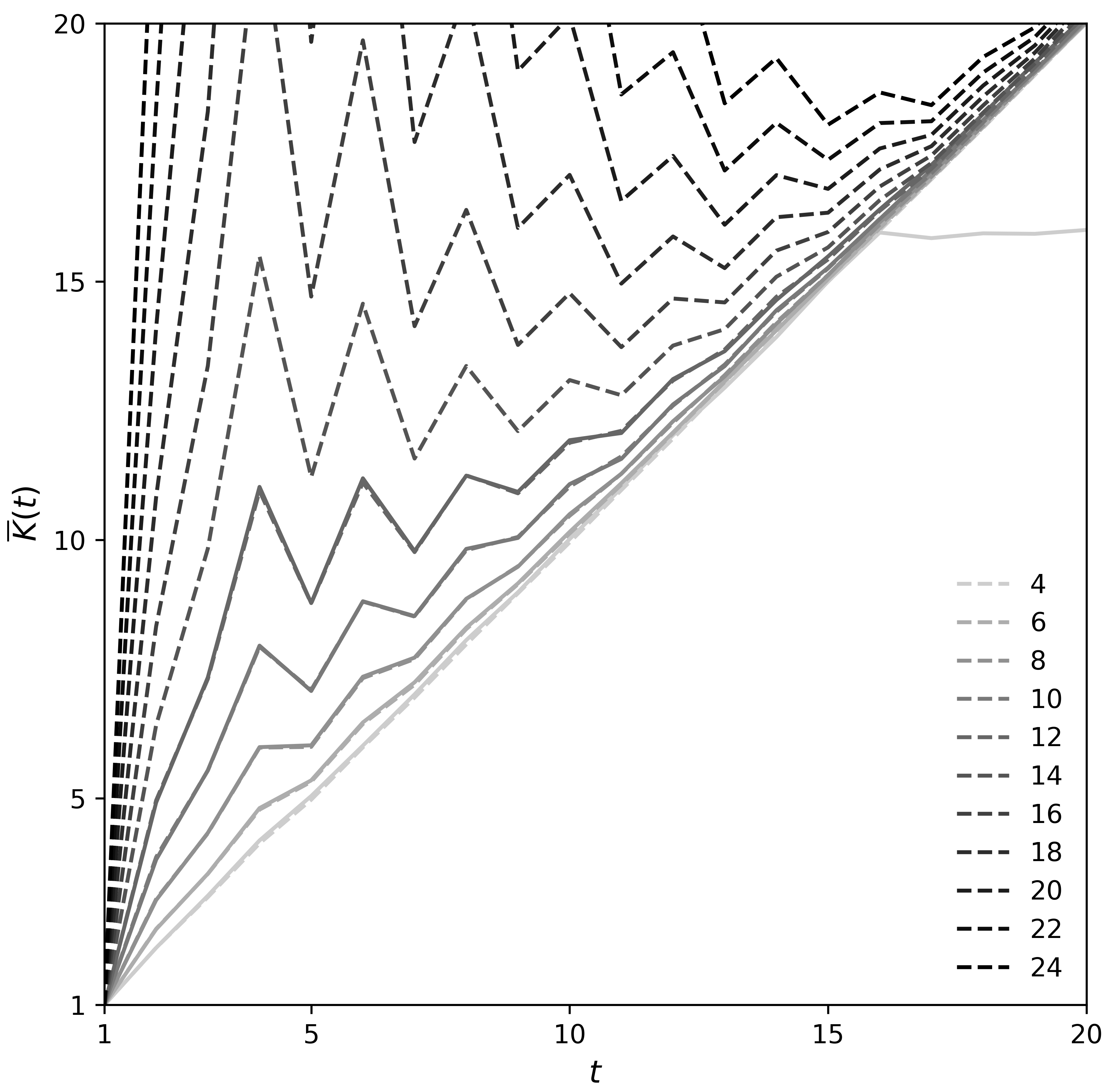}
	\caption{Calculations of the average SFF $\overline{K}(t)$ with periodic boundary conditions for various $L$ (legend), using (dashed) the leading $t$ eigenvalues $\lambda(\omega,t)$ of the average transfer matrix $\overline{\mathcal{T}}(t)$, from Fig.~\ref{fig:Kw_eigenvalue_overlap}(a), and (solid) exact diagonalisation of the Floquet operators for $L \leq 12$, from Fig.~\ref{fig:sff}. For $4 \leq L \leq 12$ we can compare the approaches, and find the difference them $|\sum_{\omega}\lambda^L(\omega,t)-\overline{K}(t)|<0.1$ for all $t < t_{\text{H}}$ shown. }
	\label{fig:sff_reconstruction}
\end{figure}

We can conduct a stringent test of the accuracy of the eigenvalues that we have extracted in Fig.~\ref{fig:Kw_eigenvalue_overlap}(a), and of the adequacy of focussing on only the $t$ leading eigenvalues. To do so we use our results to reconstruct the SFF with periodic boundary conditions, and the results are shown in Fig.~\ref{fig:sff_reconstruction}. We find remarkable agreement with exact diagonalisation from Fig.~\ref{fig:sff}(a). To demonstrate the power of the transfer matrix approach, and the exponential growth of deviations from RMT, we also extrapolate to larger systems than are directly accessible. For fixed $t$, our restriction to only the leading eigenvalues of $\overline{\mathcal{T}}(t)$ is exact in the large-$L$ limit.

Within the framework of the average transfer matrix the mechanisms giving rise to the RMT result $\overline{K}(t)=t$ in the diagonal regime are different with periodic and open boundary conditions. In the periodic case the $t$ leading eigenvalues each contribute unity. In the open case just one of these eigenvalues, in the $\omega=0$ sector, contributes to $\overline{K}(t)$, but this contribution is enhanced by a factor $\braket{\mathcal{B}_L|0,t;R}\braket{0,t;L|\mathcal{B}_R} \simeq t$.

\subsubsection{Domain wall tensions}

Having established the approximate equality ${\overline{K}(t) \simeq \sum_{\omega} \lambda^L(\omega,t)}$ in Fig.~\ref{fig:sff_reconstruction}, we will now relate the average SFF with periodic boundary conditions to properties of the domain walls studied in for example Fig.~\ref{fig:domain_wall}. At late times and for a nonzero twist $s$, the Fourier transform of Eq.~\eqref{eq:Kw} gives
\begin{align}
	\overline{Z}(s,t) \simeq (L-1) \times \frac{1}{t}\sum_{\omega} \delta \lambda(\omega,t) e^{i \omega s} + \ldots
\label{eq:Zsexpansion}
\end{align}
Here we have written $\delta \lambda(\omega,t)=\lambda(\omega,t)-1$ and used $\braket{\omega|\omega,t;R}\braket{\omega,t;L|\omega} \simeq 1$. The ellipses represent terms higher order in $\delta \lambda(\omega,t)$ and the contributions from subleading eigenvalues. With no twist we instead have $\overline{Z}(0,t) \simeq 1$ from Fig.~\ref{fig:domain_wall}. Eq.~\eqref{eq:Zsexpansion} and the observed decay of $\overline{Z}(s,t)$ in Fig.~\ref{fig:domain_wall} motivate the definition of the domain wall tensions $\varepsilon(s,t)$ through
\begin{equation}
	e^{-\varepsilon(s,t)t} = \frac{1}{t}\sum_{\omega} \delta \lambda(\omega,t)e^{i\omega s}.
\label{eq:tensions}
\end{equation}
Since $\overline{Z}(s=0,t)=1$ for large $t$, independently of $L$, we see that $\sum_{\omega}\delta \lambda(\omega,t)=0$. The statistical weight $e^{-\varepsilon_{\text{eff}} t}$ defining the effective tension $\varepsilon_{\text{eff}}$ in Eq.~\eqref{eq:effectivetension} is then simply the average over $s \neq 0$ of the statistical weights $e^{-\varepsilon(s,t)t}$. 

To write an expression for the SFF with periodic boundary conditions in terms of the domain wall tensions we invert Eq.~\eqref{eq:tensions} and use $\overline{K}(t) \simeq \sum_{\omega}\lambda^L(\omega,t)$. Expanding around the RMT result we find
\begin{equation}
	\overline{K}(t) = t + \frac{1}{2}L(L-1)t\sum_{s \neq 0} e^{-2 \varepsilon(s,t)t} + \ldots 
\label{eq:Kpbc_expansion}
\end{equation}
to second order in $\delta \lambda(\omega,t)$. We see then that the first correction to RMT with periodic boundary conditions can be interpreted as arising from pairs of many-body orbits with two domain walls, as opposed to pairs with just one domain wall in the case of open boundary conditions.

\subsection{Pairing domains}\label{sec:pairingdomains}

In this section we investigate the character of the orbit-pairing domains. We probe the space-time-local pairing of orbits across domain walls and within the different domains. Our tools are the twisted boundary conditions from Sec.~\ref{sec:twisted}, and a suitable correlator defined below.

Our results on the leading eigenvectors of $\overline{\mathcal{T}}(t)$ in Figs.~\ref{fig:Kopen_eigenvalue_overlap} and \ref{fig:Kw_eigenvalue_overlap}(b) highlight a close connection with the vectors $\ket{\omega}$, and therefore with the local diagonal orbit pairings $\ket{s}$. In particular, at late times we find $\braket{\mathcal{B}_L|0,t;R}\braket{0,t;L|\mathcal{B}_R} \simeq t$, to be compared with $\braket{\mathcal{B}_L|\omega=0}\braket{\omega=0|\mathcal{B}_R} = t$. Additionally, we have $\braket{\omega|\omega,t;R}\braket{\omega,t;L|\omega} \simeq 1$. Since $\ket{\omega,t;R}$ and $\bra{\omega,t;L}$ are vectors in a space of very high dimension for large $t$, it is quite remarkable to find that they have such large overlap with $\ket{\omega}$, and even more striking that this overlap approaches unity at large $t$. However, because the left and right eigenvectors of $\overline{\mathcal{T}}(t)$ form a biorthogonal set as opposed to being orthonormal, this does not imply equality of the leading eigenvectors and the $\ket{\omega}$ vectors.

\begin{figure}
	\hspace{-0.1in}
	\includegraphics[width=0.47\textwidth]{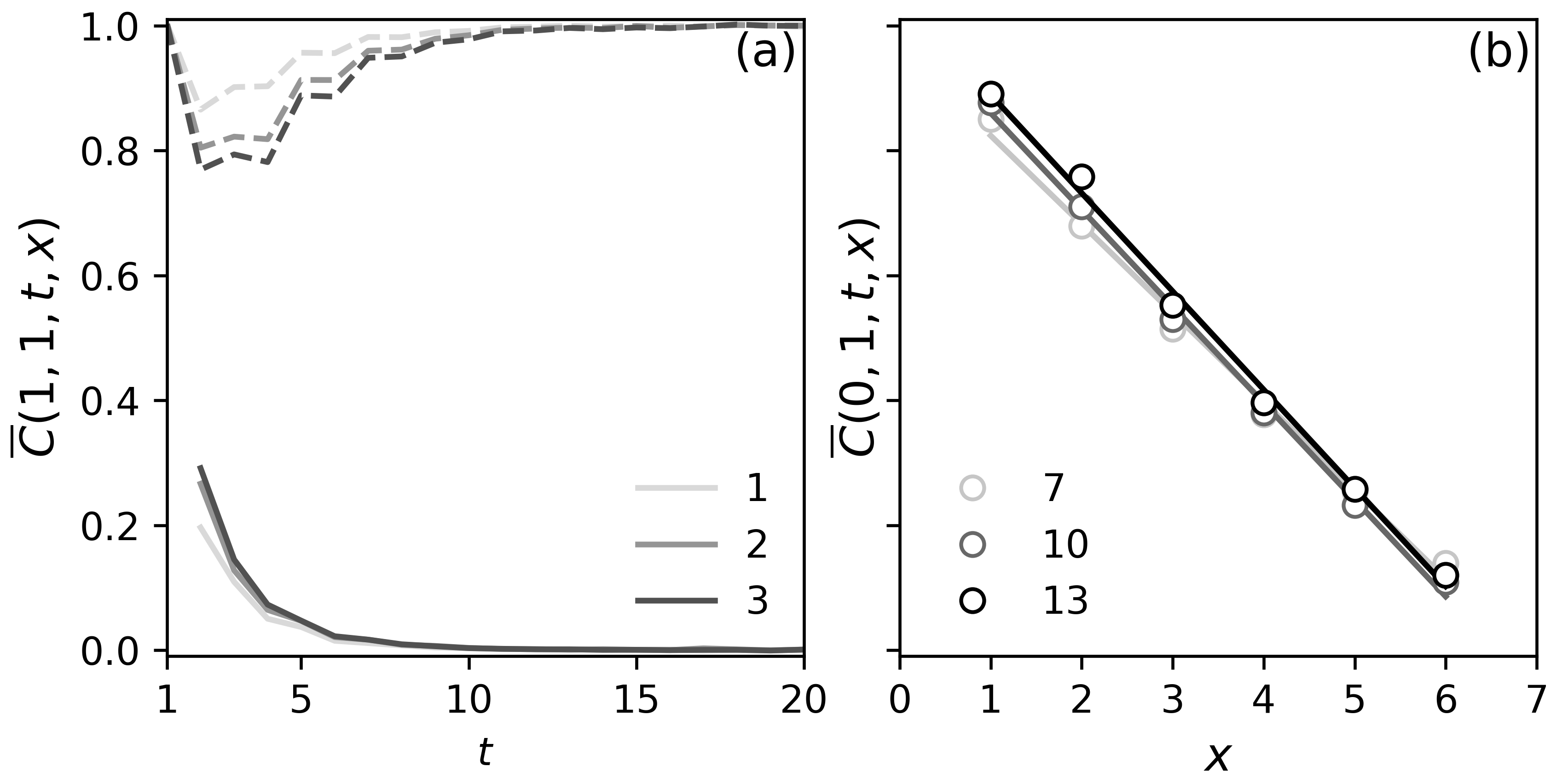}
	\caption{Correlations $\overline{C}(s_L,s_R,t,x)$ between the space-time-local pairings of orbits in a system of $L=8$ sites, with (a) a particular domain and (b) a domain wall imposed by the boundary conditions. In (a) the dashed lines correspond to $s_L=s_R=0$ and the solid lines correspond to $s_L=s_R=1$. The site $x$ is shown on the legend. In (b) $s_L=0$ and $s_R=1$, and the correlations are shown as a function of position $x$. The different sets of points correspond to different times $t$, shown on the legend, and the lines are linear fits to the data.}
	\label{fig:pairingcorrelator}
\end{figure}

To probe pairing beyond these overlaps, we construct a correlator as follows. Writing the forward orbit of site $x$ as $(a_0 b_0 \ldots a_{t-1} b_{t-1})$ and the backward orbit as $(a^*_0 b^*_0 \ldots a^*_{t-1} b^*_{t-1})$, our space-time-local pairing correlator for the step $r$ will be chosen to take the value $1$ if $a_r = a_r^*$ and $-1/q$ otherwise. In this way the correlator vanishes if $a_r$ and $a_r^*$ are uncorrelated $q$-valued random numbers. To calculate this correlator we introduce the operator $\mathcal{C}(r)$, acting in the product space of forward and backward single-site orbits,
\begin{align}
\begin{split}
	\mathcal{C}(r)&\ket{a_0 b_0 \ldots} \otimes \ket{a^*_0 b^*_0 \ldots} \\ &= \frac{\delta_{a_r a_r^*}-1/q}{1-1/q} \ket{a_0 b_0 \ldots} \otimes \ket{a^*_0 b^*_0 \ldots}.
\end{split}
\label{eq:pairingoperator}
\end{align}
Imposing the local diagonal orbit pairing $s_L$ on the site $x=0$, and $s_R$ on site $(L-1)$, the average correlator on a site with $x$ odd is given by inserting $\mathcal{C}(r)$ into a product of transfer matrices in Eq.~\eqref{eq:Zs}. If we probe the orbit pairing of a site with $x$ even we redefine $\mathcal{C} \to \mathcal{S} \mathcal{C} \mathcal{S}^T$. For an individual realisation our correlator for the site $x$ is
\begin{equation}
	C(s_L,s_R,t,x) = \frac{\braket{s_L|\mathcal{T}_{0,1}(t) \ldots \mathcal{C}\mathcal{T}_{x,x+1}(t) \ldots |s_R}}{\braket{s_L|\mathcal{T}_{0,1}(t) \ldots \mathcal{T}_{L-2,L-1}(t) |s_R}}.
\label{eq:pairingcorrelator}
\end{equation}
Here we have inserted $\mathcal{C}$ between the transfer matrices $\mathcal{T}_{x-1,x}(t)$ and $\mathcal{T}_{x,x+1}(t)$ in the numerator, and have omitted the argument of $\mathcal{C}$ which is in this case arbitrary. The denominator is $Z(s_R-s_L,t)$. We are interested in average properties of the transfer matrix, and although both the numerator and denominator are nonzero and positive after averaging, this is not the case in all individual realisations. Consequently it is not possible to average the right-hand side of Eq.~\eqref{eq:pairingcorrelator} directly. We therefore define our average correlator $\overline{C}(s_L,s_R,t,x)$ as the ratio of the average numerator to the average denominator,
\begin{equation}
	\overline{C}(s_L,s_R,t,x) = \frac{\braket{s_L|\overline{\mathcal{T}}^x(t) \mathcal{C} \overline{\mathcal{T}}^{L-1-x}(t)|s_R}}{\braket{s_L|\overline{\mathcal{T}}^{L-1}(t)|s_R}}.
\end{equation}

Consider imposing a particular pairing domain using the boundary conditions $s_L=s_R=s$, as in $Z(0,t)$. At late times if the forward and backward many-body orbits are diagonally paired we expect $\overline{C}(s,s,t,x)$ to be equal to unity if $s=0$ and equal to zero if $s \neq 0$. At early times, based on the results of Fig.~\ref{fig:domain_wall}, we anticipate domain wall contributions. The corresponding multi-domain orbit pairing configurations suppress $C(0,0,t,x)$ and enhance $C(s,s,t,x)$ for $s \neq 0$. In Fig.~\ref{fig:pairingcorrelator}(a) we show $\overline{C}(s,s,t,x)$ calculated for various $x$ and for $s=0$ and $1$ in a system of $L=8$ sites, and find exactly the behaviour expected.

To study the correlations in the pairing across a domain wall, we impose $s_L=0$ and $s_R=s \neq 0$ as in $Z(s,t)$. In this case, as we increase $x$ and thereby sweep across the domain wall, we expect our correlator $\overline{C}(0,s,t,x)$ to decrease toward zero, and in Fig.~\ref{fig:pairingcorrelator}(b) we confirm that this is indeed the case. The results of Fig.~\ref{fig:pairingcorrelator} demonstrate quite directly the existence of domains in the orbit pairing.

\subsection{Bath interpretation of $Z(0,t)$}\label{sec:bath}
It turns out that the boundary conditions used to define $Z(0,t)$ also arise for a system coupled at its ends to Markovian baths. This fact highlights a connection between thermalisation and, through the transfer matrices, the spectral statistics.

Consider first evolution under only the Floquet operator $W$. The SFF can be written ${K(t) = \text{Tr}[W(t) \otimes W^*(t)]}$, where $W \otimes W^*$ is the doubled Floquet operator, and here the trace is over the $q^{2L}$-dimensional doubled Fock space. To couple the system to a bath we act on each of the sites $x=0$ and $(L-1)$ with independent $q \times q$ Haar-random unitary matrices at each time step. Writing the composite index for the end sites $0$ and $(L-1)$ using Greek letters $\alpha,\beta$, and the composite index for the $(L-2)$ central sites using Roman letters $a,b$, the components of the doubled Floquet operator $W\otimes W^*$ are $W_{a\alpha,b\beta}W^*_{a^*\alpha^*,b^*\beta^*}$. Averaging over the bath couplings we find that the only nonzero matrix elements have $\alpha=\alpha^*$ and $\beta=\beta^*$. This average therefore forces local diagonal pairings of the orbits of each of the sites $x=0$ and $(L-1)$. For a more detailed discussion, see Appendix~\ref{sec:montecarlopairing}. 

Defining the Kraus operators $Q^{\alpha \beta}$ via their components $Q^{\alpha \beta}_{ab} = (1/q)W_{a\alpha,b\beta}$, the quantum channel describing the evolution of the central $(L-2)$ sites is represented by the operator
\begin{equation}
	\mathcal{Q} = \sum_{\alpha \beta} Q^{\alpha \beta} \otimes (Q^{\alpha \beta})^*.
\end{equation}
$\mathcal{Q}$ acts on the $q^{2(L-2)}$-dimensional doubled space of the central $(L-2)$ sites. From this
\begin{equation}
	Z(0,t) = \text{Tr}\mathcal{Q}^t,
\label{eq:K0Ct}
\end{equation}
a sum over orbits with local diagonal pairing at the ends of the system. The matrix $\mathcal{Q}$ has leading eigenvalue unity, and the corresponding eigenvector describes a density matrix proportional to the identity. In generic circuit realisations the other eigenvalues of $\mathcal{Q}$ lie within the unit circle, and so at late times $Z(0,t)$ approaches unity.

\begin{figure}
	\includegraphics[width=0.47\textwidth]{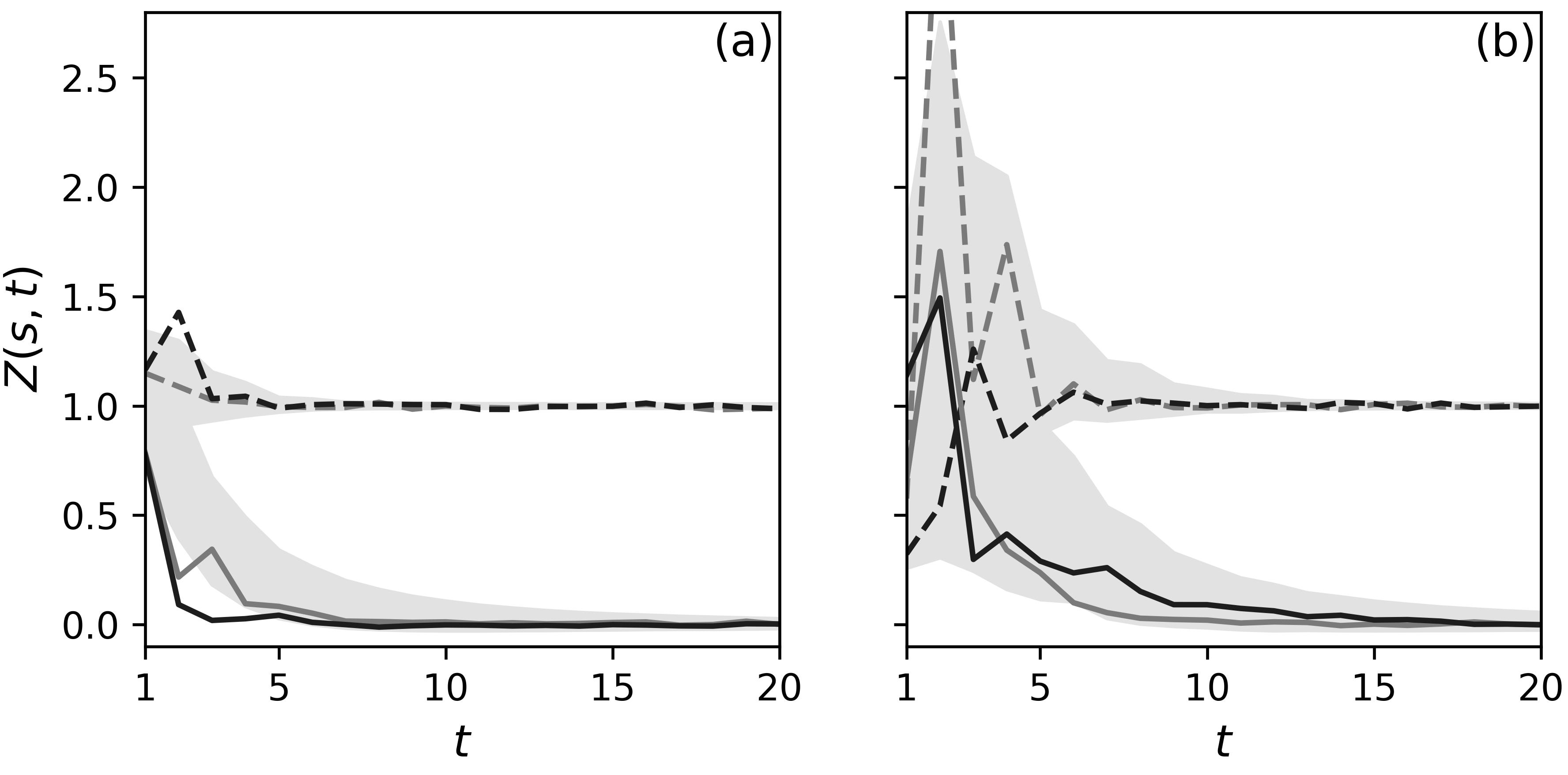}
	\caption{Approach toward a global orbit pairing in individual circuit realisations for (a) $L=4$ and (b) $L=8$ sites. The dashed and solid lines show $Z(0,t)$ and $Z(1,t)$, respectively, for two randomly selected realisations (one in grey, the other in black). The shaded areas are centred on $\overline{Z}(s,t)$ and have vertical width equal to twice the ensemble standard deviation.}
\label{fig:Ks_fix_bulk}
\end{figure}

We show this behaviour for individual circuit realisations in Fig.~\ref{fig:Ks_fix_bulk}, and find also that for $s$ nonzero, $Z(s,t)$ decays to zero with increasing time. In the language of Sec.~\ref{sec:twisted} this implies an approach toward a global orbit pairing in individual circuit realisations. From the perspective we have just described on the other hand, the statement that $Z(0,t)$ approaches unity at late times is a statement about thermalisation.

\section{Eigenstate Correlators}\label{sec:observables}

Our attention has so far been limited to spectral properties. However, the picture of local orbit pairing, and in particular the possibility of domain walls in the pairing of many-body orbits, has implications for local correlations between eigenstates. We now show that there are correlations between the diagonal matrix elements of local operators, or equivalently the reduced density matrices of eigenstates. These correlations, relative to the predictions of the ETH, grow without bound in the thermodynamic limit. We note that apparently related correlations have been observed in a recent numerical study \cite{corps2020thouless}.

We first introduce probes for these correlations. Restricting ourselves to a single site $x$, let $\tau_{j,x}$ ($j = 0 \ldots (q^2-1)$) be a set of orthonormal $q \times q$ Hermitian operators acting only on this site, which for brevity we will refer to as observables. Then $\text{Tr}[\tau_{i,x} \tau_{j,x}] = \delta_{ij}$, and we choose $\tau_{0,x} = \mathbb{1}/\sqrt{q}$ so $\tau_{j \neq 0,x}$ are traceless. The reduced density matrix of the eigenstate $\ket{n}$ on site $x$ is
\begin{align}
	\rho_x(n) = \frac{1}{q} \mathbb{1} + \sum_{j=1}^{q^2-1} \braket{n|\tau_{j,x}|n} \tau_{j,x}.
\end{align}
A correlator between diagonal matrix elements of the observables can then be defined as
\begin{align}
	\text{Tr}_x[\rho_x(n)\rho_x(m)] = \frac{1}{q} + \sum_{j=1}^{q^2-1}\braket{n|\tau_{j,x}|n}\braket{m|\tau_{j,x}|m},
\label{eq:diagonal_correlator}
\end{align}
where $\text{Tr}_x$ denotes a trace over the site $x$. In a similar way, for the off-diagonal matrix elements, consider summing $|\braket{n|\tau_{j,x}|m}|^2$ over $\tau_j$. The result is
\begin{align}
	\text{Tr}_x'[\rho_x'(n)\rho_x'(m)] = \frac{1}{q}\delta_{nm} + \sum_{j=1}^{q^2-1} |\braket{n|\tau_{j,x}|m}|^2.
\label{eq:off_diagonal_correlator}
\end{align}
Here $\rho_x'(n) = \text{Tr}_x \ket{n}\bra{n}$ is the reduced density matrix of the eigenstate $\ket{n}$ over the $(L-1)$-site complement of the site $x$, and $\text{Tr}_x'$ is a trace over that region.

\begin{figure}
\includegraphics[width=0.45\textwidth]{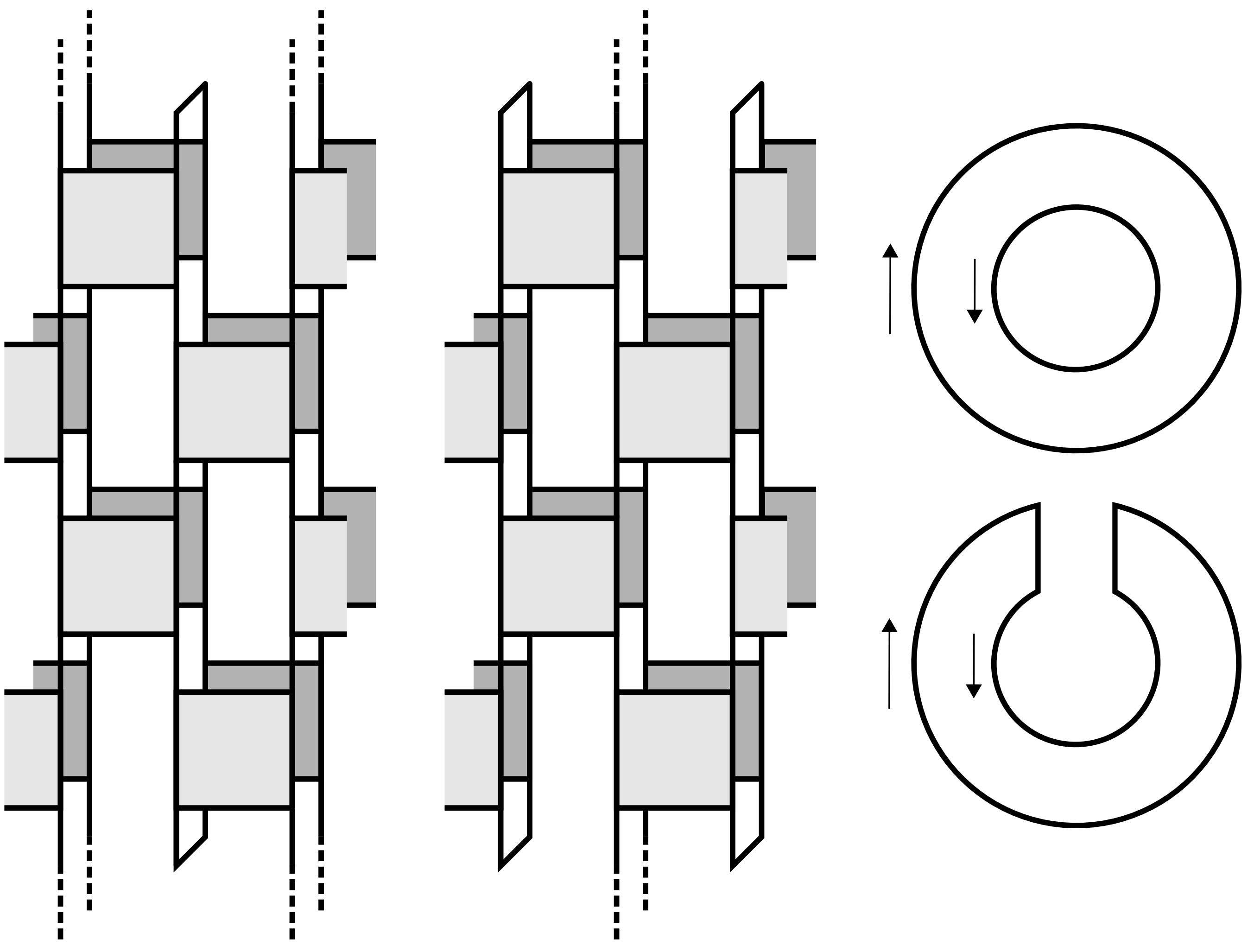}
\put(-200,178){$R_x(t)$}
\put(-120,178){$R'_x(t)$}
\put(-200,-5){$x$}
\put(-230,-5){$x-1$}
\put(-185,-5){$x+1$}
\put(-115,-5){$x$}
\put(-145,-5){$x-1$}
\put(-100,-5){$x+1$}
\caption{Circuit diagrams of $R_x(t)$ (left) and $R_x'(t)$ (right). Away from $x$, $R_x(t)$ has the same trace structure as $K(t)$. On the upper right of the figure the illustration shows the forward (outer) and backward (inner) single-site orbits, relevant to only site $x$ in $R_x'(t)$ and the other $(L-1)$ sites in $R_x(t)$. On the lower right the illustration shows the concatenation of forward and backward paths relevant to only site $x$ in $R_x(t)$ and to the other $(L-1)$ sites in $R_x'(t)$.}
\label{fig:orbit_and_correlator}
\end{figure}
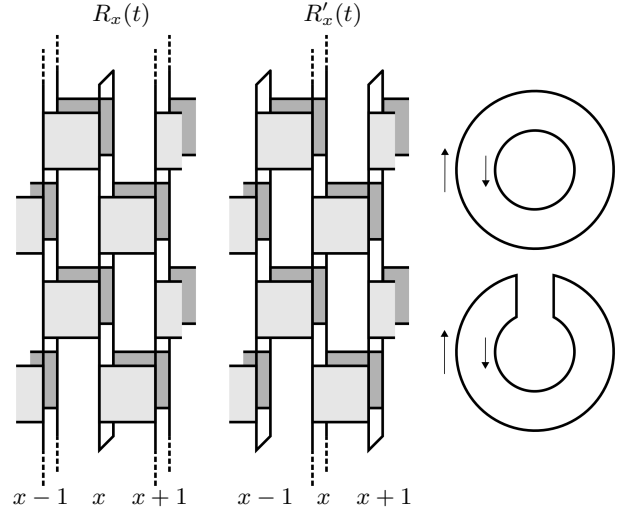

In the time domain the correlations between diagonal matrix elements are characterised by the reduced form factor (RFF)
\begin{align}
\begin{split}
	R_x(t)&= \sum_{nm}\text{Tr}_x[\rho_x(n)\rho_x(m)]e^{i(\theta_n-\theta_m)t} \\
	&= \text{Tr}_x [\text{Tr}_x' W(t) [\text{Tr}_x' W(t)]^{\dag}],
\label{eq:diagonal_correlator_time}
\end{split}
\end{align}
where in the summand we have the correlator defined in Eq.~\eqref{eq:diagonal_correlator}. Spectral structure in the off-diagonal matrix elements is instead encoded in
\begin{align}
\begin{split}
	R_x'(t) &= \sum_{nm}\text{Tr}'_x[\rho'_x(n)\rho'_x(m)]e^{i(\theta_n-\theta_m)t} \\
	&= \text{Tr}_x'[\text{Tr}_x W(t) [\text{Tr}_x W(t)]^{\dag}]
\label{eq:offdiagonal_correlator_time}
\end{split}
\end{align}
where now the object introduced in Eq.~\eqref{eq:off_diagonal_correlator} appears. Note that $R_x'(t)$ is the autocorrelation function of the operator $\tau_{j,x}$, averaged over choices of $\tau_{j,x}$, and is therefore accessible in principle to experimental measurements. In Fig.~\ref{fig:orbit_and_correlator} we show diagrams of the correlators $R_x(t)$ and $R'_x(t)$ in circuit notation, as well as the relevant orbit illustrations (as in Fig.~\ref{fig:diagonal}). In light of the picture of local orbit pairing developed through Sec.~\ref{sec:local} we see that $R_x(t)$ and $R_x'(t)$, and therefore the diagonal and off-diagonal matrix elements, behave very differently.

In $R_x(t)$ the evolution operator appears as the product of $\text{Tr}'_x W(t)$ and its Hermitian conjugate. Away from site $x$, $R_x(t)$ has the same trace structure as $K(t)$. This trace structure is associated with a freedom in the local orbit pairing, or more formally with the transfer matrices $\mathcal{T}_{x,x+1}$ discussed in Sec.~\ref{sec:local}. Therefore, just as $\overline{K}(t)$ grows exponentially with $L$, so does $\overline{R}_x(t)$. Exact calculations are straightforward for the model of Ref.~\cite{chan2018spectral} in the large-$q$ limit, yielding $\overline{R}_x(t) = (q/t)\overline{K}(t)$, to be compared with $\overline{R}_x(t)=q$ within RMT. Therefore at times $t<t_{\text{Th}}$, for which $\overline{K}(t) > t$, the average RFF $\overline{R}_x(t)$ exceeds its RMT value. The freedom in the local orbit pairing gives rise to correlations between the diagonal matrix elements, and this effect is stronger in larger systems.

In Appendix~\ref{sec:observable_transfer} we show how the correlator $R_x(t)$ can be evaluated using an extension of the transfer matrix method of Sec.~\ref{sec:local}. Here we present numerical results, obtained in the quasienergy domain using exact diagonalisation and Lanczos methods (see Appendix~\ref{sec:numericalmethods} for details), that demonstrate divergent departures from ETH. 

\begin{figure}
	\hspace{-0.1in}
	\includegraphics[width=0.47\textwidth]{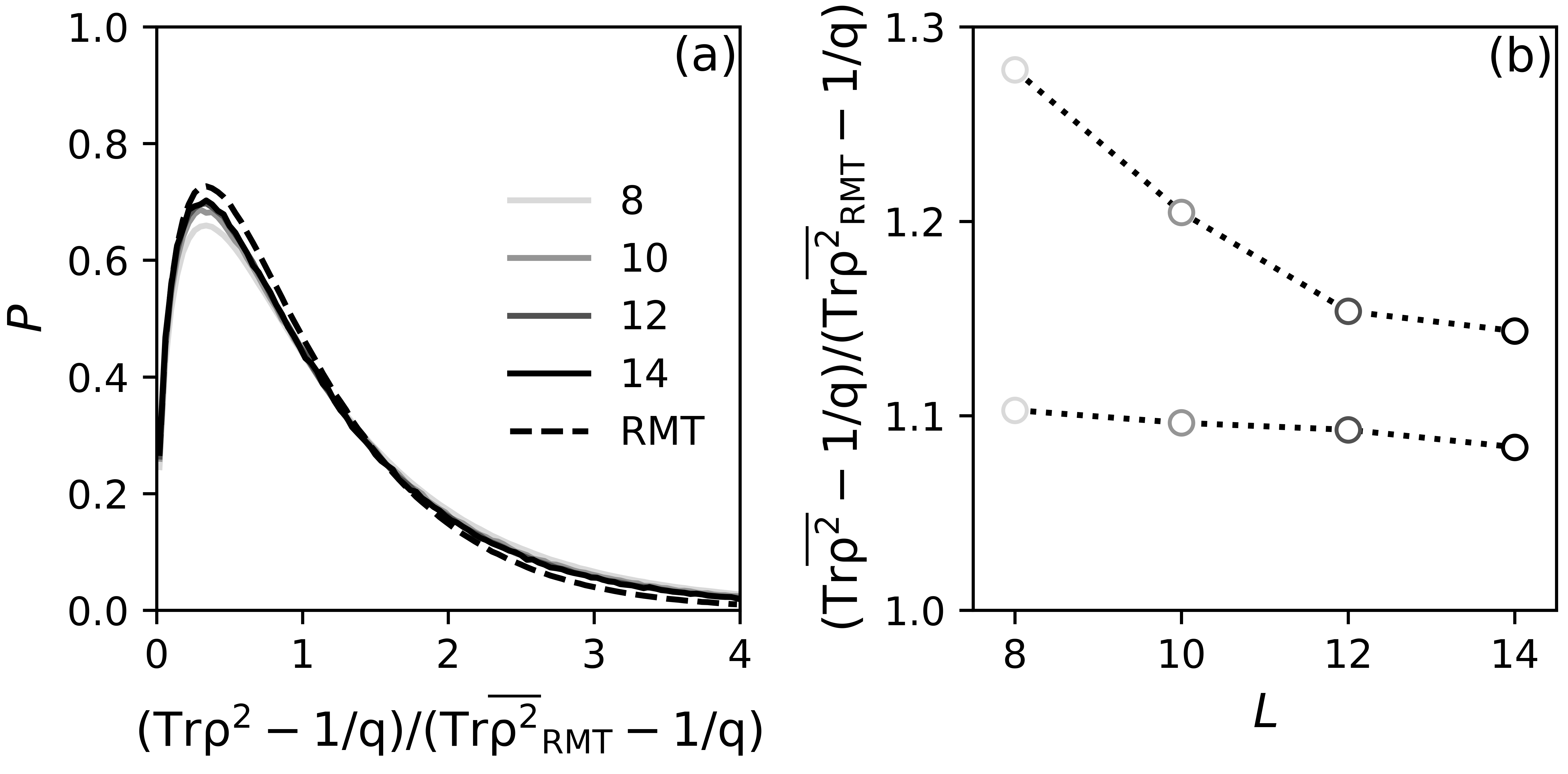}
	\caption{Fluctuations of the one-site eigenstate purities, $\text{Tr}\rho^2-1/q$, in units of the RMT-average, $\overline{\text{Tr}\rho^2}_{\rm{RMT}}-1/q$. In (a) the solid lines show the probability distribution $P$ of $(\text{Tr}\rho^2-1/q)/(\overline{\text{Tr}\rho^2}_{\rm{RMT}}-1/q)$ for systems of various sizes $L$ (legend), and with open boundary conditions. The dashed line is the distribution within RMT, calculated from Haar-random $2^8 \times 2^8$ unitary matrices. In (b) we show the mean as a function of $L$ for periodic (lower points) and open (upper points) boundary conditions. For open boundary conditions we exclude data from two sites at each of the two ends of the chain.}
\label{fig:purity}
\end{figure}

Note first that the completeness of eigenstates implies $\sum_n \rho_x(n) = q^{L-1}\mathbb{1}$, and so the correlator of diagonal matrix elements satisfies the sum rule ${\sum_{m}\text{Tr}[\rho_x(n)\rho_x(m)] = q^{L-1}}$. The RMT result for the average correlator is
\begin{equation}
	\text{Tr}[\overline{\rho_x(n)\rho_x(m)}] = \frac{1}{q} + \frac{q-q^{-1}}{q^{2L}-1}(q^L\delta_{nm}-1).
\label{eq:Trrhorho_RMT}
\end{equation}
On the right-hand side of this equation, the first term arises from the non-fluctuating component $(1/q)\mathbb{1}$ of the density matrices, while the second term characterises fluctuations in the matrix elements of operators $\tau_{j,x}$ with $j \neq 0$. The $n=m$ terms in Eq.~\eqref{eq:Trrhorho_RMT} are the eigenstate purities. We show the fluctuations of the eigenstate purities, in units of the ETH result, in Fig.~\ref{fig:purity}. Here deviations from the ETH are small and do not grow with system size. The situation is quite different for $n \neq m$.

\begin{figure}
\hspace{-0.1in}
\includegraphics[width=0.47\textwidth]{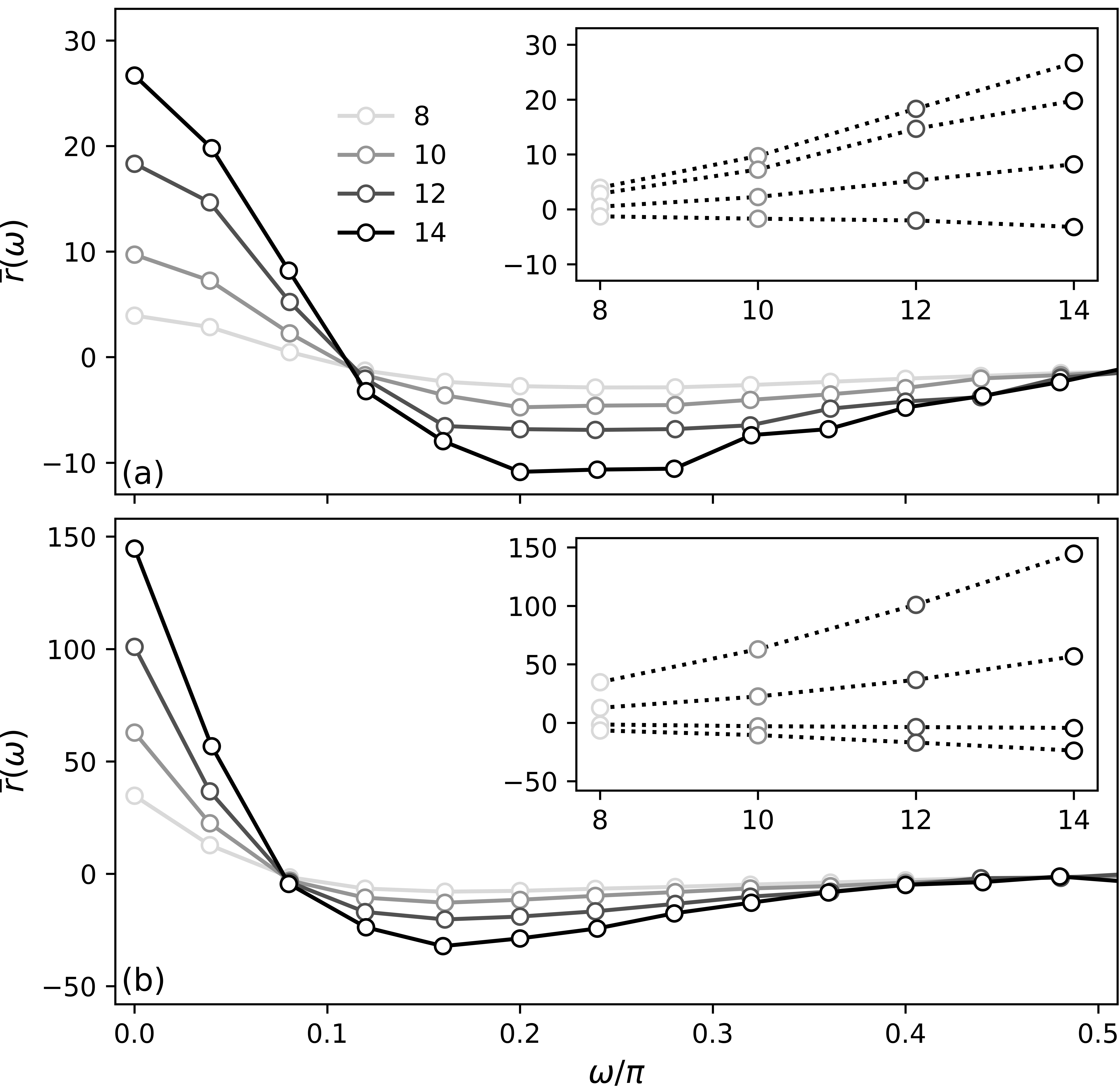}
\caption{Average correlator $\overline{r}(\omega)$ of single-site diagonal matrix elements in Haar-RFCs with (a) periodic and (b) open boundary conditions. The system size $L$ is shown on the legend, and the smallest value of $\omega$ corresponds to an average over pairs of levels drawn from a quasienergy interval of width ${\omega \simeq (2.5 \times 10^{-3})\pi}$. In the inset we show the behaviour of $\overline{r}(\omega)$ as a function of $L$ for the four smallest values of $\omega$ in the main panels. The standard error in the correlator is of order unity.}
\label{fig:diagonal_correlator}
\end{figure}

It is useful to parametrise the correlator as follows
\begin{equation}
	\text{Tr}_x[\rho_x(n)\rho_x(m)] = \frac{1}{q} + \frac{q-q^{-1}}{q^{2L}-1}(q^L \delta_{nm}+r_{x,nm}).
\label{eq:rparameter}
\end{equation}
Comparing with Eq.~\eqref{eq:Trrhorho_RMT} we see that the ETH prediction is $\overline{r}_{x,nm}=-1$. More generally, the sum rule satisfied by the correlator ensures that the average of $r_{x,nm}$ over $n$ or $m$ is $-1$. It is reasonable to expect that the ensemble average is determined only by the quasienergy difference, so $\overline{r}_{x,nm} = \overline{r}_x(\omega=|\theta_n-\theta_m|)$. We define also the average of $\overline{r}_x(\omega)$ over $x$, $\overline{r}(\omega)$, and for open boundary conditions exclude the two sites at each of the two ends of the chain from this average. $\overline{r}(\omega)$ is a correlation function for the diagonal matrix elements of local observables, averaged over position and summed over an orthonormal set of observables at each site.

In Fig.~\ref{fig:diagonal_correlator} we show $\overline{r}(\omega)$ for both open and periodic boundary conditions. The deviations from the ETH are striking, and they are most prominent at the smallest quasienergy separations, where the ETH is usually thought to be applicable. The observed low-frequency peak has its origin in the late-time decay of the leading eigenvalues of the transfer matrix, $\lambda(\omega,t)$, toward unity (see Figs.~\ref{fig:Kopen_eigenvalue_overlap} and \ref{fig:Kw_eigenvalue_overlap}(a)). From this we conclude that at large $L$ the height of the peak grows exponentially with $L$ while its width is given by the inverse of the Thouless time, which decreases with increasing $L$. With open boundary conditions and for only $L=14$ sites we find already a relative enhancement of the correlator $\overline{r}(\omega)$ by over two orders of magnitude. This result constitutes a substantial correction to the ETH for chaotic one-dimensional Floquet systems.

The picture of local orbit pairing does not suggest a significant modification to the ETH for the off-diagonal matrix elements of local observables, at least at the level of $R_x'(t)$. To see this, note that $R_x'(t)$ is constructed from the product of $\text{Tr}_x W(t)$ and its Hermitian conjugate. Based on the trace structure, there is only a freedom in the local orbit pairing at the site $x$. This fact is demonstrated by a large-$q$ calculation in the model of Ref.~\cite{chan2018spectral} where we find $\overline{R}_x'(t) = q^{L-1}(1+(t-1)e^{-2\varepsilon t})$ for $1 \leq x \leq (L-2)$, to be compared with the RMT result $\overline{R}_x'(t)=q^{L-1}$. The deviations of $\overline{R}_x'(t)$ from RMT do not grow with $L$.

Objects analogous to $R_x(t)$ and $R'_x(t)$ can also be defined for operators with arbitrary spatial support. From Eqs.~\eqref{eq:diagonal_correlator_time} and \eqref{eq:offdiagonal_correlator_time}, as well as Fig.~\ref{fig:orbit_and_correlator}, we see that $R_x(t)$ and $R_x'(t)$ are complementary. Considering operators with support on $\ell$ sites (where $\ell=1$ in the above), there is a freedom in the local orbit pairing over $(L-\ell)$ sites in the analogue of $R_x(t)$ and over $\ell$ sites in the analogue of $R_x'(t)$. The substitution $\ell \leftrightarrow (L-\ell)$ converts between the analogues of $R_x(t)$ and $R_x'(t)$. Whereas the spectral structure in the diagonal matrix elements grows with $(L-\ell)$, the spectral structure in the off-diagonal matrix elements grows with $\ell$.

\section{Spectral fluctuations}\label{sec:fluctuations}

We now turn to an investigation of the sample-to-sample fluctuations of spectral correlations within the ensemble of Haar-RFCs. In Fig.~\ref{fig:sff_averaging} we have already shown that a local average dramatically suppresses statistical fluctuations of the SFF, a global quantity. Here we set out to understand this and the distribution of spectra more generally.

Information on the distribution of spectra is buried in the moments of $K(t)$. The $n^{\text{th}}$ moment, $\overline{K^n}(t)$, is an average over the product of $n$ copies of the forward orbits $\text{Tr}W(t)$ and $n$ copies of the backward $\text{Tr}W^*(t)$. Focusing on only the second moment, in Sec.~\ref{sec:nongaussian} we make some first steps toward adapting our theory of local orbit pairing to these higher order objects, and demonstrate non-Gaussian statistics of the SFF. In Sec.~\ref{sec:entanglementmembrane} we discuss the connection with the entanglement membrane.

\subsection{Non-Gaussian statistics}\label{sec:nongaussian}

To ground the discussion, consider first the case where $W$ is a single random matrix drawn from the Haar distribution. In the limit of large matrix dimension the object $\text{Tr}W(t)$ is then normally distributed, with mean zero, in the complex plane \cite{kunz1999probability}. In calculating the second moment of the spectral form factor, $\overline{K^2}(t) = \overline{|\text{Tr}W(t)|^4}$, Wick's theorem gives a sum over the two possible pairings of copies of $\text{Tr}W(t)$ and its conjugate,
\begin{align}
\label{eq:Wick}
\begin{split}
\overline{K^2}(t) &=
	\contraction{}{\text{Tr}W(t)}{}{\text{Tr}W^*(t)}
	\contraction{\text{Tr}W(t)\text{Tr}W^*(t)}{\text{Tr}W^(t)}{}{\text{Tr}^*W(t)}
	\text{Tr}W(t)\text{Tr}W^*(t)\text{Tr}W(t)\text{Tr}W^*(t) \, \, \Big\} + \\
	&+ \contraction[2ex]{}{\text{Tr}W(t)}{\text{Tr}W(t)\text{Tr}W^*(t)}{\text{Tr}W^*(t)}
	\contraction{\text{Tr}W(t)}{\text{Tr}W(t)}{}{\text{Tr}W^*(t)}
	\text{Tr}W(t)\text{Tr}W^*(t)\text{Tr}W(t)\text{Tr}W^*(t) \, \, \Big\} -	
\end{split}
\end{align}
where we have denoted the two copy pairings {`$+$'} and {`$-$'}. Each of the {`$+$'} and {`$-$'} copy pairings contributes $\overline{K}^2(t)$, so $\overline{K^2}(t) = 2\overline{K}^2(t)$. Here the copy pairing is necessarily global since our evolution operator is just one Haar-random matrix.

\begin{figure}
	\hspace{-0.1in}
	\includegraphics[width=0.47\textwidth]{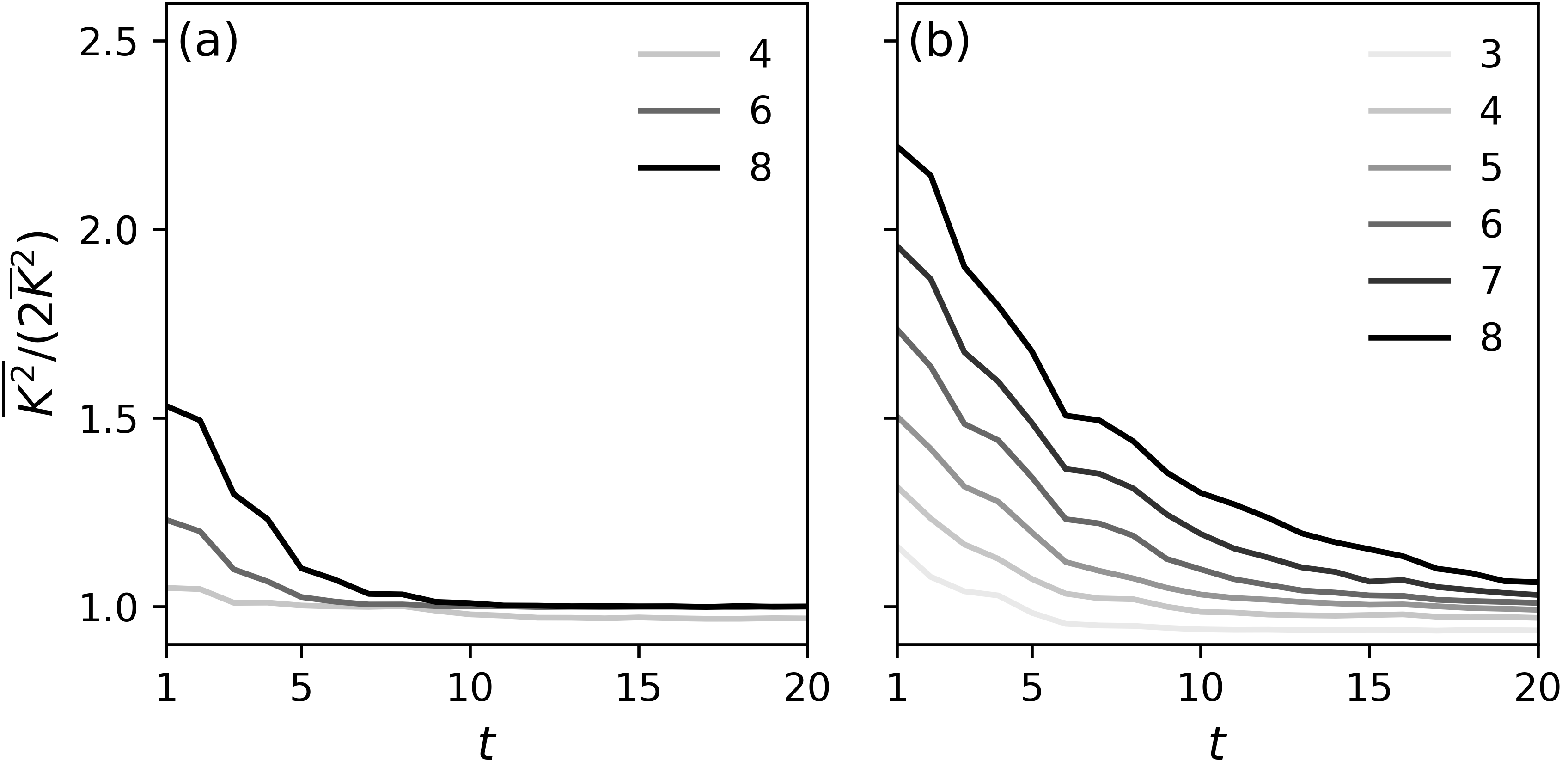}
	\caption{Statistical fluctuations of $K(t)$. Here we show $\overline{K^2}/(2\overline{K}^2)$, equal to $1$ within RMT, for (a) periodic boundary conditions and (b) open boundary conditions. The system size $L$ is shown on the legend.}
\label{fig:Kvariance}
\end{figure}

The global pairing of copies in Eq.~\eqref{eq:Wick} is to be contrasted with the situation in a circuit model. We have seen already that the local freedom in the pairing of orbits gives rise to deviations of the average SFF from the RMT prediction. In $\overline{K^2}(t)$ there is an additional local freedom in the pairings of copies of these orbits. We show in Fig.~\ref{fig:Kvariance} that this gives rise to non-Gaussian statistics of the spectral form factor at early times. Specifically we show that $\overline{K^2}/(2\overline{K}^2)$ is signficantly larger than unity even in small systems, and grows with $L$ at fixed $t$.

As a point of comparison, we now sketch the calculation of $\overline{K^2}$ for the model of Ref.~\cite{chan2018spectral} in the large-$q$ limit; see also \cite{chan2020spectral}. In $K^2(t) = |\text{Tr}W(t)|^4$ each one-site Haar-random gate appears $t$ times in each of the two copies of $\text{Tr}W(t)$, and similarly its conjugate appears $t$ times in each of the two copies of $\text{Tr}W^*(t)$. Haar-averaging a one-site gate at large $q$ we find a sum over pairings of orbits analogous to Eq.~\eqref{eq:large_q_average}, and additionally a sum over the two pairings of copies of these orbits, as in Eq.~\eqref{eq:Wick}. $\overline{K^2}(t)$ then involves a sum over all local orbit pairings, as in Eq.~\eqref{eq:phasecouplingK}, as well as a sum over all local copy pairings. Whereas a domain wall in the orbit pairing has statistical weight $e^{-\varepsilon t}$, it can be shown that a domain wall in the copy pairing here has statistical weight $e^{-2 \varepsilon t}$. With open boundary conditions, at late times
\begin{equation}
	\overline{K^2}(t) = 2\overline{K}^2(t) + 2(L-1)t^4 e^{-2 \varepsilon t} + \ldots.
\label{eq:phasecopywall}
\end{equation}
The first term on the right-hand side of Eq.~\eqref{eq:phasecopywall} involves contributions from domain walls in the orbit pairing, as in Eq.~\eqref{eq:Kobcphase}, whereas the second term arises from a domain wall in the pairing of copies of orbits. The factor $(L-1)$ is the translational entropy of this domain wall, and the factor $t^4$ arises from sums over the orbit pairing at each of the two ends of the two copies of the chain. With periodic boundary conditions the leading term in $\overline{K^2}(t)-2\overline{K}^2(t)$ at late times comes from configurations with two domain walls in the copy pairing. Fig.~\ref{fig:Kvariance} shows exactly this type of behaviour. 

A useful analogy with Sec.~\ref{sec:local} is now evident. There a great deal was learnt about $\overline{K}(t)$ via calculations in which boundary conditions were imposed to force diagonal pairings of the local orbits. In this approach the emergence of RMT level statistics on small energy scales in the Haar-RFC can be understood as arising from a global pairing of orbits at late times. Looking at Eq.~\eqref{eq:Wick}, we see that in order to recover RMT statistics of the spectral form factor beyond only the first moment, we also require a global pairing of copies of orbits.

\begin{figure}
\includegraphics[width=0.3\textwidth]{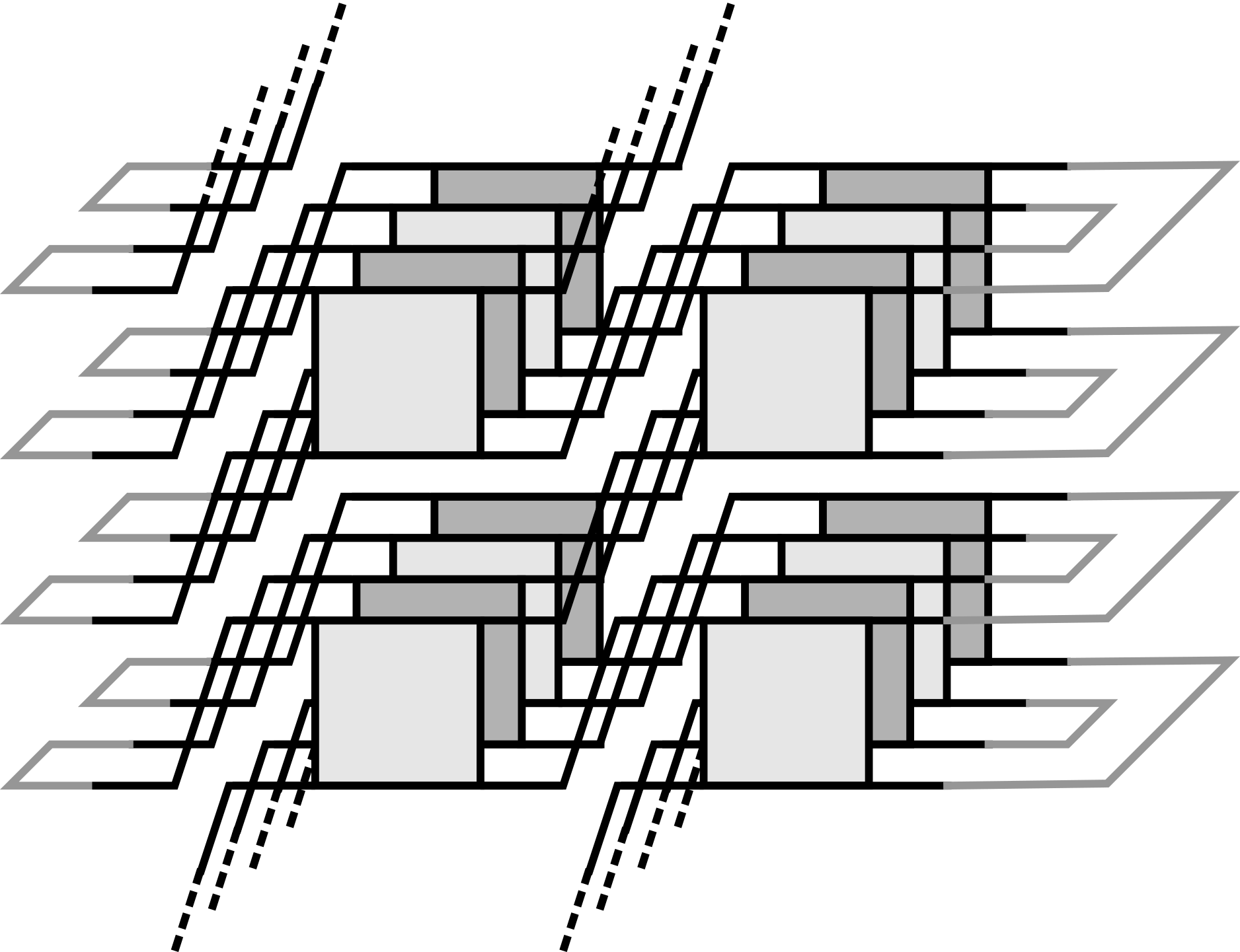}
\caption{Diagram of $Z_2(+,-,t)$, where we force a domain wall into the pairings of copies of orbits. The different circuit layers here correspond to forward (light) and backward (dark) orbits.}\label{fig:copy_twist}
\end{figure}
\begin{figure}
\hspace{-0.1in}
\includegraphics[width=0.47\textwidth]{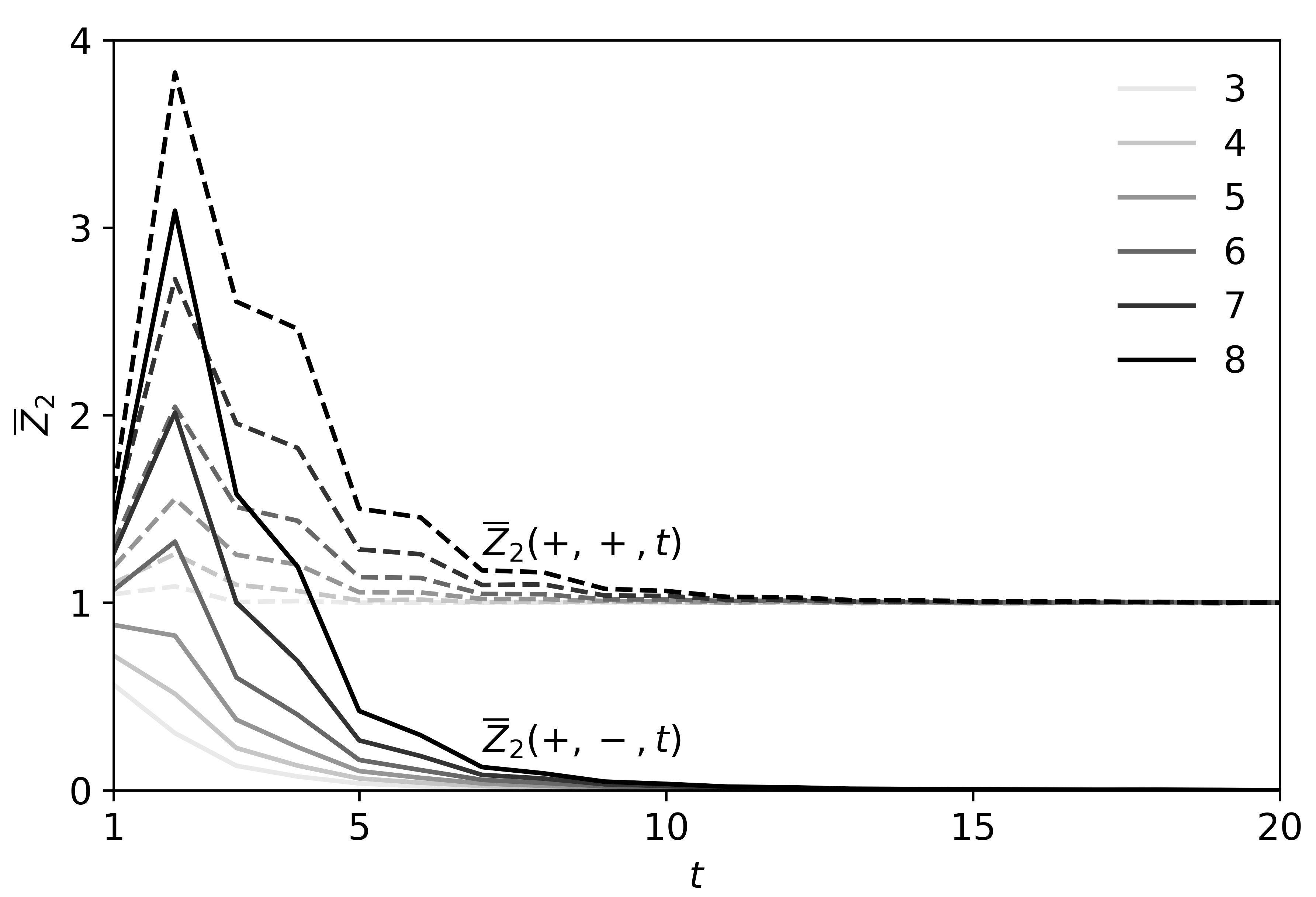}
\caption{Domain walls in the pairing of copies of orbits, relevant to the second moment of the SFF. The solid lines show the object $\overline{Z}_2(+,-,t)$, where the pairing of histories is different on the two ends of the system, and the dashed lines show $\overline{Z}_2(+,+,t)$, where the pairing is the same.}\label{fig:EM_wall}
\end{figure}

Proceeding in a similar way then to Sec.~\ref{sec:twisted}, here we investigate how, in the calculation of the second moment of the spectral form factor, a local freedom in the pairing of copies of orbits gives way to a global pairing. To do this we again impose local diagonal pairings at opposite ends of the system, but now to force a domain wall in the pairing of copies of orbits. This setup is ilustrated in Fig.~\ref{fig:copy_twist}. In principle we could also impose a relative twist on the orbit pairing, as in $Z(s \neq 0,t)$ (see Eq.~\eqref{eq:Zs} and Fig.~\ref{fig:twistdiagram}), but we choose to restrict ourselves to equal-time pairing, as in $Z(0,t)$.

In the notation of Eq.~\eqref{eq:Wick}, Fig.~\ref{fig:copy_twist} illustrates a {`$+$'} pairing on one end of the chain, and a {`$-$'} pairing on the other. The resulting object, which we denote $Z_2(+,-,t)$, is a multiple sum over two copies of the forward orbits and two copies of the backward orbits, with a domain wall in the pairing of copies running along the time direction. $Z_2(+,-,t)$ is to be contrasted with $Z_2(+,+,t)=Z^2(0,t)$, in which we impose equal-time pairings of the same copies of orbits at both ends of the system.

We show the averaged quantities $\overline{Z}_2(+,-,t)$ and $\overline{Z}_2(+,+,t)$ in Fig.~\ref{fig:EM_wall}. We see that $\overline{Z}_2(+,-,t)$ grows with $L$ at fixed $t$, and therefore so do the deviations from RMT. At late times, $t \gtrsim 10$ for $L \leq 8$, $\overline{Z}_2(+,-,t)$ goes to zero. In other words the contributions from those sets of many-body orbits with domain walls in the copy-pairing vanish. As expected, $\overline{Z}_2(+,+,t)$ approaches unity in this regime. The appearance of domain walls in the copy pairing implies non-Gaussian statistics of the SFF at times $t \lesssim 10$ for these system sizes, as revealed by direct calculation in Fig.~\ref{fig:Kvariance}. As in Sec.~\ref{sec:local}, the deviations from RMT are more prominent with open boundary conditions than with periodic because domain walls can only appear singly in the first case.

\begin{figure}
	\includegraphics[width=0.25\textwidth]{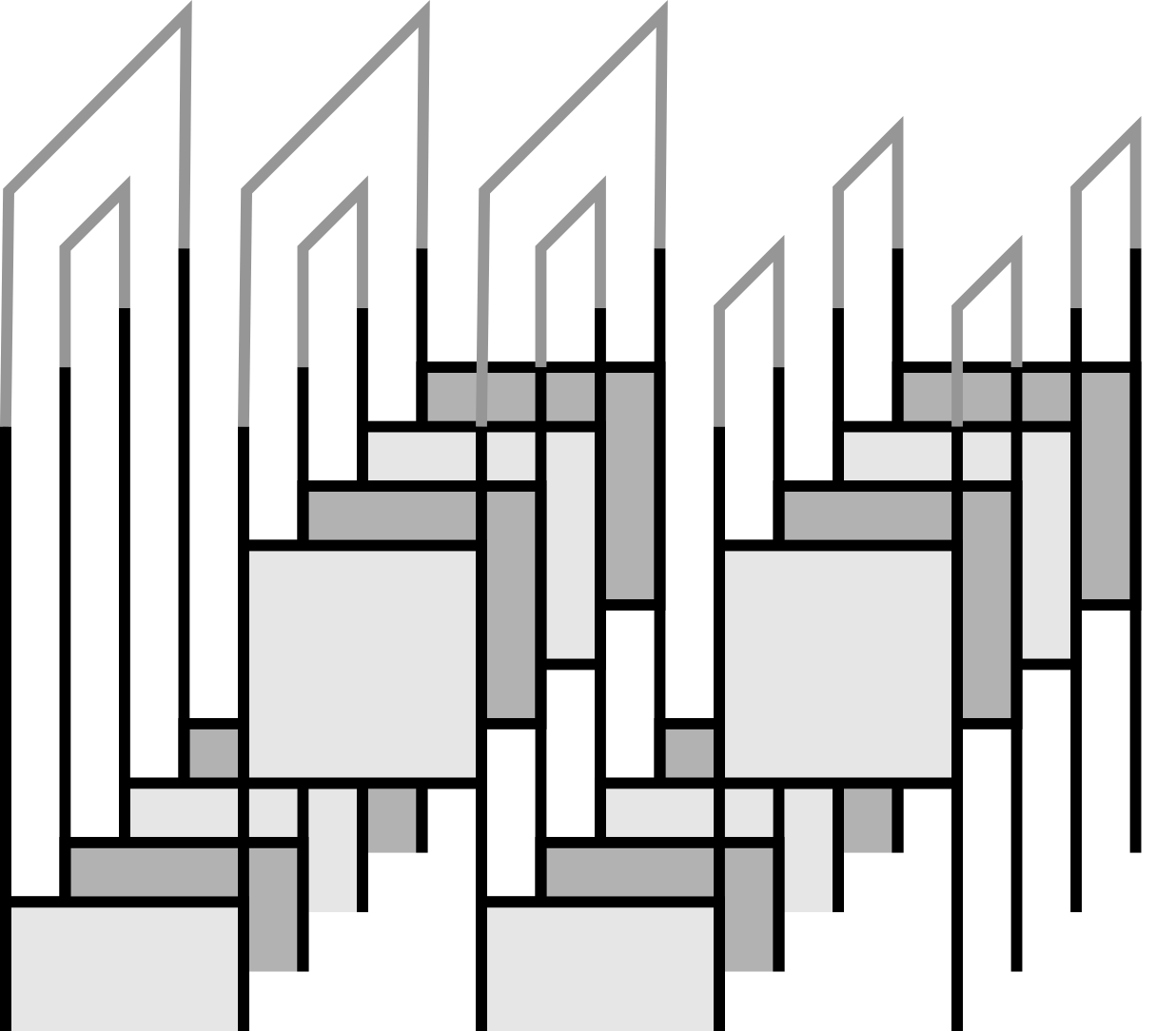}
	\put(-63,125){$e^{-S_2}$}
	\put(-90,115){$A$}
	\put(-30,115){$B$}
	\caption{Boundary conditions defining the entanglement membrane which appears in the calculation of the purity $e^{-S_2(t)}$ of subregion $A$. The subregion $A$ is on the left, where copies of the evolution operator are contracted as in the {`$-$'} pairing of Eq.~\eqref{eq:Wick}. The subregion $B$ on the right corresponds to the {`$+$'} pairing.}
	\label{fig:puritybcs}
\end{figure}

We now return to the effect of the single-gate average observed in Fig.~\ref{fig:sff_averaging}. To understand why an average of the spectral form factor over only one gate of the circuit is enough to suppress statistical fluctuations, consider performing this average and subsequently calculating the variance of the spectral form factor over the ensemble of the remaining gates. The tendency toward global copy pairing observed in Fig.~\ref{fig:EM_wall} implies that the second moment can be approximated by a sum over the two global pairings (`$+$' and `$-$') at late times, as in Eq.~\eqref{eq:Wick}. However, in performing the one-gate average beforehand, we have locally selected for just one of these pairings. Consequently the second moment is approximately halved, and so the variance is suppressed to near zero.

\subsection{Entanglement membrane}\label{sec:entanglementmembrane}

The domain walls in the copy pairing which we have discussed through this section are related to the entanglement membrane \cite{zhou2020entanglement}. To see this we recall the growth of entanglement under a unitary circuit, with or without time periodicity.

Dividing space into subregions $A$ and $B$, the purity of subregion $A$ is defined as $e^{-S_2} = \text{Tr} \rho_A^2(t)$, where $\rho_A(t)$ is the reduced density matrix in subsystem $A$ and $S_2$ is the second Renyi entropy. Writing the initial density matrix as $\rho$, we have
\begin{equation}
	e^{-S_2} = \text{Tr}_A \Big[\text{Tr}_B[W(t) \rho W^{\dag}(t)] \text{Tr}_B[W(t) \rho W^{\dag}(t)]\Big].
\label{eq:purity}
\end{equation}
Here a boundary condition at time $0$ is set by $\rho$. The boundary condition at time $t$ is determined by the trace structure in Eq.~\eqref{eq:purity}, and is illustrated in Fig.~\ref{fig:puritybcs}. The boundary condition at time $t$ imposes a domain wall, the entanglement membrane, in the pairing of the indices of the evolution operator. A similar domain wall in the pairing of copies appears in $Z_2(+,-,t)$. In that case, as shown in Fig.~\ref{fig:copy_twist}, it is imposed by boundary conditions at the left and right ends of the chain, while the boundary conditions in time are periodic.

\section{Diagonal approximation}\label{sec:moderate}

As discussed in Sec.~\ref{sec:breakdown}, in Haar-RFCs the behaviour of the SFF beyond RMT comes from the contributions of off-diagonal pairings of many-body orbits. However, deviations from RMT can also arise within the diagonal approximation through a different mechanism. This involves the subleading eigenvalues of the diagonal propagator $\mathcal{P}$. These contribute in two ways: through having nonzero average values, and through their fluctuations. In this section we discuss both types of contribution.

First we consider a general class of RFCs in which the gates can be tuned from Haar-random to the identity. Away from Haar-randomness the subleading eigenvalues of the diagonal propagator have nonzero averages. We parametrise the two-site gates $U$ of the circuit via their spectral decompositions, $U = V e^{-i \varphi} V^{\dag}$. Here $\varphi$ is a diagonal matrix, and we take the unitary matrix $V$ to be Haar random. If we choose $\varphi$ to be distributed as the eigenphases of a Haar-random unitary then $U$ is Haar random. On the other hand if $\varphi$ is proportional to the identity so is $U$. By tuning the distribution of $\varphi$ we can interpolate between these two extremes.

In a brickwork model the diagonal propagator is of the form Eq.~\eqref{eq:brickwork_propagator}. The gate acting on sites $0$ and $1$ in the first half-step appears in $\mathcal{P}_{a_0 \ldots a_{L-1},b_0 \ldots b_{L-1}}$ as $U_{c_0 c_1,b_0 b_1}U^*_{c^*_0 c^*_1,b_0 b_1}$, where the indices $c,c^*$ are to be summed over. The average over $U$ now involves an average over $V$ and an independent average over the $\varphi$-distribution. The result is
\begin{align}
\begin{split}
	\overline{U_{c_0 c_1,b_0 b_1}U^*_{c^*_0 c^*_1,b_0 b_1}} =& \nu \delta_{c_0 b_0}\delta_{c_1 b_1}\delta_{c^*_0 b_0}\delta_{c^*_1 b_1} \\ &+ \frac{1}{q^2}(1-\nu) \delta_{c_0 c_0^*}\delta_{c_1 c_1^*},
\label{eq:smoothedUUaverage}
\end{split}
\end{align}
where
\begin{align}
	\nu = \frac{\overline{|\text{Tr}U|^2}-1}{q^4-1}.
\label{eq:nuparameter}
\end{align}
Clearly for $\nu = 0$ we recover the Haar-random result Eq.~\eqref{eq:HaarUUaverage}. On the other hand for $U$ proportional to the identity we have $\nu = 1$. The average propagator, with matrix elements $\overline{\mathcal{P}}_{a_0 \ldots a_{L-1},b_0 \ldots b_{L-1}}$, is given by multiplying expressions of the form Eq.~\eqref{eq:smoothedUUaverage} and summing over indices $c,c^*$. The result is a matrix-product operator (MPO), and with periodic boundary conditions this takes the form
\begin{align}
	\overline{\mathcal{P}}_{a_0 \ldots a_{L-1},b_0 \ldots b_{L-1}} = \text{Tr}[M_{a_0 b_0} \ldots M_{a_{L-1} b_{L-1}}],
\label{eq:pMPO}
\end{align}
where $M_{ab}$ is a $2 \times 2$ matrix with components
\begin{equation}
\begin{split}
	M_{a b} = \begin{pmatrix} \nu \delta_{a b} & \frac{1}{q}\sqrt{\nu}\sqrt{1-\nu} \\ \frac{1}{q}\sqrt{\nu}\sqrt{1-\nu} & \frac{1}{q}(1-\nu) \end{pmatrix}. 
\end{split}
\end{equation}
The eigenvectors of the MPO Eq.~\eqref{eq:pMPO} are product states, which we label $\sigma_x=0 \ldots (q-1)$ for $x=0\ldots (L-1)$. Writing these eigenvectors as $v^{\sigma_0}\otimes \ldots v^{\sigma_{L-1}}$, the $q$-component vectors $v^{\sigma}$ must satisfy
\begin{equation}
	\sum_{b=0}^{q-1} M^{ij}_{ab} v^{\sigma}_b = \mu^{ij}_{\sigma} v^{\sigma}_a.
\end{equation}
This is achieved by taking $v^0$ to have constant entries, and $v^{\sigma \neq 0}$ to make up the rest of an orthogonal set (the sum of entries of any $v^{\sigma \neq 0}$ is therefore zero). The resulting $2 \times 2$ matrices $\mu_{\sigma}$, with components $\mu^{ij}_{\sigma}$, are outer products
\begin{align}
	\mu_{\sigma} = \begin{pmatrix} \sqrt{\nu} \\ \sqrt{1-\nu}\delta_{\sigma 0}\end{pmatrix} \begin{pmatrix} \sqrt{\nu} & \sqrt{1-\nu}\delta_{\sigma 0}\end{pmatrix},
\end{align}
and the eigenvalues of $\overline{\mathcal{P}}$ are $\text{Tr}[\mu_{\sigma_0} \ldots \mu_{\sigma_{L-1}}]$. The leading eigenvalue is unity as required by unitarity of $W$, and the next-to-leading eigenvalues are $\nu^2$ with multiplicity $L(q-1)$. Within the diagonal approximation the average SFF $\overline{K}(t) = t \text{Tr} \overline{\mathcal{P}}^t$ is then
\begin{align}
	\overline{K}(t) &= t\sum_{\sigma_0 \ldots \sigma_{L-1}} \text{Tr}[\mu_{\sigma_0} \ldots \mu_{\sigma_{L-1}}]^t \\ &=
	t[ 1 + L(q-1)\nu^{2t} + \ldots],
\end{align}
where in the second line we have expanded around the late-time result. For $0 < \nu < 1$, $\overline{K}(t)/t$ decreases monotonically with $t$. Writing $\nu^{2t} = e^{-2\ln(1/\nu)t}$ we see that the Thouless time is $t_{\text{Th}} = \ln L/(2\ln(1/\nu))$ within the diagonal approximation.

\begin{figure}
	\hspace{-0.1in}
	\includegraphics[width=0.47\textwidth]{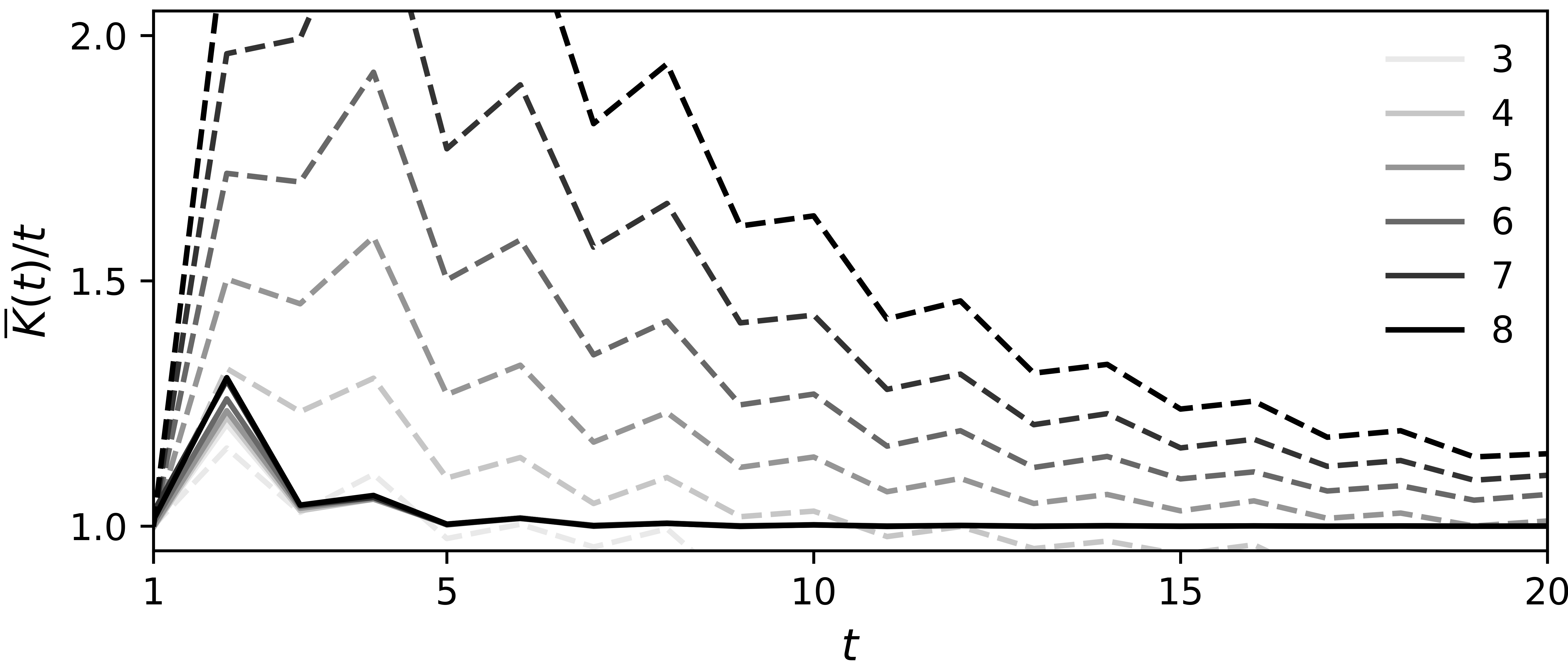}
	\caption{Comparison of the deviations of $\text{Tr}\overline{\mathcal{P}^t}$ (solid) and $\overline{K}(t)/t$ (dashed) from unity in Haar-RFCs with open boundary conditions. The system size $L$ is shown on the legend.}
	\label{fig:otherdiagapprox}
\end{figure}

As indicated below Eq.~\eqref{eq:average_propagator} a full exploration of the diagonal approximation requires a discussion not only of $\overline{\mathcal{P}}$ but also of $\overline{\mathcal{P}^t}$. We now discuss this aspect in the context of Haar-RFCs. In this case, as we have shown following Eq.~\eqref{eq:HaarPmatrix}, the average $\overline{\mathcal{P}}$ has a single eigenvalue of unity, with all subleading eigenvalues zero. Individual realisations, however, typically have non-zero subleading eigenvalues which contribute to $\overline{\mathcal{P}^t}$. We probe the consequences of this by evaluating the SFF using the approximation $\overline{K}(t) \approx t \text{Tr}\overline{\mathcal{P}^t}$. In Fig.~\ref{fig:otherdiagapprox} we compare $\text{Tr}\overline{\mathcal{P}^t}$ to $\overline{K}(t)/t$ for Haar-RFCs, and our results demonstrate that corrections to RMT arising from the subleading eigenvalues of $\mathcal{P}$ cannot account for the behaviour of Haar-RFCs.

In the general case the SFF has contributions both from subleading eigenvalues of the diagonal propagator and from domain walls. Determining which effect is dominant requires a comparison of the associated timescales. For Haar-RFCs, with $\nu=0$, we have seen in Sec.~\ref{sec:local} that the domain wall tensions $\varepsilon$ are finite. Since $\nu$ sets the proximity of the gates to the identity it is natural to expect these tensions to decrease with increasing $\nu$. On the other hand the subleading eigenvalues of the diagonal propagator vanish for $\nu=0$. The behaviour we have studied in Haar-RFCs, in which domain walls control the SFF, must therefore extend to at least a finite range of $\nu$, and so is not restricted to the Haar case.

A contrasting instance is provided by quantum circuits which are built from two-site gates but without any spatial structure imposed. It has recently been argued that models of this type have a `ramp' (or Thouless) time that is logarithmic in system size \cite{gharibyan2018onset}. In the absence of spatial structure a domain wall interpretation cannot be appropriate, and it would be interesting to investigate whether this behaviour can be attributed to subleading eigenvalues of $\mathcal{P}$.

\section{Discussion}\label{sec:discussion}

The spectral statistics of many-body quantum systems in the ergodic phase coincide with the predictions of RMT below a certain (quasi)energy scale, the inverse of our Thouless time $t_{\text{Th}}$. On timescales much greater than $t_{\text{Th}}$, but much shorter than the Heisenberg time $t_{\text{H}}$, this result can be understood through the diagonal approximation to the SFF. In this work we have determined the regime of validity of the diagonal approximation, and the form of the corrections to it, in systems having local interactions and no conserved densities. One of our main results is to reveal a generic mechanism setting the Thouless time in this context.

Our approach has been to develop a theory of local orbit pairing, centred on the properties of a transfer matrix which acts on pairs of local orbits and which generates the SFF. In large systems the average SFF is controlled by the leading eigenvalues of the average transfer matrix; we have calculated these eigenvalues and shown that there is a connection between the corresponding eigenvectors and the diagonal approximation to sums over pairs of many-body orbits. At fixed time and in the limit of large systems the dominant contributions to the SFF come from orbits which are locally paired as in the diagonal approximation, but with distinct diagonal pairings in neighbouring spatial domains. This domain structure is associated with the exponential growth of the SFF with system size at fixed time, and the corresponding divergence of the Thouless time. Conversely, in a system of fixed large size there is a wide window of time $t_{\text{Th}} \ll t \ll t_{\text{H}}$ in which the pairing for the whole system forms a single domain, so the diagonal approximation is accurate. 

We believe the structures we have uncovered are universal, in the sense that our results should apply quite generally to lattice Floquet models with local interactions, and with the form of the deviations of the SFF from RMT governed by the cost of domain walls in the orbit pairing. On timescales much greater than the Floquet period we expect that systems in the same symmetry class that share a domain-wall cost, but are different in microscopic detail, have similar spectral statistics. This kind of universality is familiar from studies of single-particle disordered conductors. In that case the form of the departure of the spectral statistics from RMT on timescales $t<t_{\text{Th}}$ is set by the diffusion constant, a coarse-grained quantity \cite{altshuler1986repulsion}.

The picture of local orbit pairing has also revealed strong correlations between the diagonal matrix elements of local observables. We have shown that, relative to the predictions of the ETH, these correlations grow exponentially with increasing system size. Our results represent a prominent correction to the ETH for Floquet models with local interactions. In the same way, we have argued that deviations from the ETH arise in the off-diagonal matrix elements of non-local operators, and that these features grow exponentially with the support of the operator as opposed to the system size. This result has implications for operator spreading, and furthermore it has been noted that the SFF can be related to the autocorrelation functions of operator strings \cite{gharibyan2018onset}. It would be interesting to understand the connection between these two approaches.

Our results should be compared against calculations within the diagonal approximation, in which the contributing many-body orbits are globally paired. In that approximation Haar-RFCs behave like Haar-random matrices, and the non-RMT behaviour we have observed in Haar-RFCs cannot be understood. There is, however, a separate mechanism, additional to the one involving pairing domains, that leads to non-RMT behaviour even from global pairing. To illustrate this mechanism we have introduced a generalised class of RFCs with gates that can be tuned between Haar randomness and the identity. Away from the Haar case, these circuits generate within the diagonal approximation a Thouless time growing logarithmically with system size $L$. The mechanism behind this, observed in Ref.~\cite{kos2018analytic}, is distinct from the domain wall contributions we have investigated here, and the effects are sub-dominant for ensembles in the vicinity of Haar randomness. 

In interacting self-dual models, however, the diagonal approximation to the average SFF is exact. This is because the transfer matrix which generates it has leading eigenvalues equal to unity \cite{bertini2019exact,braun2020transition,flack2020statistics}. Within our framework this implies an infinite domain wall tension, and therefore a vanishing contribution to the SFF from pairs of many-body orbits involving domain walls. Additionally, the subleading eigenvalues of the diagonal propagator in the Ising basis vanish at the self-dual point \cite{kos2018analytic}. The vanishing of both the domain wall tension and the subleading eigenvalues of the diagonal propagator conspire to give an SFF equal to the RMT result. Perturbing away from the self-dual point causes the leading eigenvalues of the transfer matrix to deviate from unity \cite{braun2020transition}, giving rise to behaviour similar to that observed here in Haar-RFCs, for example the exponential growth of the SFF with system size at fixed time.

It is natural to ask what other types of correction to the diagonal approximation can arise in this setting. Important candidates are provided by the class of interference effects studied by Sieber and Richter \cite{sieber2001correlations}. In systems with time-reversal symmetry these are crucial to recover the exact RMT behaviour, while they cancel in systems without time-reversal symmetry. In time-reversal symmetric Floquet quantum circuits these interference effects were studied in Ref.~\cite{kos2018analytic}, where the authors recover both the leading and dominant subleading terms in the expansion $\overline{K}(t) = 2t - 2t^2/t_{\text{H}} + \ldots$ that is expected from RMT for $t \gg t_{\text{Th}}$. The physical mechanism responsible for Sieber-Richter corrections is distinct from the pairing domains which have been our focus. In large systems these effects appear in different time regimes, $t \gg t_{\text{Th}}$ and $t < t_{\text{Th}}$, respectively.

Moving beyond the average SFF, we have initiated a similar investigation into the higher moments (see also Refs.~\cite{chan2020spectral} and \cite{flack2020statistics}). Our discussion highlights another source of deviations from RMT, associated with a freedom in the local pairing of the multiple copies of orbits which appear. The presence of domain walls in the pairing of copies of orbits enhances these higher moments, and consequently gives rise to non-Gaussian statistics of the SFF. We have discussed the connection between these domain walls and the entanglement membrane \cite{jonay2018coarse,zhou2019emergent,zhou2020entanglement}. Although our transfer matrix approach was here restricted to the first moment of the SFF, it is clear that one can construct analogous transfer matrices to generate the higher moments in generic Floquet models. Such an investigation would connect with recent work on rare region effects. Weak links for example have been shown to play an important role in entanglement growth \cite{nahum2018dynamics}. For the average transfer matrix generating the first moment of the SFF, the possibility of weak links would be captured by the decrease of the effective domain wall tension at late times.

Although our work has focused on systems without any locally conserved densities, their introduction is known to lead to prominent deviations of the SFF from RMT even within the diagonal approximation \cite{friedman2019spectral,roy2020random,moudgalya2020spectral}. With a conserved scalar charge, for example, these deviations persist up to a Thouless time which scales diffusively $t_{\text{Th}} \sim L^2$. This is far greater than the Thouless time which we have shown to arise from pairing domains, and it would be interesting to understand the interplay between these effects. There is of course an outstanding question of how to apply our ideas to Hamiltonian systems. In that case the local conservation of energy surely plays a role.

Our approach to studying the average transfer matrix involves imposing local orbit pairings on the doubled time-evolution of the system. Fixing different pairings at the two ends - our twisted boundary conditions - we forced domains walls into the many-body orbit pairing. We have also shown that the untwisted case, where the same pairing is imposed at each end, can be thought of as introducing local couplings to a bath. Applying these same boundary conditions to a many-body localised phase we are faced with its destabilisation \cite{ponte2017thermal}, and this behaviour must be reflected in the transfer matrices. A central question in that context, which we address elsewhere \cite{garratt2020manybody}, concerns the behaviour of the transfer matrix eigenvalues on crossing a phase boundary between an ergodic and a many-body localised phase.

\section*{Acknowledgements}
We thank Adam Nahum, Sid Parameswaran and Nicolas Mac\'{e} for useful discussions. This work was supported in part by EPSRC Grants EP/N01930X/1 and EP/S020527/1. Statement of compliance with EPSRC policy framework on research data: this publication is theoretical work that does not require supporting research data.

\appendix

\section{Numerical Methods}\label{sec:numericalmethods}
Here we outline the numerical methods used throughout the paper. For all numerics we use a local Hilbert space dimension $q=2$, and for all averages over the unitary Haar distribution we use Monte Carlo. We sample this distribution as follows \cite{mezzadri2007howto}. We first generate complex matrices $A$ with independent, unit-normally distributed entries. Performing the QR decomposition $A=QR$ we then define the diagonal matrix $D$ with entries $D_{ii}=R_{ii}/|R_{ii}|$. The unitary matrices $U=QD$ are then Haar-random.

In Fig.~\ref{fig:sff_averaging}(b) we have averaged the SFF over $10^4$ realisations of one gate, and in (c) over $10^6$ realisations of the Floquet operator. Our calculations of the average SFF $\overline{K}(t)$ in Fig.~\ref{fig:sff} use exact diagonalisation (ED) with $10^6$ realisations for $L \leq 8$ and with $10^5$ for $L \geq 9$. These calculations were then used for the analysis in Figs.~\ref{fig:Kopen_eigenvalue_overlap} and \ref{fig:effectivetension} in conjunction with the methods in Appendix \ref{sec:Lscaling}. Our approach for the calculations in Figs.~\ref{fig:domain_wall} and \ref{fig:Ksw} is detailed in Appendix \ref{sec:montecarlopairing}, and the analysis leading to Fig.~\ref{fig:Kw_eigenvalue_overlap} is detailed in Appendix~\ref{sec:Lscaling}.

In Figs.~\ref{fig:purity} and \ref{fig:diagonal_correlator} we have carried out calculations in the quasienergy domain using ED in system sizes $8 \leq L \leq 12$ and using a Lanczos method for $L=14$. The Lanczos method is particularly efficient since the action of the Floquet operator $W$ on many-body states is specified by its definition in terms of local unitary gates. Standard algorithms require Hermitian operators, and so instead of working with $W$ we determine first a set of eigenvectors $\ket{n}$ of $\frac{1}{2}(W+W^{\dag})$, with eigenvalues $\cos \theta_n$. There are no symmetries or degeneracies, so this approach unambiguously determines the eigenstates of $W$. The sign of $\theta_n$ is then determined by acting on the eigenvectors with $\frac{1}{2i}(W-W^{\dag})$. In all system sizes we sample eigenstates of $W$ whose quasienergies reside in bins of fixed width $20(2\pi)/(2^{14})$ centred on $\theta = 2\pi k/50$ for $k=0 \ldots 49$. For each eigenvector $\ket{n}$ we only store its eigenphase $\theta_n$ and the diagonal matrix elements of a complete set of local Hermitian operators $\tau_{x,j}$ (with $x=0 \ldots (L-1)$ and $j=0\ldots (q^2-1)$), $\braket{n|\tau_{x,j}|n}$. Square bins for the sampling of quasienergies implies triangular bins for the sampling of quasienergy differences $\omega=|\theta_n-\theta_m|$. Having fixed the bin widths, the number of circuit realisations we use varies with $L$ in such a way that we have over $10^7$ contributions to each $\text{Tr}[\rho_x(n)\rho_x(m)]$ data point in Fig.~\ref{fig:diagonal_correlator}. This number is based on using $10^3$ realisations of the Floquet operator for $L=14$.

\section{Monte-Carlo pairing}\label{sec:montecarlopairing}

In Sec.~\ref{sec:local} we have imposed local pairings between forward and backward histories, and through this we have calculated $Z(s,t)$, defined in Eq.~\eqref{eq:Zs}. A similar approach was used in the calculation of $Z(+,+,t)$ and $Z(+,-,t)$ in Sec.~\ref{sec:fluctuations}. In this Appendix we describe the numerical method used to impose these local pairings of histories. 

The local diagonal pairing $s$ is represented in the product space of forward and backward histories by the state $\ket{s}$, defined in Eq.~\eqref{eq:spairedstate}. Each of the forward and backward histories is a vector in a space of dimension $q^{2t}$, and so the product space has dimension $q^{4t}$. Since we are interested in late-time behaviour, it is not feasible to perform calculations in this space directly. One alternative, which we do not follow here, is to study the doubled Floquet operator $W \otimes W^*$, which acts on a space of dimension $q^{2L}$. Our Monte-Carlo approach instead requires us only to work with operators acting on a space of dimension $q^L$. We calculate properties of the forward and backward histories independently at first, and impose pairing through averaging.

\begin{figure}
	\includegraphics[width=0.45\textwidth]{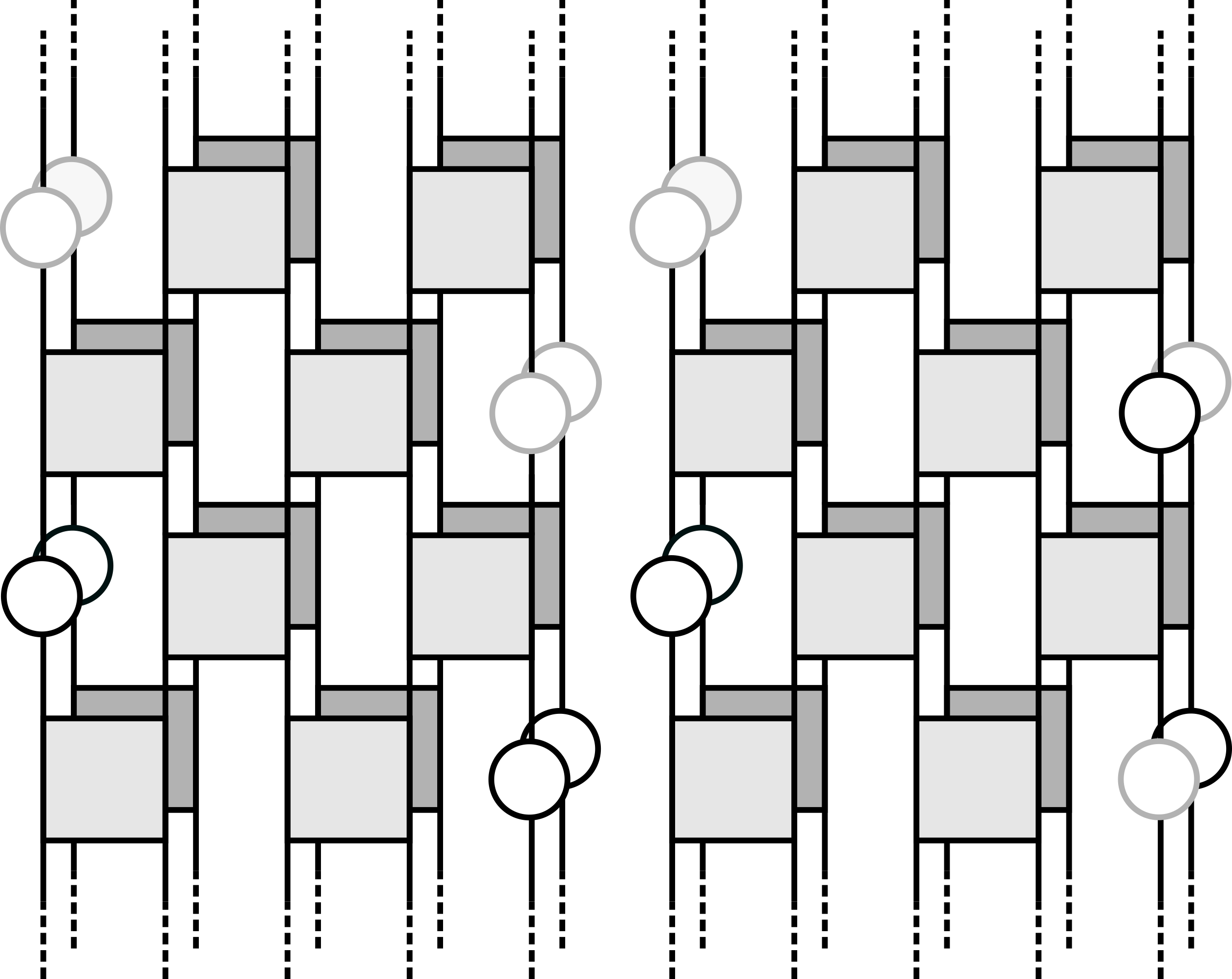}
	\caption{Configuration of single-site Haar-random matrices $u^{(r)}_{L,R}$, drawn as open circles, used to impose the local diagonal pairings at the ends of the system which define $Z(s,t)$ (see Eq.~\eqref{eq:Zsnoaverage}), here for $t=2$. Grey and black circles represent independent matrices, and furthermore the matrices at the left-hand sites and the right-hand sites are independent. Other conventions as in Fig.~\ref{fig:sff_diagram}. Averaging over the single-site matrices in the left-hand diagram fixes $s_L=0$ and $s_R=0$, so we find $Z(0,2)$. In the right-hand diagram this average fixes $s_L=0$ and $s_R=1$, so we find $Z(1,2)$.}
	\label{fig:Zmontecarlo}
\end{figure}

Focussing on a fixed Floquet operator $W$ for a chain with open boundary conditions, we discuss this construction for $Z(s,t)$. At the time step $r$, where $r$ runs from $0$ to $(t-1)$, we act on each of the left-hand and right-hand sites with independent $q \times q$ Haar-random matrices, $u^{(r)}_{L}$ and $u^{(r)}_{R}$, respectively. These matrices are also independent for different steps $r$. The forward evolution operator for step $r$ is then
\begin{align}
	W_f^{(r)} = (u^{(r)}_L \otimes \mathbb{1} \otimes \ldots \otimes \mathbb{1} \otimes u^{(r)}_R)W,
\end{align}
where we act with the identity on each of the central sites. This gives the sum over forward orbits
\begin{align}
	\text{Tr}W_f(t) = \text{Tr}[W_f^{(t-1)} \ldots W_f^{(0)}].
\end{align}
For the backward evolution, at step $r$ we instead act with $u^{(r+s_L)}_L$ on the left-hand site, and $u^{(r+s_R)}_R$ on the right-hand site, where $s_L$ and $s_R$ are integers $0\ldots(t-1)$ and addition is defined modulo $t$. The backward evolution operator for step $r$ is the conjugate of
\begin{align}
	W_b^{(r)} = (u^{(r+s_L)}_L \otimes \mathbb{1} \otimes \ldots \otimes \mathbb{1} \otimes u^{(r+s_R)}_R)W,
\end{align}
and the sum over backward orbits is 
\begin{align}
	\text{Tr}W^*_b(t) = [\text{Tr}[W_b^{(t-1)} \ldots W_b^{(0)}]]^*.
\end{align}
The double sum over forward and backward orbits is now $\text{Tr}W_f(t)\text{Tr}W^*_b(t)$. If we average this double sum over the matrices $u^{(r)}_L$ for example, using
\begin{align}
	\overline{(u^{(r)}_L)_{ab} (u^{(r')}_L)_{a^* b^*}} = \frac{1}{q} \delta^{rr'}\delta_{aa^*}\delta_{bb^*}
\end{align}
we impose the local diagonal orbit pairing $s_L$ at the left end of the system. Similarly an average over $u^{(r)}_R$ imposes the pairing $s_R$ at the right end. This procedure gives $Z(s_R-s_L,t)$. The configurations of single-site gates $u^{(r)}_{L,R}$ used to impose local diagonal pairings are illustrated in Fig.~\ref{fig:Zmontecarlo} for $Z(0,2)$ and $Z(1,2)$.

\begin{figure}
	\includegraphics[width=0.5\textwidth]{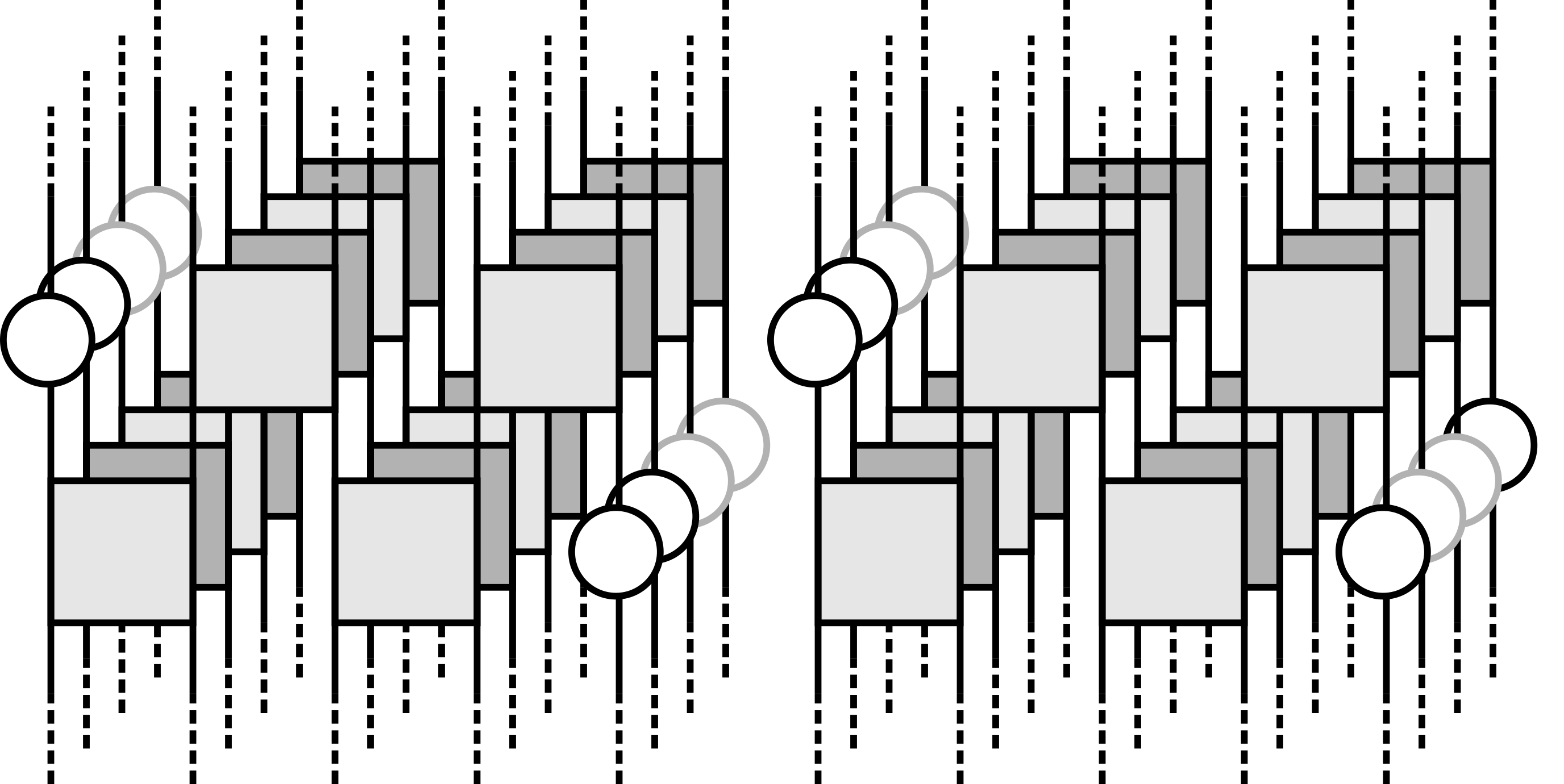}
	\caption{As in Fig.~\ref{fig:Zmontecarlo}, but here to calculate $Z_2(+,+,t)$ (left-hand diagram) and $Z_2(+,-,t)$ (right-hand diagram). Here $t=1$. The two copies of the forward evolution operator are shown light, and the two copies of the backward evolution operator are shown dark. Averaging over the single-site matrices fixes a `$+$' pairing between the copies of the many-body orbits on the left-hand site in both of the diagrams, and also on the right-hand site in the left-hand diagram. In the right-hand diagram, the average imposes a `$-$' pairing on the right-hand site.}
	\label{fig:Z2montecarlo}
\end{figure}

We use a very similar method to calculate $Z_2(+,+,t)$ and $Z_2(+,-,t)$ in Sec.~\ref{sec:fluctuations}. These objects each involve two copies of the sum over forward many-body orbits and two copies of the sum over backward many-body orbits. The different single-site Haar-random unitary matrices are then used to fix local pairings of the multiple copies of the orbits, and their arrangements are illustrated in Fig.~\ref{fig:Z2montecarlo}.

In the calculations of $Z(s,t)$ in Fig.~\ref{fig:Ks_fix_bulk}, where we consider individual realisations of the Floquet operator, we use $10^4$ realisations of the single-site unitary matrices $u^{(r)}_{L,R}$. On the other hand, in the calculations of $\overline{Z}(s,t)$ in Figs.~\ref{fig:domain_wall} and \ref{fig:Ksw}, and of $\overline{Z}_2(+,+,t)$ and $\overline{Z}_2(+,-,t)$ in Fig.~\ref{fig:copy_twist}, we perform a simultaneous Monte-Carlo average over the ensemble of Floquet operators and over the ensemble of matrices $u^{(r)}_{L,R}$. In that case we use $10^6$ independent realisations.

\section{Large-$q$ transfer matrix}\label{sec:largeq}
In this Appendix we discuss the large-$q$ limit of the average transfer matrix $\overline{\mathcal{T}}(t)$, defined in Eq.~\eqref{eq:Tmatrixaverage} in terms of Weingarten functions $\text{Wg}(\sigma \tau^{-1})$ and unnormalised permutation states $\ket{\sigma,\tau}$. 

The Weingarten functions are maximised for $\sigma=\tau$, taking the value $\text{Wg}(\mathbb{1}) = q^{-2t}$ in the large-$q$ limit \cite{samuel1980integrals,brouwer1996diagrammatic}. Furthermore, the states $\ket{\sigma, \tau}$ that maximise $\braket{\sigma,\tau|\mathcal{S}|\sigma,\tau}$ are those in which $\sigma=\tau$ are permutations mapping $r \to r+s$ modulo $t$. Normalising these states, we find $\ket{s}$ defined in Eq.~\eqref{eq:permutation_states}. This motivates the large-$q$ approximation to the average transfer matrix
\begin{align}
	\overline{\mathcal{T}}(t) = \sum_{s=0}^{t-1} \ket{s}\bra{s}.
\label{eq:Tmatrixaverage_largeq}
\end{align} 
The different states $\ket{s}$ correspond to the different local diagonal orbit pairings, and so in the large-$q$ limit the notion of local orbit pairing has a very sharply defined meaning. Note however that the states $\ket{s}$ are not orthogonal. Approximating $\overline{\mathcal{T}}(t)$ by Eq.~\eqref{eq:Tmatrixaverage_largeq} at general values of $q$, it is straightforward to calculate the SFF, and this approach reveals a number of interesting features. 

Note that Eq.~\eqref{eq:Tmatrixaverage_largeq} is the transfer matrix for a clock model: $\braket{s|s'}$ depends only on $|s-s'|$. We first diagonalise $\overline{\mathcal{T}}(t)$ by writing it in terms of the orthogonal (but not normalised) states $\ket{\omega}=(1/\sqrt{t})\sum_{s=0}^{t-1}e^{-i\omega s}\ket{s}$, where $\omega = 2 \pi n /t$ for integer $n=0\ldots (t-1)$. The result is
\begin{align}
	\overline{\mathcal{T}}(t) = \sum_{\omega=0}^{2 \pi (t-1)/t} \ket{\omega}\bra{\omega}.
\end{align}
Within this approximation the norms of the states $\ket{\omega}$ are the eigenvalues of $\overline{\mathcal{T}}(t)$
\begin{align}
	\braket{\omega|\omega} = \frac{1}{t}\sum_{s,s'}e^{i\omega (s-s')}\braket{s|s'}.
\label{eq:LDAeigenvalues}
\end{align}
We will see below that the inner products $\braket{s|s'}$ for $s \neq s'$ decay at least as quickly as $q^{-t}$, and so $\braket{\omega|\omega}$ approaches unity at late times. Also, since $\braket{s|s'}>0$, the largest eigenvalue of $\overline{\mathcal{T}}(t)$ is in the $\omega=0$ sector, as at small $q$ (see for example Fig.~\ref{fig:Kw_eigenvalue_overlap}).

From the eigenvalues $\braket{\omega|\omega}$ we can calculate the average SFF with periodic boundary conditions
\begin{equation}	
	\overline{K}(t) = \sum_{\omega=0}^{2 \pi (t-1)/t} \braket{\omega|\omega}^L.
\label{eq:KpbcLDA}
\end{equation}
With open boundary conditions on the other hand
\begin{align}
	\overline{K}(t) = t \braket{\omega=0|\omega=0}^L.
\label{eq:KobcLDA}
\end{align}
Because the eigenvalues $\braket{\omega|\omega}$ approach unity at late times, we recover the RMT result $\overline{K}(t) = t$. 

The eigenvalues $\braket{\omega|\omega}$ are given by the inner products $\braket{s|s'}$ through Eq.~\eqref{eq:LDAeigenvalues}, and these are
\begin{equation}
	\braket{s|s'} = \frac{1}{q^{2t}} \sum_{a_r b_r \atop a_r^* b_r^*} \prod_{r=0}^{t-1} \delta_{a_r a^*_{r+s}}\delta_{b_r b^*_{r+s}} \delta_{a_r a^*_{r+s'}}\delta_{b_r b^*_{r+s'}}.
\end{equation}
Evaluating the sums on the right-hand side we see that $\braket{s|s'}$ is determined by the number of cycles in the permutation mapping $r \to r+|s-s'|$ mod $t$. Denoting this cycle number $N_{t,|s-s'|}$,
\begin{equation}
	\braket{0|s} = q^{2(N_{t,s}-t)}.
\end{equation}
There is clear $s$ dependence in this expression, and therefore in the domain wall tension and $\overline{Z}(s,t)$ (defined in Eq.~\eqref{eq:Zs}), as observed for $q=2$ in Fig.~\ref{fig:Ksw}(a). This is to be contrasted with the $s$-independent domain wall tension found in the large-$q$ limit in the model of Ref.~\cite{chan2018spectral}. 

There is an interesting difference between the SFF at odd and even times. For an even value of $t$, the largest inner product $\braket{0|s}$ for $s \neq 0$ is $\braket{0|t/2} = q^{-t}$. On the other hand, for $t$ an odd multiple of $3$, the largest are $\braket{0|t/3} = \braket{0|2t/3} = q^{-4t/3}$. The deviations of $\braket{\omega|\omega}$ from unity can be significantly larger for $t$ even than for $t$ odd. From Eqs.~\eqref{eq:KpbcLDA} and \eqref{eq:KobcLDA} we see that this implies larger deviations of the SFF from RMT at even times. We have observed the same behaviour at small $q$ in Fig.~\ref{fig:sff}.

\section{Block-diagonalisation of $\overline{\mathcal{T}}(t)$}\label{sec:blockdiagonalisation}
The $q^{4t} \times q^{4t}$ matrix $\overline{\mathcal{T}}(t)$ has cokernel spanned by the vectors $\ket{\sigma, \tau}$, of which there are $(t!)^2$. For small $t$ it is then feasible to calculate $\overline{\mathcal{T}}(t)$ exactly using Eq.~\eqref{eq:Tmatrixaverage}. In this Appendix we set out the technical details required for this calculation.

The transfer matrix commutes with the full-step time-translation operators acting on each of the forward and backward orbits, $S^2 \otimes \mathbb{1}$ and $\mathbb{1} \otimes S^2$, respectively. Consequently it can be block-diagonalised. On the permutation states $\ket{\sigma,\tau}$ of Eq.~\eqref{eq:permutation_states} these translation operators act as
\begin{align}
\begin{split}
	S^2 \otimes \mathbb{1} \ket{\sigma, \tau} &= \ket{\sigma c, \tau c} \\
	\mathbb{1} \otimes S^2 \ket{\sigma, \tau} &= \ket{c^{-1}\sigma, c^{-1}\tau}.
\end{split}
\end{align}
Here $c$ is the cyclic permutation mapping $r \to (r+1)$ modulo $t$. We also have $\mathcal{S} \ket{\sigma, \tau} = \ket{\tau,c^{-1}\sigma c}$.

Starting from a permutation $\ket{\sigma, \tau}$, the states $(S^{2r_+}~\otimes~S^{2r_-})~\ket{\sigma, \tau}$ for $r_+,r_-=0 \ldots (t-1)$ define an equivalence class of maximum cardinality $t^2$: {$\ket{\sigma, \tau} \sim \ket{\mu, \nu}$} if $\ket{\mu, \nu}~=~(S^{2r_+} \otimes~S^{2r_-})~\ket{\sigma, \tau}$ for some $r_+,r_-$. This equivalence relation partitions the set of $(t!)^2$ permutation states. We label each equivalence class by one of the permutations belonging to it, which we refer to as the root. From here on the states within the class with root $\ket{\sigma, \tau}\equiv \ket{\sigma, \tau; 0,0}$ will be written
\begin{equation}
	\ket{\sigma, \tau; r_+, r_-} = (S^{2r_+} \otimes S^{2r_-})\ket{\sigma, \tau; 0,0}.
\end{equation}
This representation is not necessarily unique. For example $\ket{\mathbb{1},\mathbb{1};r_+ r_-} = \ket{\mathbb{1},\mathbb{1};r_+ +a,r_- +a}$ for integer $a$ with addition defined modulo $t$. Nevertheless, a simplification follows from the fact that if $\ket{\sigma,\tau;r_+,r_-} = \ket{\sigma,\tau;r_+',r_-'}$ then $\ket{\sigma,\tau;r_+ + a,r_- + b} = \ket{\sigma,\tau;r_+' +a,r_-'+b}$. This implies that, if we allow $r_+$, $r_-$ to run from $0\ldots(t-1)$, each state within the class appears the same number of times. We define this number as the multiplicity of the class, $m(\sigma,\tau)$. For example $m(\mathbb{1},\mathbb{1})=t$. 

With this structure in place we can separate the sum in Eq.~\eqref{eq:Tmatrixaverage} into separate sums over roots $\ket{\sigma,\tau;0,0}$ and the states within the corresponding classes. The average transfer matrix Eq.~\eqref{eq:Tmatrixaverage} involves the outer products $\mathcal{S} \ket{\sigma, \tau}\bra{\sigma,\tau}$, and having chosen the roots we can write
\begin{equation}
	\mathcal{S}\ket{\sigma,\tau;0,0} = \ket{\sigma',\tau';a_+(\sigma,\tau),a_-(\sigma,\tau)},
	\label{eq:rootmap}
\end{equation}
where the primed root $\ket{\sigma',\tau';0,0}$ is defined relative to $\ket{\sigma,\tau;0,0}$. The integers $a_+(\sigma,\tau),a_-(\sigma,\tau)$ depend on our initial choice of roots. We are now in a position to write down $\overline{\mathcal{T}}(t)$ in terms of roots and classes. We find
\begin{equation}
	\overline{\mathcal{T}} = \sum_{\sigma \tau} \frac{\text{Wg}(\sigma \tau^{-1})}{m(\sigma \tau)} \sum_{r_+,r_-} \ket{\sigma',\tau';r_+',r_-'}\bra{\sigma,\tau;r_+,r_-},
\end{equation}
where in the summand $r_{\pm}' \equiv r_{\pm}+a_{\pm}(\sigma,\tau)$ and $\sigma',\tau'$ are defined via Eq.~\eqref{eq:rootmap}. Here we have used the fact that the Weingarten function depends only on the conjugacy class of its argument.

To block-diagonalise $\overline{\mathcal{T}}$ we take the Fourier transform within each class,
\begin{align}
	\ket{\sigma,\tau;\omega_+, \omega_-} = \frac{1}{t}\sum_{r_+,r_-=0}^{t-1} & e^{-i(\omega_+ r_+ + \omega_- r_-)} \\ &\times \ket{\sigma,\tau; r_+ r_-}.
\label{eq:classFT}
\end{align}
Inverting this expression we find, for the block $(\omega_+,\omega_-)$,
\begin{align}
\begin{split}
	\overline{\mathcal{T}}(\omega_+,\omega_-) = \sum_{\sigma \tau} &\frac{\text{Wg}(\sigma \tau^{-1})}{m(\sigma \tau)}  e^{-i(\omega_+ a_+ + \omega_- a_-)} \\ & \times  \ket{\sigma',\tau';\omega_+, \omega_-} \bra{\sigma, \tau;\omega_+, \omega_-}.
\end{split}
\end{align}
To find the eigenvalues of $\overline{\mathcal{T}}$ within each sector, we first construct an orthonormal basis. Note that different sectors contain different classes. For example, the states $\ket{\mathbb{1},\mathbb{1};\omega_+,\omega_-}$ exist only in sectors with $\omega_+ +\omega_- =0$. The general procedure for constructing the basis is as follows. First we determine the class inner products
\begin{equation}
	\braket{\sigma,\tau;\omega_+,\omega_-|\sigma',\tau';\omega_+,\omega_-}
\end{equation}
which can be computed using Eq.~\eqref{eq:classFT} and $\braket{\mu, \nu|\sigma, \tau} = \braket{\mu|\sigma}\braket{\nu|\tau}$, where $\braket{\mu|\sigma} = q^{N(\sigma \mu^{-1})}$ with $N(\sigma \mu^{-1})$ the number of cycles in the permutation $\sigma \mu^{-1}$. From these we seek orthogonal (but not yet normalised) basis states
\begin{equation}
	\ket{\tilde i} = \sum_{\sigma \tau} U_{i;\sigma,\tau} \ket{\sigma,\tau;\omega_+,\omega_-},
\end{equation}
with $\braket{\tilde i|\tilde j} = 0$ for $i \neq j$. In each sector $(\omega_+,\omega_-)$ we define a matrix $J$ of class inner products via its components $J_{\sigma,\tau;\mu,\nu}=\braket{\sigma,\tau;\omega_+,\omega_-|\mu,\nu;\omega_+,\omega_-}$. Here $\sigma,\tau$ is a row index and $\mu,\nu$ a column index. Then
\begin{equation}
	\braket{\tilde i|\tilde j} = \sum_{\sigma \tau \atop \mu \nu} U_{i;\sigma,\tau}^*  J_{\sigma,\tau;\mu,\nu} U_{j;\mu,\nu},
\end{equation}
so choosing the rows of $U$ to be eigenvectors of $J$ we have $\braket{\tilde i| \tilde j} = \Lambda_i \delta_{ij}$, where $\Lambda_i$ is the $i^{\text{th}}$ eigenvalue of $J$. Since $\braket{\tilde i|\tilde i} \geq 0$, we expect $\Lambda_i \geq 0$. The orthonormal basis is then defined by $\ket{\tilde i} = \sqrt{\Lambda_i}\ket{i}$. Because $\ket{\sigma,\tau;\omega_+,\omega_-}$ are not all linearly independent in general, some of these eigenvalues will vanish: if $\Lambda_i=0$, then $\ket{\tilde i}$ does not exist.

In terms of the orthonormal basis states $\ket{i}$ we have
\begin{align}
\begin{split}
	\ket{\sigma,\tau;\omega_+,\omega_-} &= \sum_i [U^*]_{i;\sigma,\tau} \sqrt{\Lambda_i} \ket{i}.
\end{split}
\end{align}
The transfer matrix block $\omega_+,\omega_-$ is then
\begin{align}
\begin{split}
	\overline{\mathcal{T}}&(\omega_+,\omega_-) = \sum_{\sigma \tau i j}  \frac{\text{Wg}(\sigma \tau^{-1})}{m(\sigma,\tau)} \sqrt{\Lambda_i \Lambda_j} \\ \times & e^{-i(\omega_+ a_+ + \omega_- a_-)}[U^*]_{i;\sigma',\tau'} [U]_{j;\sigma,\tau} \ket{i}\bra{j}.
\end{split}
\end{align}
The different $\omega_+,\omega_-$ blocks of $\overline{\mathcal{T}}$ can be diagonalised numerically, giving exact results for the eigenvalues and eigenvectors. The leading eigenvalues computed through this method, as well as the overlaps of the corresponding eigenvectors with the local diagonal states, are shown in Fig.~\ref{fig:Kw_eigenvalue_overlap} for $t \leq 5$, and we find excellent agreement with our other approaches.

\begin{figure}[h]
	\hspace{-0.1in}
	\includegraphics[width=0.47\textwidth]{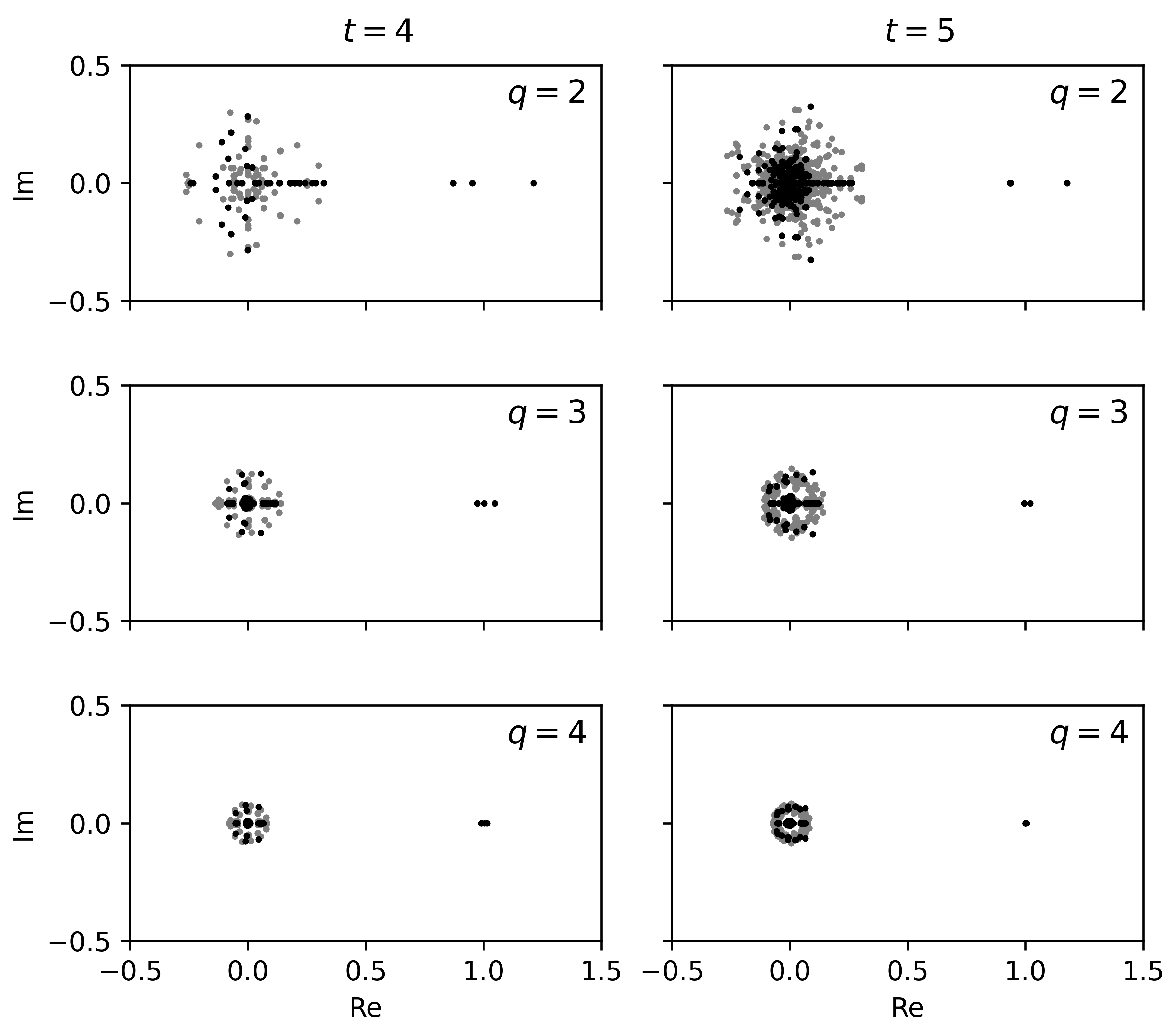}
	\caption{Distributions of eigenvalues of the average transfer matrix $\overline{\mathcal{T}}(t)$ for times $t=4$ and $t=5$, and for $q=2,3,4$, in the complex plane. The black points show eigenvalues in sectors with $\omega_+=-\omega_-$ and the grey points show eigenvalues in all other sectors. The $t$ leading eigenvalues, some degenerate, are distributed around unity.}
	\label{fig:Tsubleading}
\end{figure}

\section{Subleading eigenvalues of $\overline{\mathcal{T}}(t)$}\label{sec:subleading}
In the main text we have focused on the leading eigenvalues and eigenvectors of the average transfer matrix $\overline{\mathcal{T}}(t)$ for $q=2$, which control the domain structure in the orbit pairing. On the other hand, the subleading eigenvalues of $\overline{\mathcal{T}}(t)$ control the behaviour beyond the Heisenberg time $t_{\text{H}}$. Here, using the results of Appendix~\ref{sec:blockdiagonalisation}, we examine some aspects of their distribution.

As we have shown in Fig.~\ref{fig:Kw_eigenvalue_overlap}, the $t$ leading eigenvalues tend to unity at late times. Therefore their contribution to $\overline{K}(t)$, with periodic boundary conditions for example, is simply $t$ at late times. However, for ${t>t_{\text{H}}=q^L}$ the SFF plateaus at $\overline{K}(t)=q^L$. For $t \leq q^L$ the subleading eigenvalues appear to do nothing, whereas for $t=q^L+\delta t$ their contribution to $\overline{K}(t)$ is $-\delta t$.

\begin{figure}
\includegraphics[width=0.45\textwidth]{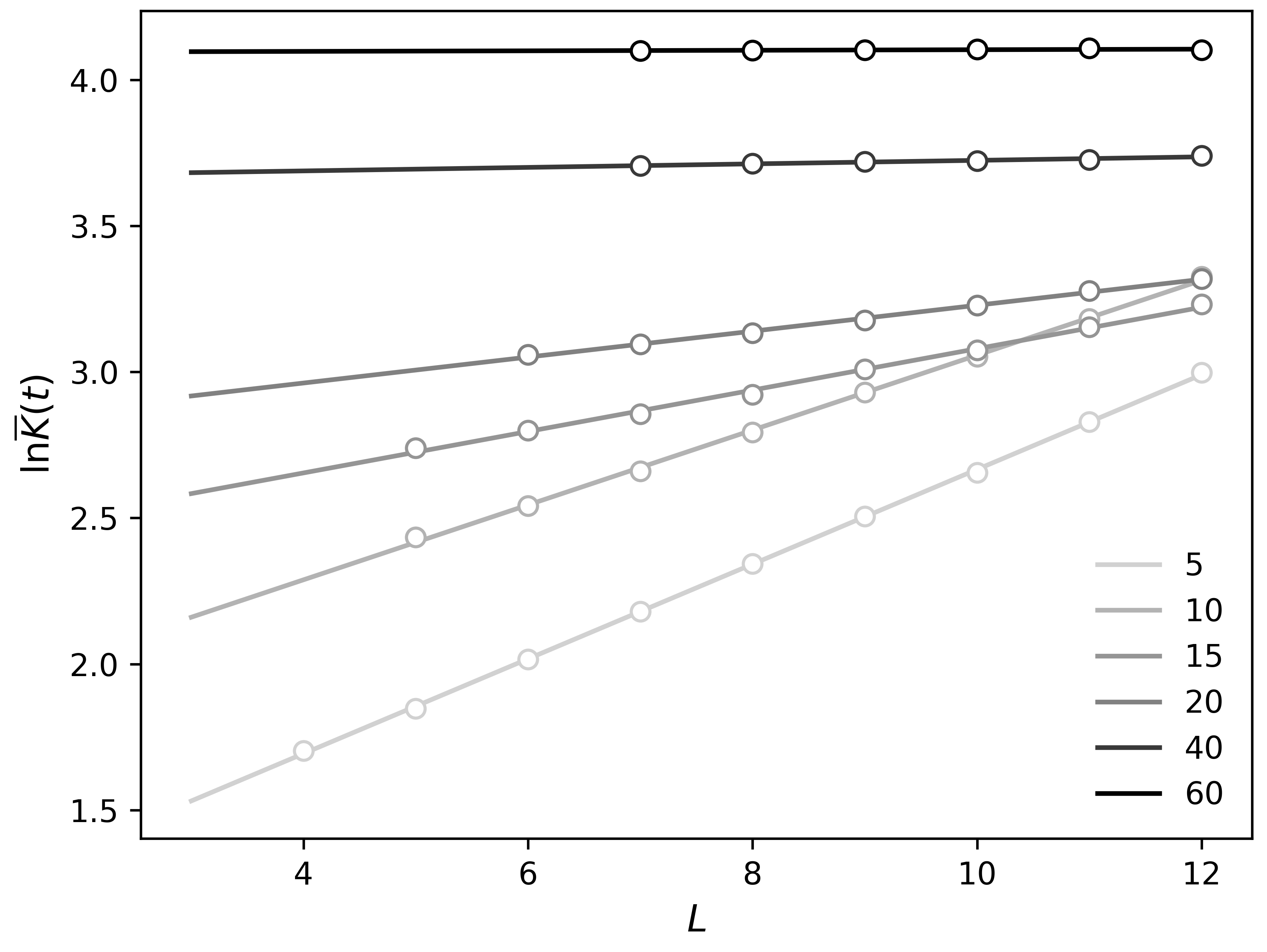}
\caption{Logarithm of the average SFF with open boundary conditions, $\ln\overline{K}(t)$, versus system length $L$ for various times $t$ (legend). From linear fits at each time we extract the leading eigenvalue in the $\omega=0$ sector, $\lambda(0,t)$, and the overlap of the corresponding left- and right-eigenvectors with the boundary states, $\braket{\mathcal{B}_L|0,t;R}\braket{0,t;L|\mathcal{B}_R}$.}
\label{fig:KobcLscaling}
\end{figure}

In practice we have access to all of the eigenvalues of $\overline{\mathcal{T}}(t)$ for $t \leq 5$, so for $q=2$ we can verify the role of the subleading eigenvalues across the Heisenberg time ($t_{\text{H}}=4$ for $L=2$). We indeed find $\overline{K}(4)=\overline{K}(5)=4$. On the other hand for $q \geq  3$, $t_{\text{H}} \geq 9$, so we cannot probe behaviour beyond the Heisenberg time. The eigenvalue distributions for times $t=4$ and $t=5$ are shown in Fig.~\ref{fig:Tsubleading}, for local Hilbert space dimensions $q=2,3,4$. At these times there is a clear gap between the magnitudes of the $t$ leading eigenvalues and the magnitudes of the subleading ones, which becomes more pronounced with increasing $q$.

\section{$L$-scaling}\label{sec:Lscaling}
In this Appendix we describe the methods used to determine the leading eigenvalues of the transfer matrix, as well as the overlaps of the associated eigenvectors with the boundary states $\bra{\mathcal{B}_L}$, $\ket{\mathcal{B}_R}$ and the local diagonal states $\ket{\omega}$.

\begin{figure}
\hspace{-0.1in}
\includegraphics[width=0.47\textwidth]{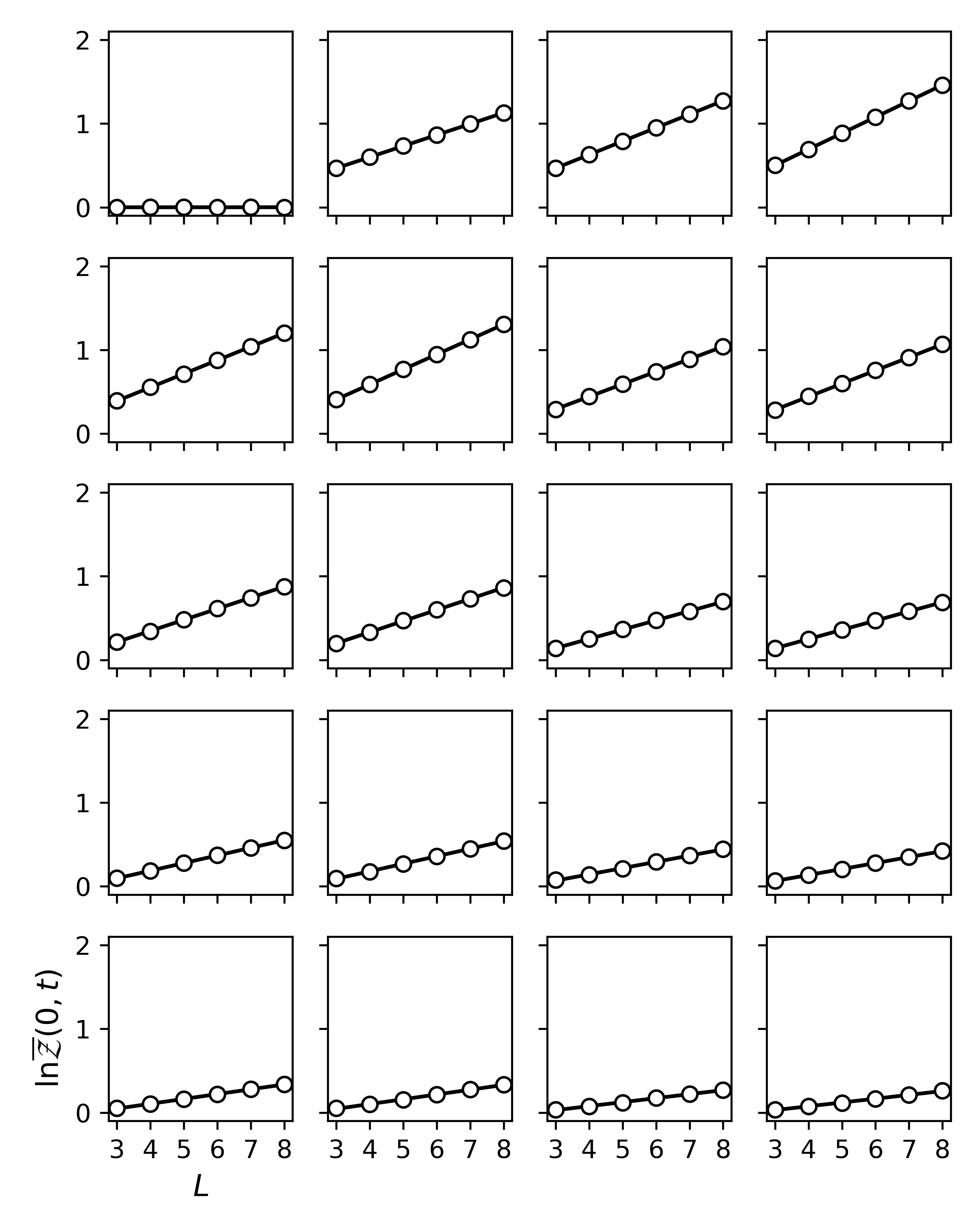}
\caption{Scaling of $\ln \overline{\mathcal{Z}}(\omega,t)$ with $L$ in sector $\omega=0$. The panels correspond to times $t=1\ldots 20$ in reading order. From linear fits we extract the leading eigenvalue $\lambda(\omega;t)$ and the overlaps of the corresponding eigenvectors with the local diagonal states, $\braket{\omega|\omega,t;R}\braket{\omega,t;L|\omega}$, in this sector.}
\label{fig:Zw0_Lscaling}
\hspace{-0.1in}
\includegraphics[width=0.47\textwidth]{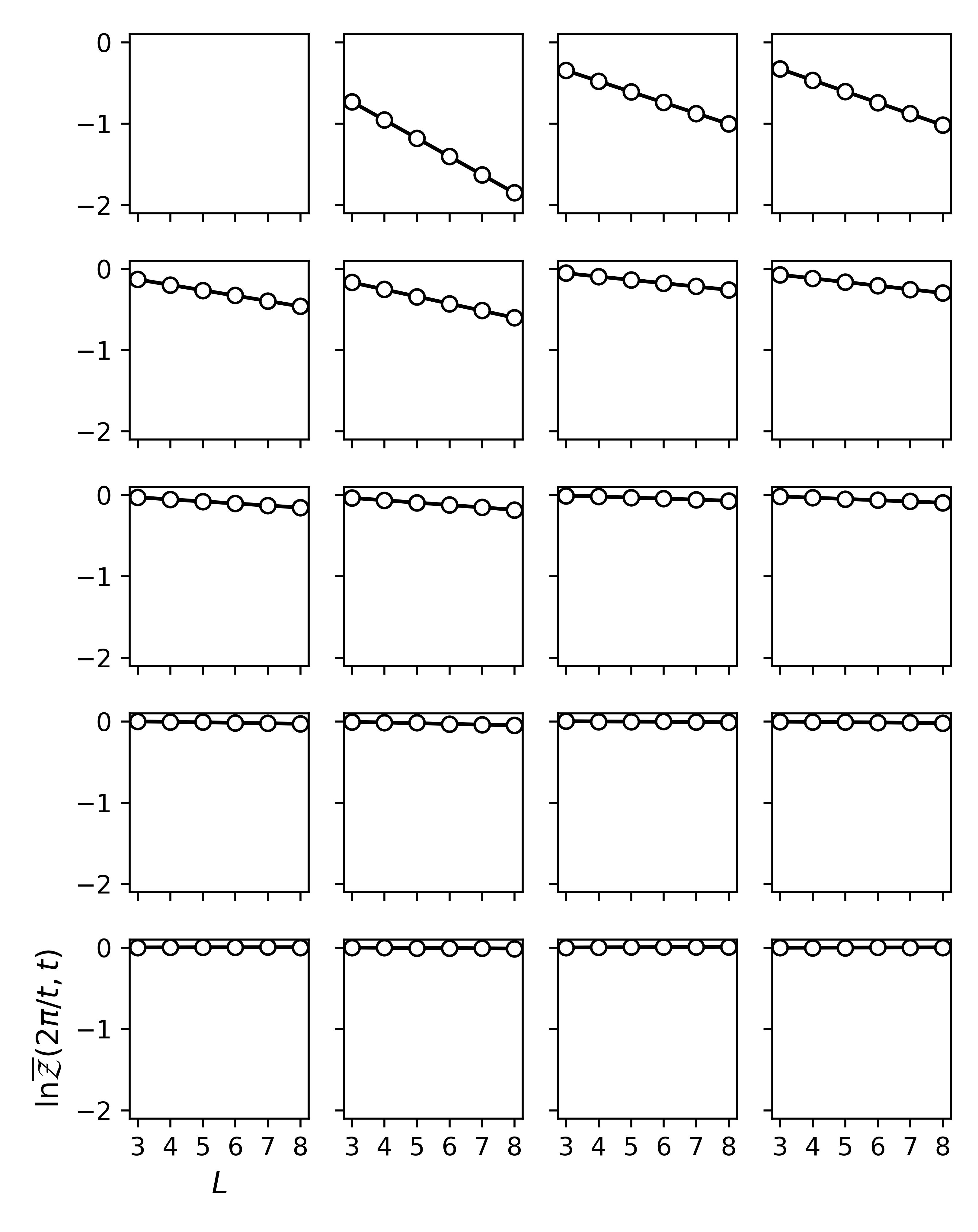}
\caption{Scaling of $\ln \overline{\mathcal{Z}}(\omega,t)$ with $L$ as in Fig.~\ref{fig:Zw0_Lscaling}, here for $\omega=2\pi/t$.}
\label{fig:Zw1_Lscaling}
\end{figure}

In Sec.~\ref{sec:open} we have shown that with open boundary conditions and large $L$ the behaviour of the leading eigenvalue of $\overline{\mathcal{T}}(t)$ in the $\omega=0$ sector, $\lambda(0,t)$, controls the behaviour of the average SFF. Taking the logarithm of Eq.~\eqref{eq:Kobc_leadingterm} we have
\begin{align}
\begin{split}
	\ln \overline{K}(t) \simeq &(L-1)\ln \lambda(0,t) \\ &+ \ln \braket{\mathcal{B}_L|\omega,t;R}\braket{\omega,t;L|\mathcal{B}_R},
\end{split}
\label{eq:logKopen}
\end{align}
where we have neglected contributions from the subleading eigenvalues. Fixing $t$ and varying $L$ we have extracted $\ln \lambda(0,t)$ and $\braket{\mathcal{B}_L|\omega,t;R}\braket{\omega,t;L|\mathcal{B}_R}$ from linear fits to $\ln \overline{K}(t)$ versus $(L-1)$, and the results are shown in Fig.~\ref{fig:Kopen_eigenvalue_overlap}. In practice, for a given $t$ we fit only to data with sufficiently large $L$ that $t<\frac{1}{2}t_{\text{H}}\equiv\frac{1}{2}q^L$. By restricting to $L$ such that $t$ is well below the Heisenberg time, we avoid contributions to $\overline{K}(t)$ from subleading eigenvalues. In Fig.~\ref{fig:KobcLscaling} we show $\ln \overline{K}(t)$ versus $L$ for various $t$, and find excellent linear scaling.

To determine the leading eigenvalues in the $\omega \neq 0$ sectors we study the objects $\overline{\mathcal{Z}}(\omega,t)$ defined in Eq.~\eqref{eq:Kw}. For sufficiently large $L$, Eq.~\eqref{eq:Kw_expansion} gives
\begin{align}
\begin{split}
	\ln \overline{\mathcal{Z}}(\omega,t) \simeq &(L-1)\ln \lambda(\omega,t) \\ &+ \ln \braket{\omega|\omega,t;R}\braket{\omega,t;L|\omega}.
\end{split}
\label{eq:logtwisted}
\end{align}
In Figs.~\ref{fig:Zw0_Lscaling} and \ref{fig:Zw1_Lscaling} we show $\ln \overline{\mathcal{Z}}(\omega,t)$ for the sectors $\omega=0$ and $\omega=2\pi/t$, respectively, as a function of $L$ and for times $1 \leq t \leq 20$. The linear relationship suggested by Eq.~\eqref{eq:logtwisted} holds very well even for systems of only $L=3$ sites (and so only two gates), and from fits we can read off $\lambda(0,t)$ and $\braket{\omega|\omega,t;R}\braket{\omega,t;L|\omega}$. The results are shown in Fig.~\ref{fig:Kw_eigenvalue_overlap}. Fits for other values of $\omega$ are of similar quality.

As we discussed through Secs.~\ref{sec:twisted} and \ref{sec:pairingdomains} the local diagonal pairings $\ket{\omega}$ are closely related to the leading eigenvectors of $\overline{\mathcal{T}}(t)$. Consequently we expect that the contributions of subleading eigenvalues to Eq.~\eqref{eq:logtwisted} are suppressed. Moreover, while there is an abrupt change in the behaviour of $\overline{K}(t)$ at $t=t_{\text{H}}$, this is not the case for $\overline{\mathcal{Z}}(\omega,t)$. For these reasons, in contrast to Fig.~\ref{fig:KobcLscaling}, we include all available data in the fits in Figs.~\ref{fig:Zw0_Lscaling} and \ref{fig:Zw1_Lscaling}. The quality of these fits is evidence that the subleading eigenvalues are playing only a minor role, if any.

\section{Heisenberg interactions}\label{sec:heisenberg}
To investigate the generality of our results, in this section we consider another Floquet model in its ergodic phase. We again use a local Hilbert space dimension $q=2$, but whereas in the main text nearest-neighbour interactions were implemented via $4 \times 4$ Haar-random unitary matrices $U$, here we instead write $U=[B \otimes B']e^{i\pi J\Sigma}[A \otimes A']$, where $A,A',B$ and $B'$ are independent $2\times 2$ Haar-random unitary matrices acting on the individual sites, $\Sigma$ is the two-site swap operator (essentially a Heisenberg coupling), and $J$ is a fixed interaction strength. The model is therefore a kicked spin-half Heisenberg chain with random fields acting at each site, and as in the Haar-random case the Floquet operator does not have time-reversal symmetry. This model is many-body localised for small $J<J_c$, as we have discussed elsewhere \cite{garratt2020manybody}, but here we restrict ourselves to $J=1/4$ deep in the ergodic phase.

\begin{figure}
\hspace{-0.1in}
\includegraphics[width=0.47\textwidth]{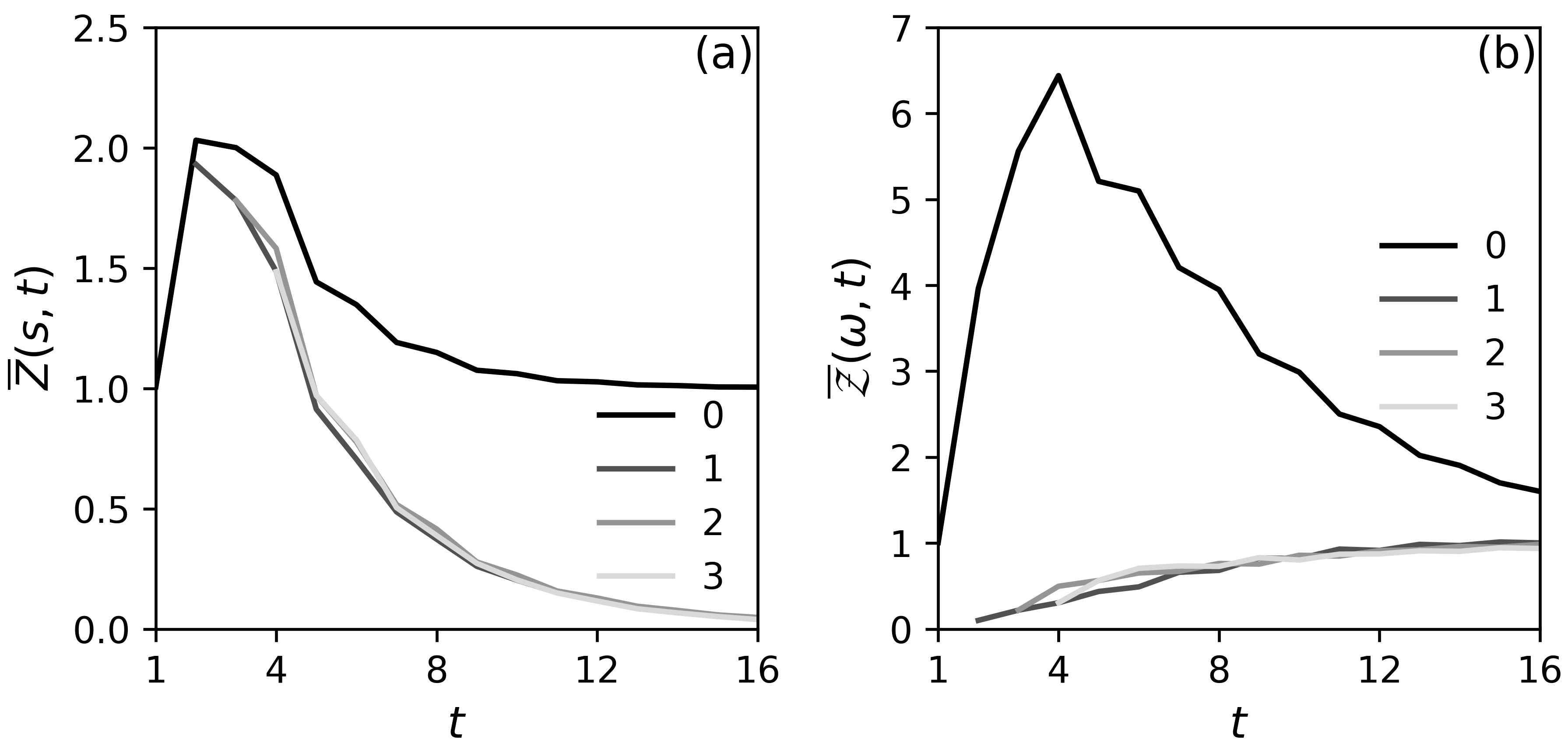}
\caption{(a) $\overline{Z}(s,t)$ and (b) $\overline{\mathcal{Z}}(\omega,t)$ for $L=8$ sites, as in Fig.~\ref{fig:Ksw} but here for the model with Heisenberg interactions, and with $J=1/4$. The legends in (a) and (b) show $s$ and $\omega t/2\pi$, respectively.}
\label{fig:Ksw_swap}
\includegraphics[width=0.47\textwidth]{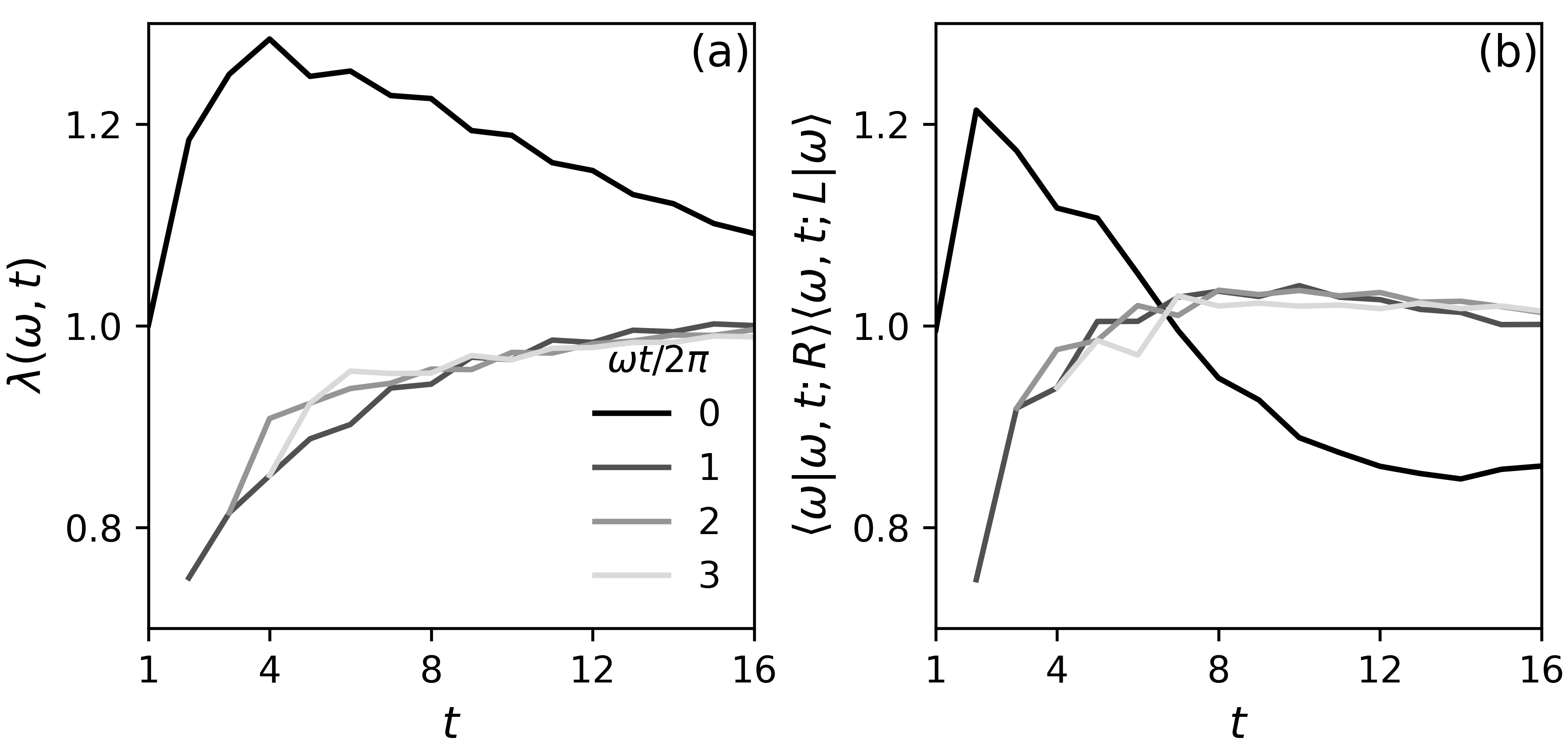}
\caption{(a) Eigenvalues of $\overline{\mathcal{T}}(t)$ and (b) overlaps of leading eigenvectors with paired states, as in Fig.~\ref{fig:Kw_eigenvalue_overlap}, here with Heisenberg interactions and $J=1/4$.}
\label{fig:Kw_eigenvalue_overlap_swap}
\end{figure}

In Fig.~\ref{fig:Ksw_swap} we calculate $\overline{Z}(s,t)$ and $\overline{\mathcal{Z}}(\omega,t)$ for $L=8$ sites and different values of $s$ and $\omega$. As in Fig.~\ref{fig:Ksw}(a) of the main text, $\overline{Z}(0,t)$ approaches unity at late times whereas $\overline{Z}(s \neq 0,t)$ decays to zero. This implies the observed behaviour of $\overline{\mathcal{Z}}(\omega,t)$. From the scaling of $\overline{\mathcal{Z}}(\omega,t)$ with $L$ we then extract the leading eigenvalues of the average transfer matrix $\overline{\mathcal{T}}(t)$ for this model, as well as the overlaps of the corresponding eigenvectors with the paired states $\ket{\omega}$. The results, shown in Fig.~\ref{fig:Kw_eigenvalue_overlap_swap}, are similar to those for the original model in Fig.~\ref{fig:Kw_eigenvalue_overlap}.

\section{Transfer matrix for observables}\label{sec:observable_transfer}

In this Appendix we show how to construct the form factor giving information on correlations between the diagonal matrix elements of a local observable $\tau$, $|\text{Tr}[\tau W(t)]|^2$, in terms of the transfer matrices $\mathcal{T}_{x,x+1}$ which also generate the SFF. The time-domain approach here is complementary to the calculations in the quasienergy domain discussed in Sec.~\ref{sec:observables}.

For a $q \times q$ operator $\tau$ acting only on site $x$, we define $\tilde \tau$ acting on the $q^{2t}$-dimensional space of forward orbits at the site $x$ via its matrix elements
\begin{equation}
	\braket{a_0 b_0 a_1 \ldots |\tilde \tau|a_0' b_0' a_1' \ldots} = \tau_{a_0' a_0} \prod_{r=1}^{t-1} \delta_{a_r a_r'}\delta_{b_r b_r'}.
\end{equation}
For $x$ even we redefine $\tilde \tau \to S \tilde \tau S^{\rm{T}}$. We can then write an expression for $\text{Tr}[\tau W(t)]$ analogous to Eq.~\eqref{eq:TrWtperiodic} or \eqref{eq:TrWtopen}. With periodic boundary conditions and $x=0$ for example
\begin{equation}
	\text{Tr}[\tau W(t)] = \text{tr}[\tilde \tau S \tilde U^{\otimes t}_{0,1} \ldots S \tilde U^{\otimes t}_{L-1,0}(S^{\rm{T}})^L].
\label{eq:TrtauWtperiodic}
\end{equation}
It will be convenient to symmetrise $\tilde \tau$, making use of the cyclic property of the trace ${\text{Tr}[\tau W(t)] = \text{Tr}[W(r)\tau W(t-r)]}$. We therefore introduce
\begin{equation}
	\tilde \tau^{(t)} = \frac{1}{t}\sum_{r=0}^{t-1} S^{2r} \tilde \tau (S^{\rm{T}})^{2r},
\label{eq:tausymmetrised}
\end{equation}
and it can be straightfowardly verified that the right-hand side of Eq.~\eqref{eq:TrtauWtperiodic} is unchanged by the replacement of $\tilde \tau$ by $\tilde \tau^{(t)}$. 

\begin{figure}
\includegraphics[width=0.3\textwidth]{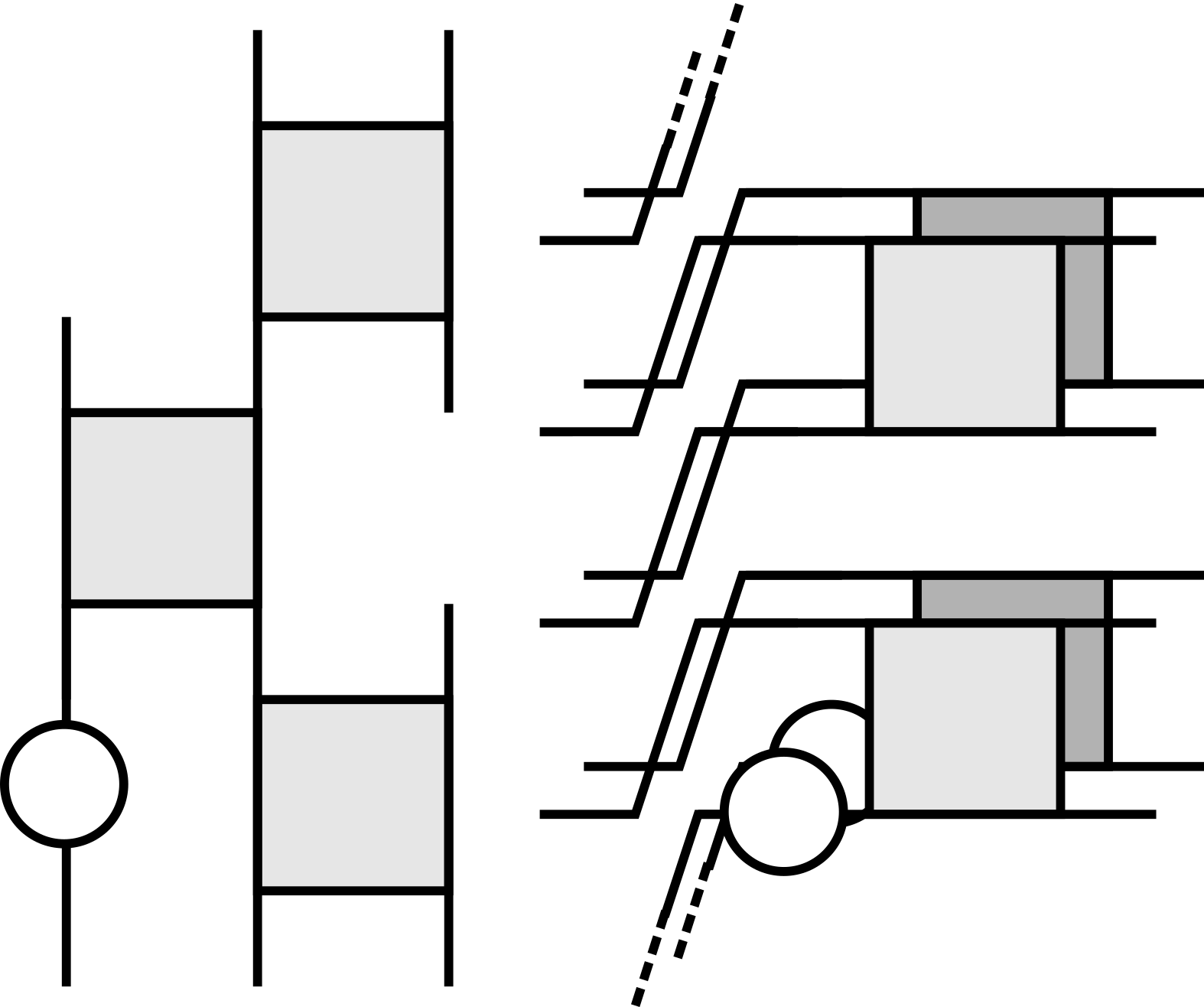}
\caption{The construction of $|\text{Tr}[\tau W(t)]|^2$ for single-site observable $\tau$, represented by an open circle. On the left we show how $\tau$ appears within the circuit, and on the right illustrate how $\tau$ appears in the product of transfer matrices which generates $|\text{Tr}[\tau W(t)]|^2$, here in the case where $\tau$ acts on a site $x$ with $x$ even.}
\label{fig:observable_transfer}
\end{figure}

From Eq.~\eqref{eq:TrtauWtperiodic} and its conjugate, the form factor $|\text{Tr}[\tau W(t)]|^2$ can be constructed as
\begin{align}
\begin{split}
	|\text{Tr}[\tau W(t)]|^2 = \text{tr}[&(\tilde \tau^{(t)} \otimes [\tilde \tau^{(t)}]^*)\mathcal{T}_{0,1} \\ &\ldots \mathcal{T}_{L-1,0} (\mathcal{S}^{\rm{T}})^L], \\
\end{split}
\label{eq:unaveraged_observable_form_factor}
\end{align}
where we have used the symmetrised operator Eq.~\eqref{eq:tausymmetrised}. The appearances of the operator $\tau$ in a circuit, and in a product of transfer matrices, are illustrated in Fig.~\ref{fig:observable_transfer}. The ensemble average of Eq.~\eqref{eq:unaveraged_observable_form_factor} is
\begin{align}
	\overline{|\text{Tr}[\tau W(t)]|^2} &= \text{tr}[(\tilde \tau^{(t)} \otimes [\tilde \tau^{(t)}]^*)\overline{\mathcal{T}}^L (\mathcal{S}^{\rm{T}})^L] \label{eq:observable_form_factor} \\
	&= \sum_{\omega_+ \omega_- \alpha} \lambda^L(\omega_+,\omega_-,\alpha,t)e^{-i(L/2)(\omega_+ + \omega_-)} \notag\\ \times &\braket{\omega_+,\omega_-,\alpha_L;t|(\tilde \tau^{(t)} \otimes [\tilde \tau^{(t)}]^*)|\omega_+,\omega_-,\alpha_R;t}, \notag
\end{align}
where in the second line we have used the spectral decomposition of the average transfer matrix. This result highlights the exponential growth with system length of correlations between the diagonal matrix elements of local observables. 

\bibliographystyle{apsrev4-2}
\bibliography{pairing_bibliography}

\end{document}